\documentclass[11pt]{article}

\usepackage[left=1in,top=1in,right=1in,bottom=1.5in,head=.1in]{geometry}

\usepackage{setspace,url,bm,amsmath} 

\usepackage{titlesec} 
\titlelabel{\thetitle.\quad} 

\usepackage{graphicx} 
\usepackage{bbm}
\usepackage{latexsym}
\usepackage{caption}
\usepackage[margin=20pt]{subcaption}
\usepackage{hyperref, enumerate}
\usepackage[all,import]{xy}
\usepackage{amsfonts}
\usepackage{cite}
 
\usepackage{amssymb}

\usepackage[table]{xcolor}

\newcommand{\GG}[1]{}

\usepackage{amsthm}
\usepackage{comment}
\theoremstyle{definition}

\newtheorem*{theorem*}{Theorem}
\newtheorem{theorem}{Theorem}
\newtheorem{proposition}{Proposition}
\newtheorem{lemma}{Lemma}

\newtheorem{remark}{Remark}

\newtheorem{definition}{Definition}

\newtheorem*{corollary*}{Corollary}

\newtheorem{condition}{Condition}

\usepackage{natbib} 
\bibpunct{(}{)}{;}{a}{}{,} 

\usepackage{etoolbox} 
\apptocmd{\sloppy}{\hbadness 10000\relax}{}{} 

\usepackage{color}
\usepackage{listings}

\DeclareMathOperator*{\argmin}{arg\,min}

\usepackage{algorithm}
\usepackage{algorithmic}

\def\T{{ \mathrm{\scriptscriptstyle T} }}

\def\LPR{\textsc{LPR}}
\def\rBLPR{r\textsc{BLPR}}
\def\pBLPR{p\textsc{BLPR}}
\def\pr{\textnormal{pr}}

\def\ZZ{{LDPE}}
\def\JM{{JM}}

\newcommand\JJL[1]{{#1}}
\newcommand{\JL}[1]{{#1}}
\newcommand{\sign}{\operatorname{sign}}
\newcommand\liu[1]{{#1}}
\newcommand\hanzhong[1]{{#1}}
\newcommand\hanzhongg[1]{{#1}}
\newcommand\hanzhongsep[1]{{#1}}
\newcommand\hanzhongsepsec[1]{{#1}}
\newcommand\revise[1]{{{\color{black} #1}}}
\newcommand\secrevise[1]{{{\color{black} #1}}}
\newcommand\finalrevise[1]{{{\color{black} #1}}}

\DeclareFontFamily{U}{mathx}{\hyphenchar\font45}
\DeclareFontShape{U}{mathx}{m}{n}{
      <5> <6> <7> <8> <9> <10>
      <10.95> <12> <14.4> <17.28> <20.74> <24.88>
      mathx10
      }{}
\DeclareSymbolFont{mathx}{U}{mathx}{m}{n}
\DeclareFontSubstitution{U}{mathx}{m}{n}
\DeclareMathAccent{\widecheck}{0}{mathx}{"71}
\DeclareMathAccent{\wideparen}{0}{mathx}{"75}

\begin{document}

\title{A Bootstrap Lasso + Partial Ridge Method to \\ Construct Confidence Intervals for Parameters in \\ High-dimensional Sparse Linear Models}
\author{
	Hanzhong Liu\thanks{Center for Statistical Science and Department of Industrial Engineering, Tsinghua University, Beijing, 100084, China}
	\and
	Xin Xu\thanks{Department of Statistics, Yale University, New Haven, Connecticut, 06520, U.S.A.}
	\and
	Jingyi Jessica Li\thanks{Department of Statistics, University of California, Los Angeles, California, 90095, U.S.A.}
	 \thanks{Department of Human Genetics, University of California, Los Angeles, California, 90095, U.S.A.}
	 \thanks{To whom correspondence should be addressed. Email: jli@stat.ucla.edu}
}
\maketitle

\begin{abstract}

Constructing confidence intervals for the coefficients of high-dimensional sparse linear models remains a challenge, mainly because of the complicated limiting distributions of the widely used estimators, such as \JL{the lasso}. Several methods have been developed for constructing such intervals. Bootstrap lasso+ols is notable for its technical simplicity, good interpretability, and performance that is comparable with that of other more complicated methods.  However, bootstrap lasso+ols depends on the beta-min assumption, a theoretic criterion that is often violated in practice. Thus, we introduce \JL{a new method, called} bootstrap lasso+partial ridge, to relax this assumption. Lasso+partial ridge is a \JL{two-stage} estimator. First, the lasso is used to select features. Then, the partial ridge is used to refit the coefficients. \JL{Simulation results} show that bootstrap lasso+partial ridge outperforms bootstrap lasso+ols \JL{when there exist} small, but nonzero coefficients\JL{, a common situation that violates the beta-min assumption}. For \JL{such} coefficients, \JL{the confidence intervals constructed using} bootstrap lasso+partial ridge \JL{have}, on average, $50\%$ larger coverage probabilities than those of bootstrap lasso+ols. Bootstrap lasso+partial ridge also has, on average, $35\%$ shorter confidence interval lengths than those of \hanzhong{the de-sparsified lasso methods,} \JL{regardless of whether the linear models are misspecified. Additionally}, we provide theoretical guarantees for bootstrap lasso+partial ridge under appropriate conditions, and implement it in the R package ``HDCI."
\ \\ 

\noindent \textit{ Key words and phrases:}
\noindent Bootstrap, Confidence interval, High-dimensional inference, \JL{Lasso+partial ridge}, Model selection consistency.

\end{abstract}

\section{Introduction}

The proliferation of high-dimensional data in fields such as information technology, astronomy, neuroscience, and bioinformatics has necessitated new analysis methods. Data are high dimensional if the number of predictors $p$ is comparable to, or much larger than, the sample size $n$. Over the past two decades, statistical \JJL{and} machine learning theory, methodologies, and algorithms have been developed to \JL{tackle} high-dimensional data problems under certain \JJL{sparsity} constraints, \JJL{such as} the number of nonzero \JL{linear model coefficients} \JJL{$s$} being much smaller than the sample size \JJL{$n$}. Regularization is required to perform sparse \JJL{estimation} \JL{under} this \JL{regime}. For example, the lasso \citep{Tibshirani1996} uses $l_1$ regularization to perform model selection and parameter estimation simultaneously in a high-dimensional sparse linear regression. Previous works have focused on recovering a sparse parameter vector (denoted by $\beta^0 \in R^p$), based on criteria such as (i) model selection \JJL{consistency}, (ii) the $l_q$ estimation error $||\hat \beta - \beta^0||_q$\JL{,} where $\hat \beta$ is an estimate of $\beta^0$ \JL{and} $q$ \JL{is typically equal} to one or two, and (iii) the prediction error $||X\hat \beta - X\beta^0||_2$, \JL{with} $X$ \JL{as} the design matrix. \JL{The book  \citep{Buhlmannbook} and the review paper \citep{FanLv2010}} give a thorough summary of the recent advances in high-dimensional statistics. 


\JL{An important question in research on high-dimensional statistics is how to perform statistical inference, that is, constructing confidence intervals and hypothesis tests for individual coefficients in linear models}. Inference is crucial when the \JL{purpose of statistical modeling is to understand scientific principles beyond those of prediction. However, inference is difficult for high-dimensional model parameters, because the limiting distributions of the widely used estimators, \revise{such as} the lasso, are complicated and difficult to compute in high dimensions. To address this challenge, we develop a novel and practical inference procedure called bootstrap lasso+partial ridge (LPR)}, which is based on \JL{three} canonical methods\JL{:} the bootstrap, lasso, and ridge. \JL{Before presenting our method, we briefly review several existing high-dimensional inference methods.}

\JL{There is a growing body of statistical literature on high-dimensional inference problems. Existing methods are divided into several categories, including the sample-splitting-based methods, bootstrap-based methods, de-sparsified lasso methods, post-selection inference \JL{methods}, and knockoff filter.} In particular, Wasserman and Roeder proposed \JL{a} sample-splitting method \citep{Wasserman2009} that \JL{splits} $n$ data points into two halves. \JL{The first half is used for model selection (say, by the lasso), and the second half is used to construct confidence intervals or $p$-values for the parameters in the selected model}. For a fixed \JL{dimension} $p$, Minnier et~al. developed a perturbation resampling-based method to approximate the distribution of penalized regression estimates under a general class of loss functions \citep{Minnier2009}. Chatterjee and Lahiri proposed a modified residual bootstrap lasso method \citep{Chatterjee2011}, which is consistent in estimating the limiting distribution of a modified lasso estimator. For \JL{scenarios in which $p$ goes to infinity at a polynomial rate of $n$}, Chatterjee and Lahiri showed that a residual bootstrap adaptive lasso estimator can consistently estimate the limiting distribution of the adaptive lasso estimator under \JL{several} intricate conditions \citep{Chatterjee2012}. \JL{Two of these conditions are similar to the irrepresentible condition and the beta-min condition (the beta-min condition means that the minimum absolute value of the nonzero regression coefficients is much larger than $n^{-1/2}$), which together guarantee the model selection consistency of the lasso}. Liu and Yu proposed another residual bootstrap method based on a two-stage estimator (lasso+ols), showing its consistency under the irrepresentible condition, beta-min condition, and other regularity conditions \citep{Liu2013}. Here, lasso+ols denotes using the lasso method to select a model, and then using the ordinary least squares (OLS) method to refit the coefficients in the selected model. \JL{However, a common issue with these methods is that they all }require the rather restrictive beta-min condition\JL{, which should be relaxed} in high-dimensional inference, if possible.

The de-sparsified lasso, proposed by \cite{Zhang}, and \JL{later} investigated by \cite{Buhlmann2014}, \cite{JM}, is \JL{another type of method. This method aims to} remove the \JL{biases} of \JL{the} lasso \JL{estimates} and produce an asymptotically normal estimate for each parameter. \JL{Specifically, we refer to the popular de-sparsified lasso methods developed by \cite{Zhang} and \cite{JM}} as \ZZ{}  and \JM{}, respectively. These methods \JL{do not rely on} the beta-min condition, but do require that we estimate the precision matrix of predictors using the graphical lasso method \citep{Zhang,Buhlmann2014}, or some other convex optimization procedure \citep{JM}. \revise{There are two main issues with these methods. First, they reply heavily on the sparse linear model assumption and, thus, may exhibit poor performance for misspecified models. Second, the computational costs of these methods are quite high. For example, constructing confidence intervals for all entries of $\beta^0$ requires solving $(p+1)$ separate quadratic optimization problems.} Despite these drawbacks, the methods can serve as a \hanzhong{theoretically proven} benchmark for high-dimensional inference. Other new tools include \JL{the} post-selection inference \JL{methods} \citep{posiBerk,posi}, knockoff filter \citep{Knockoff}, covariance \JL{test} \citep{SignificantTestlasso}, group-bound confidence intervals \citep{Meinshausen2013}, bootstrapping ridge regression \citep{Lopes2014}, \JL{and} ridge projection and bias correction \citep{BuhlmannRidge}, among others; see \cite{Nicolai2014} for a comprehensive review of high-dimensional inference methods. 
	
\hanzhongg{According to the results of simulation studies in \JL{an independent assessment} \citep{Nicolai2014}, \revise{the bootstrap lasso+ols method} produces \JL{confidence intervals with} coverage probabilities and lengths that are comparable \JL{with} those of other existing \JL{methods} when \JL{the} beta-min condition \JL{holds}. \JL{Bootstrap lasso+ols} is built on three \JL{canonical statistical techniques (i.e., the bootstrap, lasso, and OLS), all of which are well known to a broad audience and, hence, easily accessible to data scientists.} However, as mentioned, \JL{the main drawback} of bootstrap lasso+ols is the \JL{rather restrictive} beta-min condition, which results in poor coverage probabilities for the confidence intervals of small, but nonzero coefficients (e.g., $95\%$ confidence intervals with coverage probabilities lower than $50\%$). This is because these small coefficients are seldom selected by the lasso and, hence, are not refitted by the OLS, resulting in coefficient estimates of zero in most bootstrap runs. Therefore, the confidence intervals produced by bootstrap lasso+ols have lengths and coverage probabilities that are close to zero. Intuitively, it seems advantageous to adopt a different second-step procedure after the lasso to replace the OLS. Ideally, this procedure should not place a penalty on the coefficients selected by the lasso, in order to reduce the bias. However, it should place a small, but nonzero $l_2$ penalty on the unselected coefficients in order to recover them. We call this the LPR estimator.} \revise{An independent work by \cite{Gao2017} proposes a post-selection ridge estimator similar to our LPR estimator. However, their aim is to improve the prediction performance, which they do by adding a thresholding step. \cite{Lava} proposes a penalization-based estimation strategy called Lava to deal with ``sparse + dense" coefficients. However, they also focus on improving the prediction performance rather than the quality of the inference.}

\hanzhongg{In this paper, we propose a new inference procedure \JL{called} bootstrap LPR as an improvement over the bootstrap lasso+ols method.} \JL{The problem setting is to construct} confidence intervals for individual regression coefficients $\beta^0_j$, for $j=1,\ldots,p$, in a high-dimensional linear regression model, where $\beta^0$ is weakly sparse \citep{unifiedpaper2009journal}. That is, its elements can be divided into two groups: ``large" coefficients, with absolute values $\gg n^{-\frac{1}{2}}$, and ``small" coefficients, with absolute values $\ll n^{-\frac{1}{2}}$. We define this type of sparsity as the {\em cliff-weak-sparsity}, which means that if we order the absolute coefficients from the largest to the smallest, there exits a cliff-like drop that divides the coefficients into two groups. \JL{Obviously, cliff-weak-sparsity is a weaker assumption than hard (or exact) sparsity ($\beta^0$ has at most $s$ ($s\ll n$) nonzero elements) and the beta-min condition.} 

Inference for small coefficients has been investigated by \cite{Shi2016}, who proposed a two-step inference procedure to identify weak signals (small coefficients). Their method is designed for an orthogonal design matrix, and is based on a combination of the asymptotic normality of a bias-corrected adaptive lasso estimator (for large coefficients) and the least squares estimator (for small coefficients) instead of the bootstrap. However, their method performs well only when $p \ll n$, whereas our method, based on the bootstrap, can be used when $p \gg n$. 

\revise{\cite{boot_ldpe} and \cite{ZhangCheng2017} combine the bootstrap and de-sparsified lasso methods to deal with nonGaussian and heteroscedastic errors. We refer to this method as the bootstrap version of \ZZ{} (BLDPE), and include it in the method comparison in our simulation and real-data studies.}

\textbf{Our contributions} to the literature are summarized as follows:

First, our proposed bootstrap \LPR{} method relaxes the beta-min condition required by the bootstrap lasso+ols method. We provide conditions under which the bootstrap \LPR{} method can consistently estimate the distribution of the \LPR{} estimator and, therefore, is valid for constructing a confidence interval for each coefficient.
	
Second, we conduct comprehensive simulation studies to evaluate the finite-sample performance of the bootstrap \LPR{} method for both sparse linear models and misspecified models. Our main findings are as follows. First, compared with bootstrap lasso+ols, bootstrap \LPR{} improves the coverage probabilities of the $95\%$ confidence intervals by about $50\%$, on average, for small nonzero regression coefficients. However, this improvement incurs a $15\%$ heavier computational burden for $n = 200, \ p = 500$. Second, compared with the two de-sparsified lasso methods, \ZZ{} and \JM{}, bootstrap \LPR{} produces good coverage probabilities for large and small regression coefficients. In some cases, it even outperforms these methods by producing confidence intervals with lengths that are more than $50\%$ shorter, on average. Third, bootstrap \LPR{} is more than $30\%$ faster than the two de-sparsified lasso methods, and is robust to model misspecification. We also demonstrate the performance of bootstrap \LPR{} on \revise{two real data sets: functional magnetic resonance imaging (fMRI) data, and neuroblastoma gene expression data.}


	
Third, we extend the model selection consistency of the lasso from the hard sparsity case \citep{ZhaoYu2006, Wainwright2009} to a more general {\em cliff-weak-sparsity} case. Under the irrepresentable condition and other reasonable conditions, we show that the lasso can correctly select all \JL{``large"} elements of $\beta^0$, while shrinking all ``small" elements to zero.
	
Fourth, we develop \JL{an} R package ``HDCI" to \JL{implement} the bootstrap lasso, bootstrap lasso+ols, and proposed bootstrap \LPR{} methods. This package makes these methods easily accessible to practitioners.

Fifth, our method is not limited to using the lasso in the selection stage, but can be extended to any other model selection criteria, such as stability selection \citep{MeinshausenBuhlmann2010}, the smoothly clipped absolute deviation (SCAD) estimator \citep{FanLi2001}, the Dantzig selector \citep{CandexTao2007}, \revise{and post-double selection \citep{post-double} that does not require the beta-min condition. If we replace the lasso with this method, the resulting confidence intervals may achieve better coverages for medium-sized coefficients. This is an interesting research direction that is worth further investigation, because the methodology, computation, and theory will differ from those of the current work in many respects.}


\revise{ \hanzhongg{The remainder of this paper proceeds as follows. In Section 2, we define the \LPR{} estimator and introduce the residual bootstrap \LPR{} (\rBLPR{}) and \JL{the} paired bootstrap \LPR{} (\pBLPR{}) methods. In Section 3, we investigate \JL{the} theoretical properties of the proposed method. \JL{In Section 4, we} conduct comprehensive simulation studies to compare the finite-sample performance of \rBLPR{}, \pBLPR{}, bootstrap lasso+ols, and three de-sparsified lasso methods (\JL{\ZZ{}, \JM{}}, and BLDPE). \revise{In Sections 5 and 6, we present two real-data case studies}. \JL{Section 7 concludes the paper}. All \JL{relevant proofs and additional simulation results} can be found in the Appendix.} }

\setcounter{equation}{0} 
\section{Framework and definitions}

\subsection{Overview and background}

\hanzhongg{In this section, we begin by introducing \hanzhong{high-dimensional} sparse linear model\JL{s}. We \JL{next define} the cliff-weak-sparsity and the \LPR{} estimator. Finally, we propose two bootstrap procedures \JL{(residual bootstrap and paired bootstrap), based on the \LPR{} estimator}, to construct confidence intervals for \JL{individual} regression coefficients.}  

\JL{We assume that data} are generated from the following \JL{linear model}:
\begin{equation}
 Y = X \beta^0 +\epsilon,
 \label{eqn:lrm}
\end{equation}
where $\epsilon = (\epsilon_1,\ldots,\epsilon_n)^\T$ is a vector of independent and identically distributed (i.i.d.) random error variables, with mean 0 and variance $\sigma^2$\JL{,} $Y=(y_1,\ldots,y_n)^\T \in \mathbb R^n$ is an $n$-dimensional response vector, and $X = (x_1,\ldots,x_n)^\T = (X_1,\ldots,X_p) \in \mathbb R^{n \times p}$ is a deterministic or random design matrix. \hanzhongg{\JL{Without loss of generality, we assume that every} predictor is centered, that is, $\sum_{i=1}^n x_{ij}/n =0,$ for  $j=1,\ldots,p$, and there is no intercept term in the \JL{linear} model.} \JL{Denoting $\beta^0\in \mathbb R^p$ as a vector of coefficients, we assume} that $\beta^0$ satisfies the cliff-weak-sparsity.
\begin{definition}[Cliff-weak-sparsity] $\beta^0$ satisfies the cliff-weak-sparsity if its elements can be divided into two groups. The first group has $s$ $(s \ll n)$ large elements, with absolute values much larger than $n^{-1/2}$, and the second group contains $p-s$ small elements, with absolute values much smaller than  $n^{-1/2}$.
\label{def:cliffweak}
\end{definition}

\hanzhongg{We are interested in constructing a confidence interval for each coefficient $\beta^0_j$, $j=1,\ldots,p$.} We consider the high-dimensional setting where both $p$ and $s$ grow with $n$. Here, and in what follows, $Y$, $X$, and $\beta^0$ are all indexed by $n$, but we omit the index $n$ whenever this does not cause confusion.


The lasso estimator \citep{Tibshirani1996} \hanzhongg{is a useful tool for enforcing sparsity when estimating high-dimensional parameters. The estimator is} defined as follows:
\begin{equation}
 \hat \beta_{\textnormal{lasso}} = \argmin \limits_{\beta} \left\{ || {Y - X\beta } ||_2^2/(2n)  + \lambda_{1} || {\beta } ||_1 \right\},
 \label{eqn:lasso}
\end{equation}
where $\lambda_{1} \geq 0$ is the tuning parameter controlling the amount of regularization applied to the estimate. In general, $ \lambda_1$ depends on $n$, but we omit this dependence in the notation, for simplicity. \hanzhongg{The limiting distribution of the lasso is complicated \citep{KnightFu2000}, and the usual residual bootstrap lasso fails to construct valid} confidence intervals \citep{Chatterjee2010}. \revise{Various modifications have been proposed to form a valid inference procedure, but these \JL{rely} on two restrictive assumptions: the hard sparsity and beta-min condition.} In order to \JL{relax} these two \JL{often} unrealistic assumptions, we propose the \LPR{} estimator with two associated bootstrap procedures (the \rBLPR{} and \pBLPR{}).


\subsection{The \LPR{} estimator}

\hanzhongg{In this \JL{subsection}, we first \JL{describe the rationale of} the \LPR{} estimator and then formally define \JL{it.} \JL{We argue that this \LPR{} estimator is useful for} weakly sparse linear models, \JL{the} coefficients of which have many small, but nonzero elements decaying at a certain rate, \JL{satisfying} the {\em cliff-weak-sparsity}.}

In case of  the {\em cliff-weak-sparsity}, existing bootstrap methods, such as bootstrap lasso+ols, give very poor coverage probabilities for the small, but nonzero regression coefficients \revise{because they are seldom selected by the lasso. Hence, a large fraction of the bootstrap lasso+ols estimates are zero, producing zero-length noncoverage confidence intervals, such as $[0,0]$. To fix this problem, we need to increase the variance of our estimates for small coefficients with corresponding predictors that are missed by the lasso. This is the motivation for the \LPR{} estimator proposed in this paper.}

The \LPR{} estimator is a two-stage estimator. It adopts \JL{the} lasso to select the predictors, and then \hanzhongg{refits the coefficients using \JL{the} partial ridge. The latter is defined to minimize the empirical $l_2$ loss with no penalty \JL{on the selected predictors, but with an $l_2$ penalty on the unselected predictors. This reduces the bias of the coefficient estimates of the selected predictors, while increasing the variance of the coefficient estimates of the unselected predictors}.} \liu{The $l_2$ penalty (as used in a ridge regression \citep{Hoerl1970}) is used because it regularizes the coefficient estimates without imposing sparsity.} Formally, \hanzhongsepsec{let $S = \{ j \in \{1,\ldots,p\}:\ \beta^0_j \neq 0 \}$ be the support set of $\beta^0$}, and let $\hat S = \{ j \in \{1,\ldots,p\}:\ (\hat \beta_{\textnormal{lasso}})_j \neq 0 \}$ be the set of predictors selected by the lasso. Then, we define the \LPR{} estimator as
\begin{equation}
 \hat \beta_{\textnormal{\LPR{}}} = \argmin\limits_{\beta}  \left\{ \frac{1}{2n}  || {Y - X\beta } ||_2^2 + \frac{\lambda_{2}}{2} \sum_{j \notin  \hat S} \beta_j^2 \right\}.
 \label{eqn:lasso-partial-ridge}
\end{equation} Here, $\lambda_2$ is a tuning parameter that, in general, depends on $n$, but we omit the dependence in the notation, for simplicity. Our simulations in Section \ref{sec:simulation} show that fixing $\lambda_{2}$ at $O(1/n)$ works quite well for a range of error variance levels. For the sake of simplicity, we set $\lambda_2 = 1/n$, with the understanding that further research should be done on the selection of $\lambda_{2}$.

\hanzhongg{In the next two \JL{subsections}, we discuss two commonly used bootstrap procedures} \JL{for} the \LPR{} estimator, and \JL{explain} how to \JL{use them to} construct a confidence interval for each coefficient, respectively.

\subsection{The \rBLPR{} method}

For a deterministic design matrix $X$ in a linear regression model, the residual bootstrap is a standard method \JL{used to construct} confidence intervals. \hanzhongg{In this \JL{subsection}, we introduce the \rBLPR{} procedure.}


\hanzhongg{We first need to appropriately define residuals \JL{so that their} empirical distribution can \JL{well} approximate the true distribution of the error, $\epsilon_i$. In a high-dimensional linear regression, there are multiple ways to obtain residuals\JL{. For example, we can calculate} the residuals} using estimation methods such as \JL{the} lasso, lasso+ols, and \LPR{}. Simulations suggest that the residuals obtained from the lasso+ols approximate the true distribution of $\epsilon_i$ best
and, hence, are adopted in this study. Note that, when the beta-min condition is not satisfied, lasso+ols could fail to select all nonzero coefficients correctly. That is, it is not consistent for model selection, but its prediction performance could still be good (i.e., it has a smaller mean squared error than that of the lasso). Let $\hat \beta_{\textnormal{lasso+ols}}$ denote the lasso+ols estimator, 
\begin{equation}
 \hat \beta_{\textnormal{lasso+ols}} = \argmin\limits_{\beta:\  \beta_{\hat S^c}=0}  \left\{ \frac{1}{2n}  || {Y - X\beta } ||_2^2  \right\}, \ \textnormal{where},  \ \beta_{\hat{S}^c} = \{\beta_j: j \not\in \hat{S}\}.
 \label{eqn:lasso+ols}
\end{equation}

The residual vector is defined as $ \hat \epsilon = (\hat \epsilon_1,\ldots,\hat \epsilon_n)^\T = Y - X \hat \beta_{\textnormal{lasso+ols}}. $
Consider the centered residuals at the mean $\{\hat \epsilon_i - \tilde{\epsilon},\ i=1,\ldots,n\}$, where $\tilde{\epsilon} =  \sum_{i=1}^{n} {\hat \epsilon_i}/n$. For the residual bootstrap, we obtain $\epsilon^*=(\epsilon_1^*,\ldots,\epsilon_n^*)^\T$ by resampling, with replacement, from the centered residuals $\{\hat \epsilon_i - \tilde{\epsilon},\ i=1,\ldots,n\}$, and then construct the residual bootstrap (``rboot") version of $Y$:
\begin{equation}
 Y^*_{\textnormal{rboot}} = X \hat \beta_{\textnormal{lasso+ols}} + \epsilon^*.
 \label{eqn:resampleY}
\end{equation}
Then, based on the residual bootstrap sample $(X,Y^*_{\textnormal{rboot}})$, we can compute the residual bootstrap lasso (rBlasso) estimator $\hat \beta^*_{\textnormal{rBlasso}}$, as in \eqref{eqn:res-boot-lasso} (replacing $Y$ in equation \eqref{eqn:lasso} with $Y^*_{\textnormal{rboot}}$), and its selected predictor set $\hat S^*_{\textnormal{rBlasso}}= \left\{ j \in \{1,\ldots,p \right \}:\ (\hat \beta^*_{\textnormal{rBlasso}})_j \neq 0 \}$. We can also compute the \rBLPR{} estimator $\hat \beta^*_{\textnormal{\rBLPR{}}}$, as in \eqref{eqn:res-boot-lasso-partial-ridge}, in the same way as in equation~\eqref{eqn:lasso-partial-ridge}, except that we replace $Y,\hat S$ with $Y^*_{\textnormal{rboot}},\hat S^*_{\textnormal{rBlasso}}$, respectively:
\begin{equation}
\hat \beta^*_{\textnormal{rBlasso}} = \argmin\limits_{\beta} \left\{ \frac{1}{2n} || {Y^*_{\textnormal{rboot}} - X\beta } ||_2^2  + \lambda_{1} || {\beta } ||_1 \right\},
\label{eqn:res-boot-lasso}
\end{equation}
\begin{equation}
\hat \beta^*_{\textnormal{\rBLPR{}}} = \argmin\limits_{\beta}  \left\{ \frac{1}{2n}  || {Y^*_{\textnormal{rboot}} - X\beta } ||_2^2 + \frac{\lambda_{2}}{2} \sum_{j \notin  \hat S^*_{\textnormal{rBlasso}}} \beta_j^2 \right\}.
\label{eqn:res-boot-lasso-partial-ridge}
\end{equation}
If the conditional distribution (given $\epsilon$) of $T_n^*= \sqrt{n}(\hat \beta^*_{\textnormal{\rBLPR{}}}- \hat \beta_{\textnormal{lasso+ols}})$ from the bootstrap is a good approximation of the distribution of $T_n = \sqrt{n}(\hat \beta_{\textnormal{\LPR{}}}-\beta^0)$, then we can use the residual bootstrap to construct asymptotically valid confidence intervals; see Algorithm \ref{alg:residual-boot} for the complete procedure. 

\begin{algorithm}
\caption{\hspace{0.2cm} Residual Bootstrap LPR (rBLPR) procedure for confidence interval construction}
\label{alg:residual-boot}
\begin{algorithmic}[1]\vspace{0.2cm}
\REQUIRE
    Data $(X,Y)$; Confidence level $1-\alpha$; Number of replications $B$. \vspace{0.2cm}
\ENSURE
    Confidence interval $[l_j, u_j]$ of $\beta^0_j$, for $j=1,\ldots,p$. \vspace{0.2cm}
\STATE Compute the Lasso+OLS estimator $\hat \beta_{\textnormal{Lasso+OLS}}$, given data $(X,Y)$;
\STATE Compute residual vector $\hat \epsilon = (\hat \epsilon_1,\ldots,\hat \epsilon_n)^T = Y - X \hat \beta_{\textnormal{Lasso+OLS}}$;
\STATE Re-sample from the empirical distribution of the centered residual $\{\hat \epsilon_i - \bar{\hat\epsilon}, i=1,\ldots,n\}$, where $\bar{\hat\epsilon} = \frac{1}{n} \sum\limits_{i=1}^{n} {\hat \epsilon_i}$, to form $\epsilon^*=(\epsilon_1^*,\ldots,\epsilon_n^*)^T$;
\STATE Generate residual Bootstrap response $Y^*_{\textnormal{rboot}} = X \hat \beta_{\textnormal{Lasso+OLS}} + \epsilon^*$;
\STATE Compute the residual Bootstrap LPR, $\hat \beta_{\textnormal{rBLPR}}^*$, based on  $(X,Y^*_{\textnormal{rboot}})$ as in equations \eqref{eqn:res-boot-lasso} and \eqref{eqn:res-boot-lasso-partial-ridge};
\STATE Repeat steps 3-5 for $B$ times, and obtain $\hat \beta_{\textnormal{rBLPR}}^{*(1)},\ldots,\hat \beta_{\textnormal{rBLPR}}^{*(B)}$;
\STATE For each $j=1,\ldots,p$, compute the $\alpha/2$ and $1-\alpha/2$ quantiles of $\left\{ ( \hat \beta^{*(b)}_{\textnormal{rBLPR}})_j  \right\}_{b=1}^{B}$, and denote them as $a_j$ and $b_j$, respectively; let $l_j= (\hat \beta_{\textnormal{LPR}})_j + (\hat \beta_{\textnormal{Lasso+OLS}})_j - b_j$ and $u_j= (\hat \beta_{\textnormal{LPR}})_j + (\hat \beta_{\textnormal{Lasso+OLS}})_j - a_j$;
\RETURN $1-\alpha$ confidence interval $[l_j, u_j]$, for $j=1,\ldots,p$.
\end{algorithmic}
\end{algorithm}

\subsection{The \pBLPR{} method}

\hanzhongg{In this \JL{subsection}, we introduce \JL{the} \pBLPR{} procedure.} Paired bootstraps are widely used  \JL{for the inference in linear models}. In this procedure, we generate a resample $\left\{ (x_i^*,y_i^*), i=1,\ldots,n \right\}$ from the empirical distribution of $\{ (x_i,y_i), i=1,\ldots,n \}$, and then compute the paired bootstrap lasso (pBlasso) estimator
\begin{equation}
 \hat \beta^*_{\textnormal{pBlasso}} = \argmin\limits_{\beta} \left\{ \frac{1}{2n} || {Y^*_{\textnormal{pboot}} - X^*_{\textnormal{pboot}}\beta } ||_2^2  + \lambda_{1} || {\beta } ||_1 \right\},
 \label{eqn:paired-boot-lasso}
\end{equation}
where $Y^*_{\textnormal{pboot}}=(y^*_1,\ldots,y^*_n)^\T$ and $X^*_{\textnormal{pboot}} = \left(x^*_1,\ldots,x^*_n \right)^\T$ \hanzhongg{\JL{denote} the paired bootstrap samples}. Let $\hat S^*_{\textnormal{pBlasso}} = \{ j \in \{1,\ldots,p\}:\ (\hat \beta^*_{\textnormal{pBlasso}})_j \neq 0 \}$ be the set of predictors selected by the paired bootstrap lasso, and define the \pBLPR{} estimator as
\begin{equation}
 \hat \beta^*_{\textnormal{\pBLPR{}}} = \argmin\limits_{\beta}  \left\{ \frac{1}{2n}  || {Y^*_{\textnormal{pboot}} - X^*_{\textnormal{pboot}}\beta } ||_2^2 + \frac{\lambda_{2}}{2} \sum_{j \notin  \hat S^*_{\textnormal{pBlasso}}} \beta_j^2 \right\}.
 \label{eqn:paired-boot-lasso-partial-ridge}
\end{equation}

\hanzhongg{\JL{The \pBLPR{} procedure for constructing confidence intervals} is summarized in \JL{Algorithm} \ref{alg:paired-boot}.}

\begin{algorithm}
\caption{\hspace{0.2cm} Paired Bootstrap LPR (pBLPR) procedure for confidence interval construction}
\label{alg:paired-boot}
\begin{algorithmic}[1]\vspace{0.2cm}
\REQUIRE
    Data $(X,Y)$; Confidence level $1-\alpha$; Number of replications $B$. \vspace{0.2cm}
\ENSURE
    Confidence interval $[l_j, u_j]$ of $\beta^0_j$, for $j=1,\ldots,p$. \vspace{0.2cm}
\STATE Generate a Bootstrap sample $(X^*_{\textnormal{pboot}},Y^*_{\textnormal{pboot}})=\left \{ (x_i^*,y_i^*), i=1,\ldots,n \right\}$ from the empirical distribution of $\{ (x_i,y_i), i=1,\ldots,n \}$; \label{code:fram:extract}
\STATE Based on $(X^*_{\textnormal{pboot}},Y^*_{\textnormal{pboot}})$, compute the paired Bootstrap Lasso estimator, $\hat \beta^*_{\textnormal{pBLasso}}$, as in equation \eqref{eqn:paired-boot-lasso} and its selected predictor set, $\hat S^*_{\textnormal{pBLasso}}$; and then compute the paired Bootstrap LPR estimator, $\hat \beta^*_{\textnormal{pBLPR}}$, as in equation \eqref{eqn:paired-boot-lasso-partial-ridge};
\STATE Repeat steps 1-2 for $B$ times and obtain $\hat \beta_{\textnormal{pBLPR}}^{*(1)},\ldots,\hat \beta_{\textnormal{pBLPR}}^{*(B)}$;
\STATE For each $j=1,\ldots,p$, compute the $\alpha/2$ and $1-\alpha/2$ quantiles of $\left\{ (\hat \beta^{*(b)}_{\textnormal{pBLPR}})_j  \right\}_{b=1}^{B}$, and denote them as $l_j$ and $u_j$, respectively;
\RETURN $1-\alpha$ confidence interval $[l_j, u_j]$, for $j=1,\ldots,p$.
\end{algorithmic}
\end{algorithm}


\setcounter{equation}{0} 
\section{Theoretical results}

\subsection{Overview}

In this section, we investigate the theoretical properties of the \rBLPR{} method. In particular, we first show that, under the cliff-weak-sparsity and other reasonable conditions, the lasso exhibits model selection consistency, in the sense that it correctly identifies all large components of $\beta^0$, while shrinking all small components to zero; see Theorem~\ref{thm:selection-consistency-lasso}. \hanzhongg{\JL{Second}, and} more interestingly, we show \hanzhongg{in Theorem~\ref{thm:selection-consistency-bootlasso-resid}} that, under one further condition, the residual bootstrap lasso estimator \JL{achieves the same} kind of model selection consistency. Based on these properties, we provide the conditions under which the limiting distribution of $\sqrt{n} u^\T T_n^*= \sqrt{n}u^\T (\hat \beta^*_{\textnormal{\rBLPR{}}}- \hat \beta_{\textnormal{lasso+ols}})$, conditional on $\epsilon$, is the same as the (unconditional) limiting distribution  of $\sqrt{n} u^\T T_n = \sqrt{n}u^\T(\hat \beta_{\textnormal{\LPR{}}}-\beta^0)$, for a general class of $u \in R^p$; see Theorem~\ref{thm:valid-resid-boot}.

\subsection{Model selection consistency of the lasso under the {\em cliff-weak-sparsity}}

\hanzhongg{In this \JL{subsection}, we extend the model selection consistency of the lasso from \JL{the} hard sparsity case to the \JL{more general} {\em cliff-weak-sparsity} case, where $\beta^0$ has many small, \liu{but nonzero} elements.}

\liu{\citep{ZhaoYu2006, Wainwright2009} showed that the lasso is sign-consistent (i.e., $\pr(\sign(\hat \beta_{\textnormal{lasso}})= \sign(\beta^0) ) \rightarrow 1$ as $ n\rightarrow \infty$, which implies model selection consistency) under appropriate conditions, including the irrepresentable condition, beta-min condition, and hard sparsity.}

\begin{definition}[\cite{ZhaoYu2006}]
	If an estimator $\hat\beta$ is equal in sign to the true $\beta^0$, we write $\hat\beta =_s \beta^0$, which is equivalent to $\sign(\hat\beta) = \sign(\beta^0)$, where $\sign(\cdot)$ maps positive entries to one, negative entries to -1, and zero entries to zero.
\end{definition}

We extend this result to the {\em cliff-weak-sparsity} case. Without loss of generality, we assume $\beta^0=(\beta^0_1,\ldots,\beta^0_s,\beta^0_{s+1},\ldots,\beta^0_p)$, with $\beta^0_j \gg n^{-1/2}$ for $j=1,\ldots,s$, and $\beta^0_j \ll n^{-1/2} $ for $j=s+1,\ldots,p$. Let $S=\{1,\ldots,s\}$ and $\beta^0_S = (\beta^0_1,\ldots,\beta^0_s)$. Assuming the columns of $X$ are ordered in accordance with the components of $\beta^0$, we write $X_S$ and $X_{S^c}$ as the first $s$ and the last $p-s$ columns of $X$, respectively. Let $C= X^\T X/n$, which can be expressed in block-wise form, with four blocks, $C_{11}=X_S^\T X_S/n$, $C_{12}=X_S^\T X_{S^c}/n$, $C_{21}=X_{S^c}^\T X_S/n$, and $C_{22}=X_{S^c}^\T X_{S^c}/n$. Let $\Lambda_{\min} (A)$ and $\Lambda_{\max} (A)$ denote the smallest and largest eigenvalues of a matrix $A$. To obtain model selection consistency, we require the following assumptions:

\begin{condition}
	\label{cond:subgaussian}
	 $\epsilon_i$ are i.i.d. sub-Gaussian random variables.
\end{condition}

\begin{condition}
	\label{cond:standard}
	The predictors are standardized, that is,
	\begin{equation*}
	\frac{1}{n} \sum\limits_{i=1}^{n} {x_{ij}} = 0, \quad \ \frac{1}{n} \sum\limits_{i=1}^{n} {x_{ij}^2}=1, \quad j=1,\ldots,p.
	\label{eqn:standardized}
	\end{equation*}
\end{condition}

\begin{condition}
	\label{cond:smallesteigen}
	There exists a constant $\Lambda>0$, such that $\Lambda_{\min} (C_{11})  \geq \Lambda.$
\end{condition}

Conditions \ref{cond:subgaussian} and \ref{cond:standard} are fairly standard in the sparse linear regression literature; see, for example, \citep{ZhaoYu2006,Huang2008,HuangZhang2008}. Theorems \ref{thm:selection-consistency-lasso},  \ref{thm:selection-consistency-bootlasso-resid}, and \ref{thm:valid-resid-boot} hold if we replace Condition \ref{cond:standard} with a bounded second-moment condition. However, to simplify our argument, we use Condition \ref{cond:standard}. Condition \ref{cond:smallesteigen} ensures that the smallest eigenvalue of $C_{11}$ is bounded away from zero, such that its inverse behaves well.

\begin{condition}
	\label{cond:scaling}
	There exist constants $0< c_1<1$ and $0<c_2<1-c_1$, such that
	\begin{equation}
	s=s_n=O(n^{c_1}) \ , \ p=p_n = O(e^{n^{c_2}}).
	\label{eqn:sparsity}
	\end{equation}
\end{condition}

\begin{condition}[Irrepresentable condition \citep{ZhaoYu2006}]
	\label{cond:irrepresentable}
	There exists a constant vector $\eta$ with entries in $(0,1]$, such that
	$
	|C_{21}C_{11}^{-1}\textnormal{sign}(\beta^0_S)|\leq \mathbf{1}-\eta,
	$
	where $\mathbf{1}$ is a $(p-s) \times 1$ vector with entries equal to one, and the inequality holds, element-wise.
\end{condition}

\begin{remark}
	The irrepresentable condition is implied by the slightly stronger condition, $ |C_{21}C_{11}^{-1}|\leq \mathbf{1}-\eta. $ This condition basically imposes a regularization constraint on the regression coefficients of the unimportant covariates (with small coefficients) on the important covariates (with large coefficients): the absolute value of any unimportant covariate's regression coefficient, represented by the important covariates, is strictly smaller than one. \hanzhongsep{This condition can be weakened if we use other model selection criteria, such as stability selection.}
\end{remark}

\begin{condition}
	\label{cond:cliff-weak}
	There exist constants $c_1+c_2<c_3\leq 1$ and $M>0$, such that
	\begin{equation}
	n^{\frac{1-c_3}{2}} \mathop {\min}\limits_{1\leq i \leq s}|\beta^0_i| \geq M; \quad  n^{\frac{1+c_1}{2}} \mathop {\max}\limits_{s < j \leq p }|\beta^0_j| \leq M.
	\label{eqn:beta-min}
	\end{equation}
\end{condition}

\begin{condition}
	\label{cond:lambda1}
	There exists a constant $c_4$ $(c_2<c_4<c_3-c_1)$, such that the tuning parameter $\lambda_{1}$ in the definition of the lasso in equation \eqref{eqn:lasso} satisfies $\lambda_{1}  \propto n^{(c_4-1)/2}$. \hanzhongsepsec{Based on empirical evidence from the simulation results (see subsection 4.2), we assume  the tuning parameter $\lambda_2 \propto n^{-1}$}. 
	\end{condition}

\begin{condition}
	\label{cond:extra}
	Let $c_4$ be the constant defined in Condition \ref{cond:lambda1}, and suppose that	
	\begin{equation}
	||\sqrt{n} C_{11}^{-1} C_{12} \beta^0_{S^c} ||_\infty = O(1); \ || \sqrt{n} ( C_{21}C_{11}^{-1}C_{12} - C_{22} ) \beta^0_{S^c} ||_\infty = o(n^{\frac{c_4}{2}}).
	\label{eqn:beta2}
	\end{equation}
	
\end{condition}

Condition \ref{cond:scaling} implies that both the number of larger components of $\beta^0$ (i.e., $s$) and the number of predictors (i.e., $p$) diverge with the sample size $n$. In particular, $s$ is allowed to diverge much more slowly than $n$, and $p$ can grow much faster than $n$ (up to exponentially fast), which is standard in almost all of the high-dimensional inference literature. Although this assumption is stronger than the typical one $(s \log p) /n \rightarrow 0$, it has been used in previous works \citep{ZhaoYu2006}. Condition \ref{cond:cliff-weak} is the cliff-weak-sparsity assumption on $\beta^0$, which allows the existence of small, but nonzero coefficients, and is thus weaker than the hard sparsity and beta-min condition. Conditions \ref{cond:subgaussian}--\ref{cond:irrepresentable}, the first half of the \hanzhongsepsec{statement of Condition \ref{cond:cliff-weak} on $\mathop {\min}_{1\leq i \leq s}|\beta^0_i|$, and the first half of the statement of Condition \ref{cond:lambda1} on $\lambda_1$} are the same as those used in \citep{ZhaoYu2006} to show the sign-consistency of the lasso. Condition \ref{cond:extra} is a technical assumption stating that the projection of small effects (i.e., $X_{S^c}\beta^0_{S^c}$) onto the linear subspace spanned by the predictors corresponding to the large coefficients (i.e., the predictors in $S$) \hanzhongsepsec{decays} at a certain rate. In the Appendix, we present examples where this condition holds. Conditions \ref{cond:subgaussian}--\ref{cond:irrepresentable} and \ref{cond:lambda1} are also assumed in \citep{Liu2013} to show the validity of the residual bootstrap lasso+ols.

\hanzhongsepsec{An interesting fact is} that both the lasso and the residual bootstrap lasso are model selection consistent under the cliff-weak-sparsity, and appropriate conditions.

\begin{theorem}\label{thm:selection-consistency-lasso}
Under Conditions \ref{cond:subgaussian} -- \ref{cond:extra}, we have
\[  \pr \left( (\hat \beta_{\textnormal{lasso}})_S =_s \beta^0_S,\ (\hat \beta_{\textnormal{lasso}})_{S^c}=\textbf{0} \right) = 1 - o(e^{-n^{c_2}}) \rightarrow 1  \ \ \textnormal{as} \ \ n\rightarrow \infty.  \]
\end{theorem}

\begin{remark}
	Theorem~\ref{thm:selection-consistency-lasso} shows that, under suitable conditions, the probability that the lasso correctly identifies the large coefficients of $\beta^0$, while shrinking the small ones to zero, goes to one at an exponential rate. This is a natural generalization of the sign consistency of the lasso from the hard sparsity to the cliff-weak-sparsity. We adopt the analytical techniques in \citep{ZhaoYu2006}, with necessary modifications to account for the cliff-weak-sparsity. The proof  \hanzhongsepsec{is provided} in the Appendix.
\end{remark}

\subsection{Weak convergence of the \rBLPR{} method}


\begin{condition}
	\label{cond:s-square}
	The number of large coefficients $s$ satisfies $s^2/n \rightarrow 0.$
\end{condition}

\begin{condition}
	\label{cond:s-square-1}
	There exists a constant $D>0$, such that
	\begin{equation}
	\label{cond:x-S}
	\mathop {\max}\limits_{1\leq i \leq n}||x_{i,S}||_2^2 = o(\sqrt{n}); \ \mathop {\max}\limits_{1\leq i \leq n}|x_{i,S^c}^\T \beta^0_{S^c}| < D,\  \textnormal{where}, \ x_{i,S} = (x_{i1},\ldots,x_{is})^\T. \nonumber
	\end{equation}
\end{condition}

\hanzhongsepsec{Condition \ref{cond:s-square} is stronger than Condition \ref{cond:scaling} because it requires $0<c_1< 1/2$.} Without considering model selection, \cite{BickelFreedman1983} showed that a residual bootstrap OLS fails if $p^2/n$ does not tend to zero. Therefore, Condition \ref{cond:s-square} cannot be weakened easily. This condition is weaker than $(s \log p)/\sqrt{n} \rightarrow 0$, as required by the de-sparsified lasso \citep{Zhang,Buhlmann2014,JM}. The first part of Condition \ref{cond:s-square-1} is not very restrictive, because the length of the vector $x_{i,S}$ is $s\ll \sqrt{n}$, and it holds, for example, when the predictors corresponding to the large coefficients are bounded by a constant $M$; that is, $|x_{ij}|\leq M$, for $\ i=1,\ldots,n, \ j=1,\ldots,s$. This condition is also assumed in \citep{Huang2008} to obtain the asymptotic normality of the bridge estimator. The second part of Condition \ref{cond:s-square-1} assumes that the small effects, $\{x_{i,S^c}^\T \beta^0_{S^c}, i=1,\ldots,n\}$, are bounded from above by a constant. 

Theorem \ref{thm:selection-consistency-bootlasso-resid} shows that the residual bootstrap lasso estimator also has sign-consistency under the cliff-weak-sparsity and other appropriate conditions. The proof of this theorem is given in the Appendix.

\begin{theorem}\label{thm:selection-consistency-bootlasso-resid}
Under Conditions \ref{cond:subgaussian} -- \ref{cond:s-square-1} , the residual bootstrap lasso estimator has the sign-consistency; that is,
\[  \pr\left( (\hat \beta^*_{\textnormal{rBlasso}})_S =_s \beta^0_S,\ (\hat \beta^*_{\textnormal{rBlasso}})_{S^c}=\textbf{0}  \mid \epsilon \right) = 1 - o_p(e^{-n^{c_2}}).  \]
\end{theorem}

\hanzhongsep{
\begin{remark}
By Theorem~\ref{thm:selection-consistency-bootlasso-resid}, the residual bootstrap lasso correctly identifies the large coefficients and shrinks the small ones to zero, with probability approaching one. The proposed bootstrap \LPR{} method uses the partial ridge regression to recover these small, but nonzero coefficients.
\end{remark}
}

\hanzhongsepsec{Using Theorems \ref{thm:selection-consistency-lasso} and \ref{thm:selection-consistency-bootlasso-resid} and Condition~\ref{cond:orthogonal}, we can show that the \rBLPR{} procedure can consistently estimate the distribution of $\hat \beta_{\textnormal{\LPR{}}}$ and, thus, construct asymptotically valid confidence intervals for the regression coefficients $\beta^0$.}

Let $I$ be a $(p-s) \times (p-s)$ identity matrix, and denote the matrix $C_{\lambda_2}$ as
\begin{equation}       
C_{\lambda_2}= \left(            
\begin{array}{ll}   …  C_{11} & C_{12} \\… C_{21} & C_{22} + \lambda_2 I \\
\end{array}
\right) .                
\end{equation}

\begin{condition}
	\label{cond:orthogonal}
	Let $u \in R^p$ be a fixed vector, with $||u||_2=1$. Assume that
	 $$ 
	 \sigma_1^2 = \lim_{n \rightarrow \infty} \left(  u^\T  C_{\lambda_2}^{-1}   C  (C_{\lambda_2}^{-1})^\T u \right) \sigma^2 < \infty,
	 $$
	\begin{equation}
	\label{eqn:small-effect}
	\max \left\{ \left( \beta^0_{ S^c} \right)^\T C_{22}  \left( \beta^0_{ S^c} \right),  \ \max_{1 \leq k \leq n}  \frac{ \left|  u^\T C_{\lambda_2}^{-1} x_k  \right| }{\sqrt{n}}, \  \frac{  u^\T C_{\lambda_2}^{-1} \left( 
	 \textbf{0}^\T, 
	 \left(\beta^0_{ S^c} \right)^\T 
	 \right)^\T } {  \sqrt{n} } \right\} = o(1). \nonumber
	\end{equation}
\end{condition}

\begin{remark}
The first statement $ \left( \beta^0_{ S^c} \right)^\T C_{22}  \left( \beta^0_{ S^c} \right) = o(1)$ is used to guarantee that the conditional variance of $\epsilon_i^*$, \hanzhongsepsec{given $\epsilon$}, converges to $\sigma^2$, the variance of $\epsilon_i$ and, \hanzhongsepsec{thus}, the conditional distribution of $\epsilon_i^*$ is a valid approximation of the distribution of $\epsilon_i$. \revise{The other two statements  are a Linderberg-type condition and a technical condition, respectively, used to obtain asymptotic normality.}
\end{remark}
\begin{remark}
For an orthogonal design matrix (i.e., $(1/n)X^\T X = I$), in which there are no correlations between predictors and $p\leq n$, $\sigma_1^2 = \sigma^2$, and Condition \ref{cond:orthogonal} reduces to the following, much simpler form:
$ \mathop {\max}\limits_{1\leq k \leq n} |u^\T X_k| = o(\sqrt{n}) $. 
\revise{When $u=e_j$, a basis vector with the $j$th element equal to one and other elements equal to zero, this condition is equivalent to $\mathop {\max}_{1\leq k \leq n} |x_{kj}|  = o(\sqrt{n})$, which is not a strong condition, and is expected to hold in many practical situations. The conclusion is still true when the correlation between two covariates satisfies cor$(X_i, X_j) = \rho^{| i - j | }$, with $\rho < 1/5$ (see Section S3 for more detail).} 
\end{remark}

\begin{theorem}\label{thm:valid-resid-boot}
Under Conditions \ref{cond:subgaussian} -- \ref{cond:orthogonal},  we have
\begin{equation}\label{theorem:boot}
 \sqrt{n}u^\T(\hat \beta_{\textnormal{\LPR{}}}-\beta^0) = U + o_p(1); \quad \sqrt{n}u^\T(\hat \beta^*_{\textnormal{\rBLPR{}}}- \hat \beta_{\textnormal{lasso+ols}})  = U^* + o_p(1). \nonumber
\end{equation}
Both $U$ and $(U^* \mid \epsilon)$ converge in distribution to the normal distribution  $N(0,  \sigma_1^2)$.
\end{theorem}

\begin{remark} Theorem \ref{thm:valid-resid-boot} shows that, under appropriate conditions, the limiting distribution of \hanzhongsepsec{$\sqrt{n}u^\T(\hat \beta^*_{\textnormal{\rBLPR{}}}- \hat \beta_{\textnormal{lasso+ols}})$, conditional on $\epsilon$, is the same as the (unconditional) limiting distribution  of $\sqrt{n}u^\T(\hat \beta_{\textnormal{\LPR{}}}-\beta^0)  $. Thus, the unknown distribution of $\sqrt{n}u^\T(\hat \beta_{\textnormal{\LPR{}}}-\beta^0)  $ can be approximated by the conditional distribution of  $\sqrt{n}u^\T(\hat \beta^*_{\textnormal{\rBLPR{}}}- \hat \beta_{\textnormal{lasso+ols}})$, which can be estimated using the bootstrap. Based on the estimated conditional distribution of $\sqrt{n}u^\T(\hat \beta^*_{\textnormal{\rBLPR{}}}- \hat \beta_{\textnormal{lasso+ols}})$} , we can construct asymptotically valid confidence intervals for the linear combination $u^\T \beta^0$. \hanzhongg{Specifically, by setting $u = e_j$, we can construct an asymptotically valid confidence interval for an individual coefficient $\beta_j^0$.}
\end{remark}

We can also show the model selection consistency of the paired bootstrap lasso estimator (similar to Theorem \ref{thm:selection-consistency-bootlasso-resid}). However, even in the orthogonal design matrix case, the design matrix $X^*$ of the paired bootstrap samples is no longer orthogonal, making \JL{the components of the \pBLPR{} estimates, $(\hat\beta^*_{\textnormal{\pBLPR{}}} )_S$ and $(\hat\beta^*_{\textnormal{\pBLPR{}}})_{S^c}$, dependent on each other and, thus, have complicated forms. Hence, it becomes difficult} to verify the convergence property of the \pBLPR{} estimator using techniques similar to those used to prove Theorem \ref{thm:valid-resid-boot} for \hanzhongsepsec{the \rBLPR{}} estimator. Our simulation studies in the following section indicate that the \pBLPR{} method can work as well as the \rBLPR{} method. We leave the theoretical analysis of the \pBLPR{} method to future research.

\section{Simulation studies}\label{sec:simulation}


We perform simulation studies to evaluate the finite-sample performance of two bootstrap \LPR{} methods, \rBLPR{} and \pBLPR{}. We compare our method with the bootstrap lasso+ols method \revise{ and three de-sparsified lasso methods (\ZZ{}, \JM{}, and BLDPE)} in terms of their coverage probabilities and confidence interval lengths. The main conclusions are summarized as follows:


(1) Setting $\lambda_2 = O(1/n)$ works well for a wide range of noise levels.

(2) \revise{\pBLPR{} is slightly better than \rBLPR{}, in most cases}. 


(3) Under the setting of normal design matrices, bootstrap lasso+ols has the shortest confidence interval lengths, with good coverage probabilities for large coefficients. However, for small, but nonzero coefficients, \rBLPR{} and \pBLPR{} have the shortest confidence interval lengths, with good coverage probabilities.


(4) \ZZ{} and \JM{} are more \JL{robust to low signal-to-noise ratios (SNRs),} whereas \JL{\rBLPR{} and \pBLPR{} do} not \JL{perform well} when the \JL{SNRs are low, that is, no greater than one. This is mainly because} the lasso cannot select \JL{all} of the important predictors correctly. The \JL{\rBLPR{} and \pBLPR{} methods produce} much better confidence intervals when the \JL{SNRs are high, that is, larger} than five\JL{: with comparable coverage probabilities, their interval lengths are $50\%$ shorter than those of \ZZ{} and \JM{}, on average}.

 (5) \JL{With regard to the point estimates of the linear model coefficients, the \LPR{} estimator has smaller biases for most coefficients than those of \ZZ{} and \JM{}. However, its standard deviations are larger than those of \ZZ{} and \JM{} for large coefficients, and are smaller for small coefficients. Overall, its root mean squared errors (RMSEs) are $60\%$ smaller than those of \ZZ{}, but $42\%$ larger than those of \JM{}.}	

(6) When the predictors are generated from \JL{a Student's} $t$ distribution with two degrees of freedom, the methods all fail \JL{to produce valid confidence intervals}.  \JL{New statistical techniques are needed for inference} in this case.	


(7) Our \rBLPR{} and \pBLPR{} methods are robust to model misspecification, and the \JL{confidence intervals constructed using our methods are more than $50\%$ shorter, on average, than those produced by \ZZ{} and \JM{}.}

\revise{(8) BLDPE has the best coverage probabilities of the considered methods. Its confidence interval lengths are close to the better ones of LDPE and JM, but are still larger than those of \pBLPR{} and \rBLPR{}.}


\revise{\hanzhongg{The simulation section is organized as follows.\JL{ Subsection 4.1 introduces} the simulation setups. \JL{Subsection} 4.2 studies the impact of the partial ridge tuning parameter $\lambda_2$ on the coverage probabilities and \JL{the} mean interval lengths of \JL{the confidence intervals constructed by} the \rBLPR{} and \pBLPR{} methods. \JL{In Subsection 4.3}, we compare the performance of the \rBLPR{} and \pBLPR{} methods with that of the bootstrap lasso+ols method. \JL{Subsection} 4.4 presents the comparison results of \rBLPR{}, \pBLPR{}\JL{, \ZZ{}, \JM{}}, and BLDPE. We investigate the robustness of the \rBLPR{} and \pBLPR{} methods by \JL{varying signal-to-noise ratios} in Subsection 4.5. \JL{In Subsection 4.6, we present the comparison results of different methods under a misspecified model.}}}

\subsection{Simulation setups}
\label{sec:simu_setups}

We use R package ``glmnet" to compute the lasso solution path and select the tuning parameter, $\lambda_{1}$, by 5-fold Cross Validation cv(lasso+ols); see Algorithm~\ref{alg:cv} in the Appendix for details. \hanzhongsep{The number of replications in the bootstrap is $1000$, that is, $B = 1000$.} We consider two generative models for data simulation.

(1) Linear regression model. The simulated data are drawn from the linear model:
	\begin{equation}\label{eqn:lm}
	y_i=x_i^\T \beta^0 + \epsilon_i,\quad \epsilon_i \sim N(0,\sigma^2), \quad i=1,\ldots,n.
	\end{equation}
We fix $n=200$ and $p=500$. We generate the design matrix $X$ in three scenarios, using the R package ``mvtnorm". In Scenarios 1 and 2, we choose $\sigma$ such that the Signal-to-Noise-Ratio equals ten, that is, SNR $=||X\beta^0||_2^2/(n\sigma^2)=10$. We also examine other values of $n,p$ and $\sigma$, but they are not reported here because the conclusions are similar. 

Scenario\ 1 (Normal): Predictor vectors $x_i$, for $i=1,\ldots,n$, are generated independently from a multivariate normal distribution $N(\mathbf{0},\Sigma)$ with covariance matrix $\Sigma$. We consider three types of $\Sigma$, following the setup in \citep{Nicolai2014}.
\begin{eqnarray}
\textnormal{Toeplitz}:&  & \Sigma_{ij}=\rho^{|i-j|}, \ \textnormal{with} \ \rho=0.5, 0.9, \nonumber \\
\textnormal{Exponential decay}:&  & (\Sigma^{-1})_{ij}=\rho^{|i-j|}, \ \textnormal{with} \ \rho=0.5,0.9, \nonumber \\
\textnormal{Equal correlation}:&  & \Sigma_{ij}=\rho, \ \textnormal{with} \ \rho=0.5,0.9. \nonumber
\end{eqnarray}
			
Scenario\ 2 ($t_2$): Predictor vectors $x_i$, for $i=1,\ldots,n$, are generated independently from a multivariate $t$ distribution, with two degrees of freedom, $t_2(\mathbf{0},\Sigma)$, where $\Sigma$ is a Toeplitz-type matrix: $\Sigma_{ij}=\rho^{|i-j|}$, with $\rho=0.5,0.9$.

Scenario\ 3 (fMRI data): A $200 \times 500$ design matrix $X$ is generated by random sampling, without replacement, from the real $1750 \times 2000$ design matrix in the fMRI data (see \JL{Section~\ref{real_data}} for more details on this data). Every column of $X$ is normalized to have zero mean and unit variance, and we choose $\sigma$, such that \finalrevise{$\textnormal{SNR}= 1, 5$ or $10$.}

We also consider two cases to generate $\beta^0$. 

Case\ 1 (hard sparsity): $\beta^0$ has 10 nonzero elements whose indices are randomly sampled, without replacement, from $\{1,\ldots,p\}$, and whose values are generated from $U[1/3,1]$, a uniform distribution on the interval $[1/3,1]$. The remaining $490$ elements are set to be zero. 

Case\ 2 (weak sparsity): The setup is similar to that in \citep{Zhang}. $\beta^0$ has $10$ large elements whose indices are randomly sampled, without replacement, from $\{1,\ldots,p\}$, and whose values are generated from a normal distribution, $N(1,0.001)$. The remaining $490$ elements decay at a rate of $1/(j+3)^2$, that is, $\beta^0_j = 1/(j+3)^2$.

The values of $x_i$ and $\beta^0$ are generated once and then kept fixed. \revise{The average absolute correlations among the covariates with large coefficients are $0.08$, $0.06$, and $0.47$ for the normal design with a Toeplitz type covariance matrix, normal design with an Exponential decay type covariance matrix, and $t_2$ design with a Toeplitz type covariance matrix, respectively.} After $X=(x_1^\T,\ldots,x_n^\T)^\T$ and $\beta^0$ are generated, we simulate $Y=(y_1,\ldots,y_n)^\T$ from the linear model~\eqref{eqn:lm} by generating independent error terms for $1000$ replications. Then we \JL{construct confidence intervals for each regression coefficient, and compute their} coverage probabilities and \JL{mean interval lengths}.

(2) Misspecified linear model. The simulation is based on a real data set: fMRI (see \JL{Section~\ref{real_data}} for more details). Let $X$ and $Y^f$ (distinguished from the simulated response $Y$ below) denote the design matrix (with $n=1750$ observations and $p=2000$ predictors) and the actual response (of the ninth voxel) in the fMRI data set. The original design matrix in the fMRI data set has $10921$ predictors, but we first removed the predictors with variances no more than $1e^{-4}$ and selected $p = 2000$ predictors that have the largest absolute correlations with the response. We compute the lasso+ols estimator $\beta^f_{\textnormal{lasso+ols}}$ (selecting the tuning parameter $\lambda_{1}$ by 5-fold cross validation on lasso+ols):
	\[  \beta_{\textnormal{lasso}}^{f} = \argmin\limits_{\beta} \left\{ \frac{1}{2n} || {Y^f - X\beta } ||_2^2  + \lambda_{1} || {\beta } ||_1 \right\}, \]
	\[  \beta^f_{\textnormal{lasso+ols}} = \argmin\limits_{\beta: \beta_j=0, \  j \notin S}  \frac{1}{2n}  || {Y^f - X\beta } ||_2^2,  \]
	where $S = \{j: (\beta_{\textnormal{lasso}}^{f})_j \neq 0 \}$ is the set of relevant predictors. We re-ordered the predictors by sorting the values of $\beta^f_{\textnormal{lasso+ols}}$ in a decreasing order, such that the first four predictors corresponds to the largest 4 nonzero elements of $\beta^f_{\textnormal{lasso+ols}}$. Then we generate the simulated response $Y=(y_1,\ldots,y_n)^\T$ from the following model:
	\begin{equation}
	\label{eqn:yi-mis}
	y_i = E(y_i|x_i) +\epsilon_i, \ \  \epsilon_i \sim N(0,\sigma^2),
	\end{equation} 
	\begin{equation*}
	E(y_i|x_i) = x_i^\T \beta^f_{\textnormal{lasso+ols}} + \sum_{j=1}^{4} \alpha_{j} x_{ij}^2 + \sum_{1\leq j < k \leq 4}\alpha_{jk} x_{ij}x_{ik} ,
	\end{equation*} 
	where $\alpha_{j}$, for $j=1,..,4$, and $\alpha_{jk}$, for $1\leq j \neq k \leq 4$, are independently generated from a uniform distribution, $U(0,0.1)$. The values of $\alpha_{j}$ and $\alpha_{jk}$ are generated once and then kept fixed. We set $\sigma$ such that $\textnormal{SNR}= \sum_{i=1}^{n}E(y_i|x_i)^2/(n\sigma^2)=0.5,\ 1$ or $5$. Since the quadratic and interaction terms are not included in the design matrix $X = (x_1^\T,\ldots,x_n^\T)^\T$, a linear model between $Y$ and $X$, $y_i = x_i^\T \beta^0 + \epsilon_i$, is misspecified. In this misspecified linear model, the parameter vector $\beta^0$ we are interested in is the projection coefficient of $E(Y \mid X)$ onto the subspace spanned by the relevant predictors:
	\[ \beta^0_{S} = (X_S^\T X_S)^{-1}X_S^\T E(Y \mid X); \quad \beta^0_{S^c} = \mathbf{0}. \]
Again, in order to compute the coverage probabilities and mean confidence interval lengths, we generate $Y$ by simulating independent error terms $\epsilon_i$'s in equation \eqref{eqn:yi-mis} for $1000$ times. \hanzhongg{The confidence level is \JL{set to} $95\%$.}

\subsection{Selection of the partial ridge tuning parameter $\lambda_{2}$}
\label{sec:simu_select_lambda2}

\begin{figure}[ht]
\centerline{\includegraphics[width=.9\textwidth]{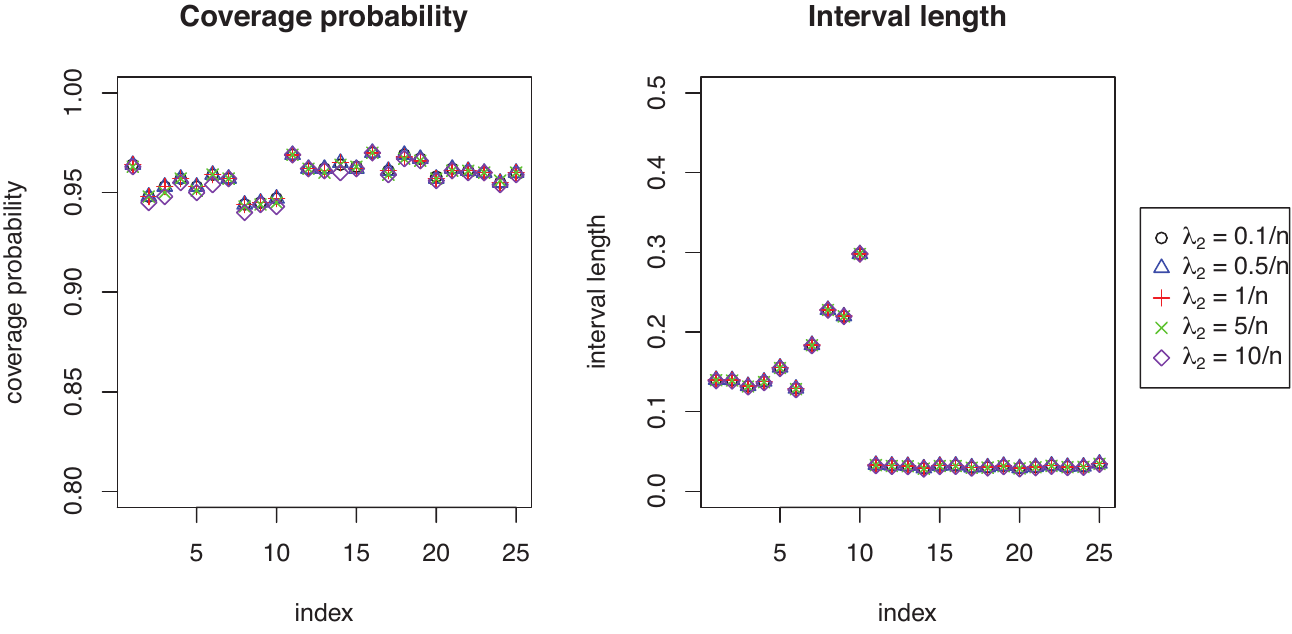}}
\caption[The effects of $\lambda_{2}$ on coverage probabilities and mean confidence interval lengths]{The effects of $\lambda_{2}$ on coverage probabilities and mean confidence interval lengths. The predictors are generated from a Normal distribution in Scenario 1 with a Toeplitz \finalrevise{type} covariance matrix, and $\rho=0.5$. The coefficient vector $\beta^0$ is hard sparse.}\label{fig:paired-resid-prob-all}
\end{figure}

We first study the effects of the partial ridge tuning parameter $\lambda_{2}$ on the performance of the bootstrap \LPR{} methods (\rBLPR{} and \pBLPR{}). Figure~\ref{fig:paired-resid-prob-all} compares the coverage probabilities and mean confidence interval lengths produced by different values of $\lambda_{2}$, based on the following simulation setup: the predictors are generated from a Normal distribution as in Scenario 1, with a Toeplitz type covariance matrix corresponding to $\rho=0.5$, and $\beta^0$ is hard sparse. We also compare the results for other simulation setups, but the conclusions are essential the same and are not reported here. In order to give a better view, in the following figures without further emphasizing, we sort the elements of $\beta^0$ in a decreasing order (in absolute value) and only plot the results for the largest $25$ elements of $\beta^0$. We can see that both the coverage probabilities and mean confidence interval lengths are very stable with respect to a large range of $\lambda_{2}$ values. Our simulation experiments show that fixing $\lambda_{2}$ at $1/n$ works quite well for a wide range of noise levels. For the sake of simplicity, we take $\lambda_{2}=1/n$ in this study, but acknowledging that further research is needed to find a more systematic approach for selecting $\lambda_2$.


\subsection{Comparison of bootstrap lasso+ols and bootstrap \LPR{} methods} \label{section:comp-lols-lpr}
\begin{figure}[htp]
	\centerline{\includegraphics[width=\textwidth]{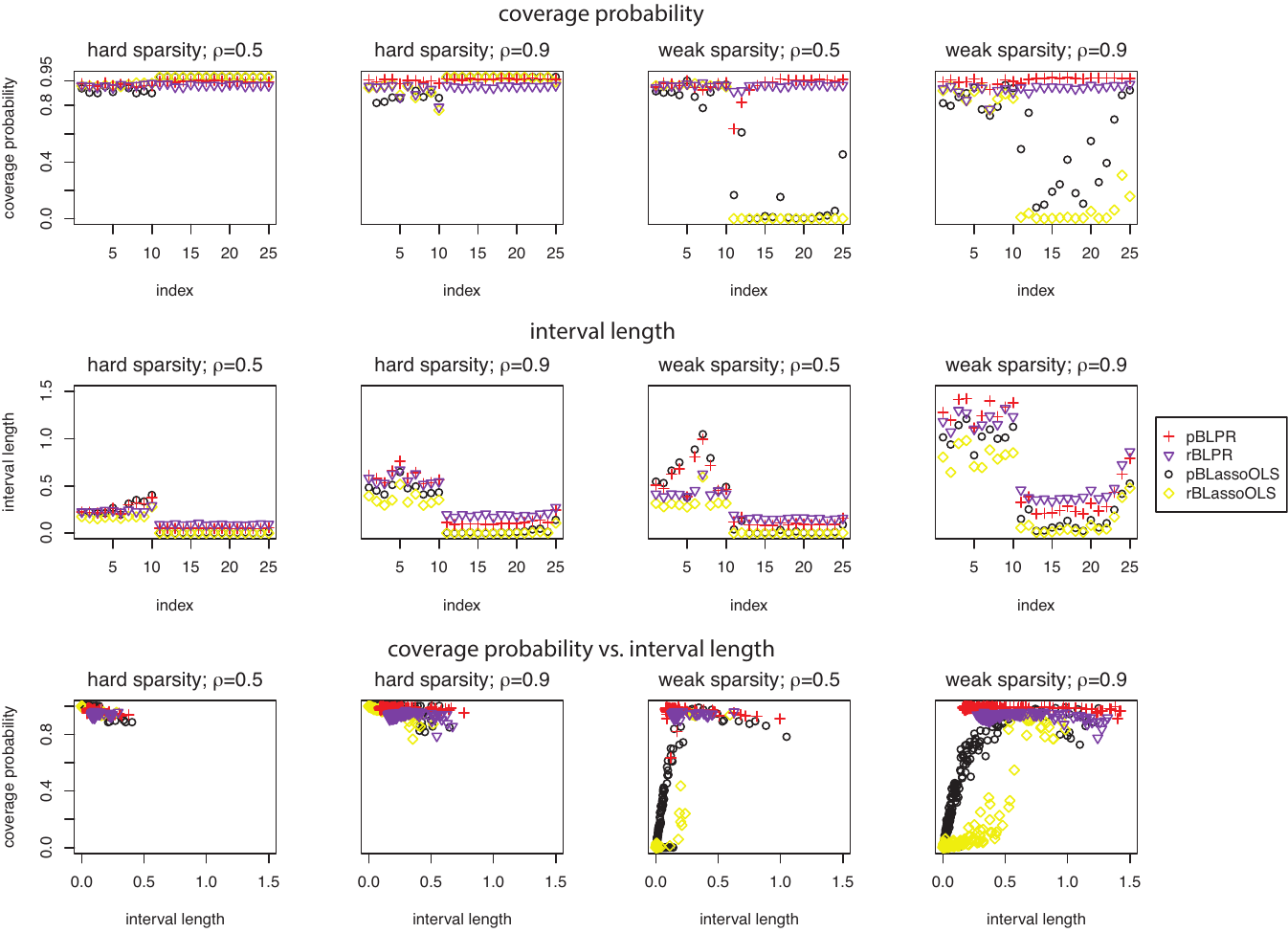}}
	\vspace*{0.05in}
	\caption[Comparison of rBLPR and pBLPR -- Normal design with a Toeplitz type covariance matrix]{Comparison of coverage probabilities (the first row) and mean confidence interval lengths (the second row) produced by four methods: rBLPR, pBLPR, residual bootstrap lasso+OLS (denoted by rBLassoOLS) and paired bootstrap lasso+OLS (denoted by pBLassoOLS). The third row shows the coverage probabilities v.s. mean interval lengths. The design matrix is generated from a Normal distribution with a Toeplitz type covariance matrix.}\label{fig:paired-resid-prob-length-N1}
\end{figure}

We now compare the performance of the \rBLPR{} and \pBLPR{} methods with that of the bootstrap lasso+ols method. Figure~\ref{fig:paired-resid-prob-length-N1} shows the comparison results in terms of coverage probabilities and mean confidence interval lengths for the Normal distributed design matrix in Scenario 1 with a Toeplitz type covariance matrix corresponding to $\rho = 0.5$ or $0.9$, and for $\beta^0$ with hard or weak sparsity. For other design matrices, the conclusions are similar. We see that the \rBLPR{} and \pBLPR{} have similar performance, while the latter performs slightly better, therefore, we only present the results for \pBLPR{} in the following contents. In the hard sparsity cases, all the methods work very well. In the weak sparsity cases, however, the bootstrap lasso+ols method gives very poor coverage probabilities \hanzhongg{(less than $50\%$ for $95\%$ confidence intervals)} for the small, but nonzero elements of $\beta^0$. This is because these elements are seldom selected by the lasso and, therefore, a large proportion of their bootstrap lasso+ols estimates are zero, producing noncoverage confidence intervals, such as $[0,0]$. The \pBLPR{} method dramatically improve the performance of the bootstrap lasso+ols method. It produces promising coverage probabilities, at the price of slightly increasing the confidence interval lengths. However, for medium-size components of $\beta^0$, \pBLPR{} has problems covering true values even when design matrices are generated from a normal distribution (The coverage probability for one particular such component is only $63\%$). This is because the lasso cannot identify these medium-size components with high probability. 

\subsection{Comparison of bootstrap \LPR{} and de-sparsified methods}
\label{sec:comp-lpr-desparsified}

\begin{figure}[htp]
	\centerline{\includegraphics[width=\textwidth]{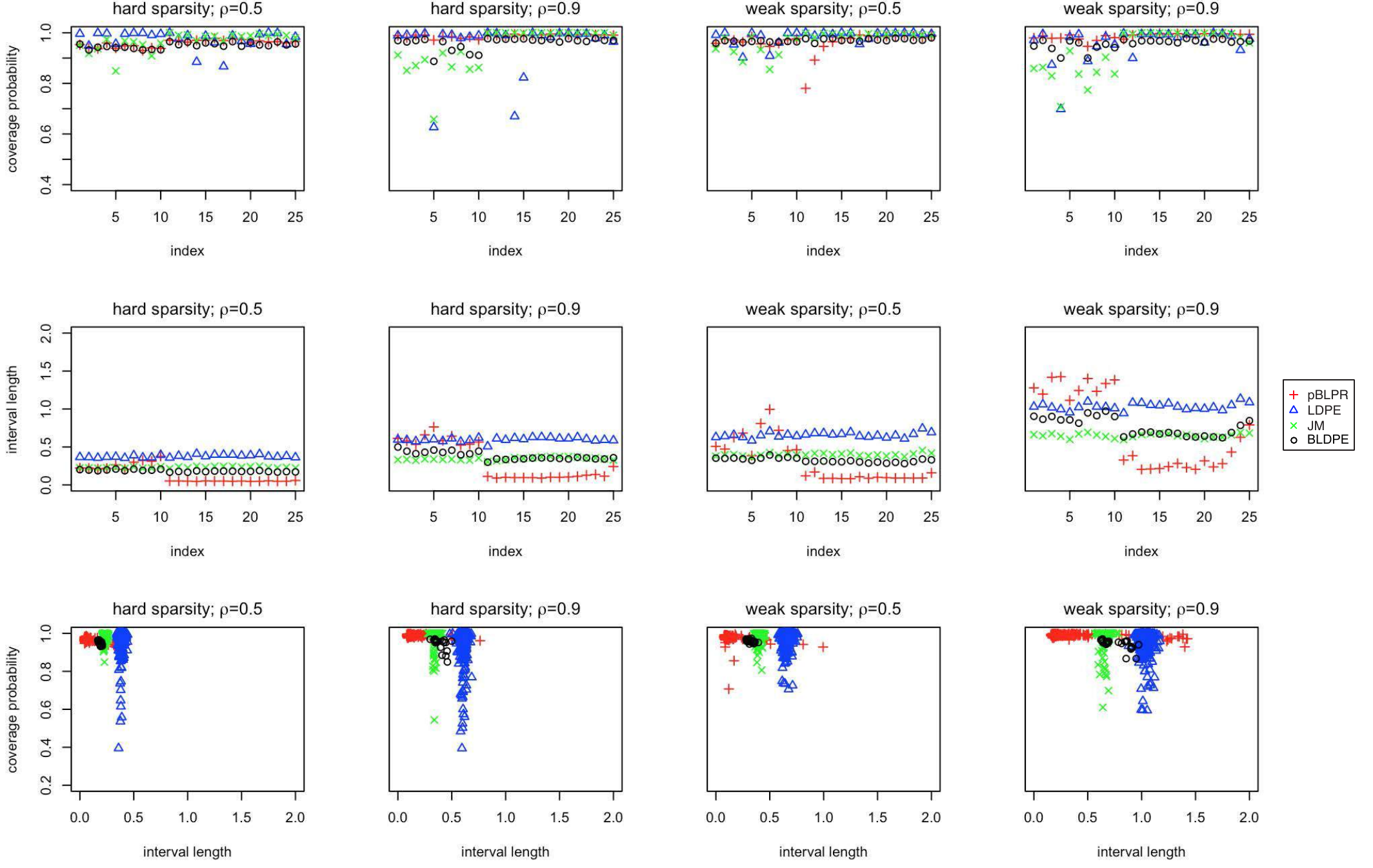}}
	\caption[Comparison of \pBLPR, \ZZ{}, \JM{} and BLDPE -- Normal design with a Toeplitz type covariance matrix]{Comparison of coverage probabilities (the first row) and mean interval lengths (the second row) produced by pBLPR, \ZZ{}, \JM{}, and BLDPE. The third \finalrevise{row} shows the coverage probabilities v.s. mean interval lengths. The design matrix is generated from a Normal distribution with a Toeplitz type covariance matrix.}\label{fig:paired-prob-length-N1}
\end{figure}

\begin{figure}[htp]
	\centerline{\includegraphics[width=\textwidth]{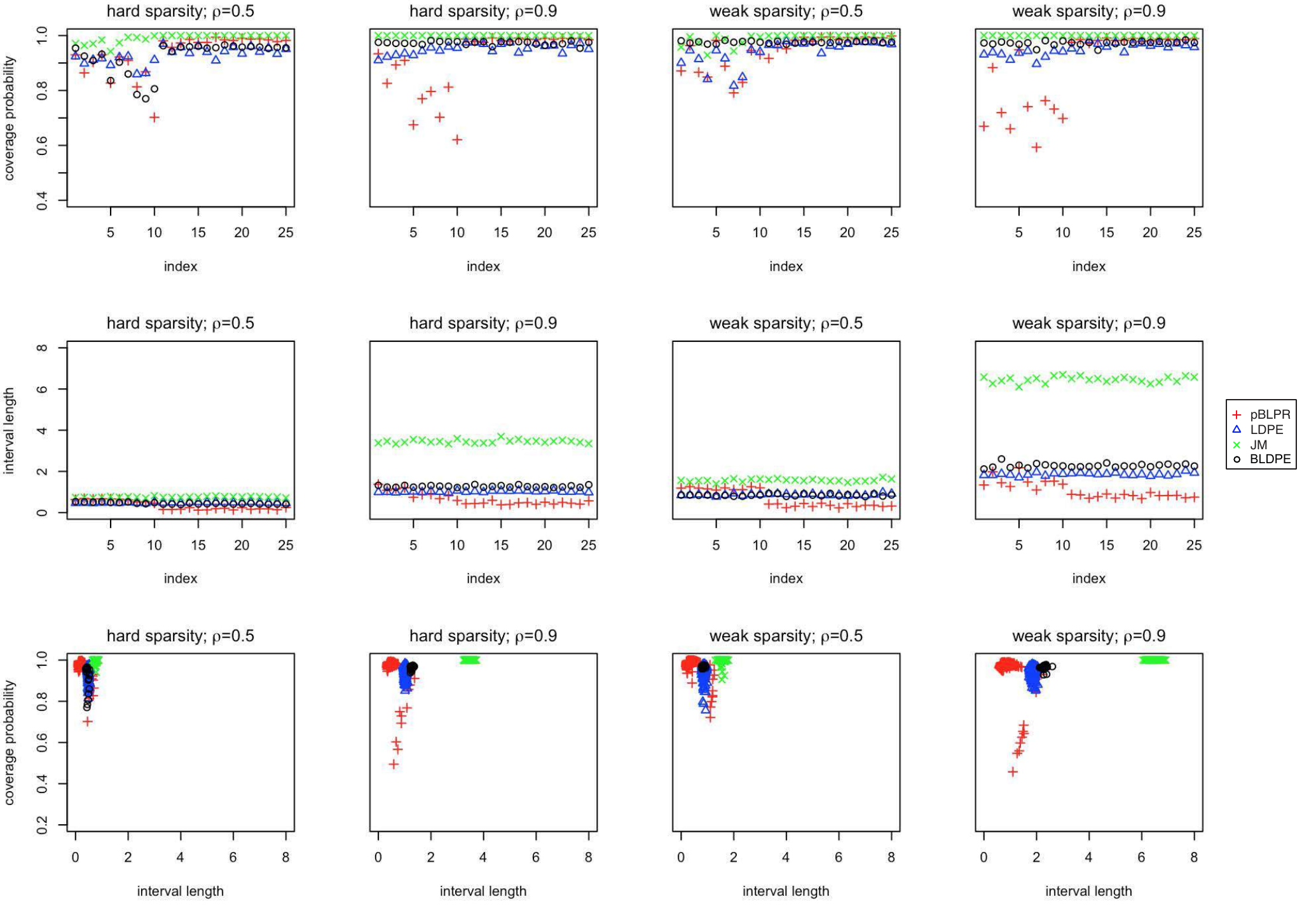}}
	\caption[Comparison of pBLPR, \ZZ{}, \JM{} and BLDPE -- Normal design with an Equi.corr type covariance matrix]{See caption of Figure~\ref{fig:paired-prob-length-N1} with the only difference being that the covariance matrix is an Equi.corr type.}\label{fig:paired-prob-length-N3}
\end{figure}

\begin{figure}[htp]
	\centerline{\includegraphics[width=\textwidth]{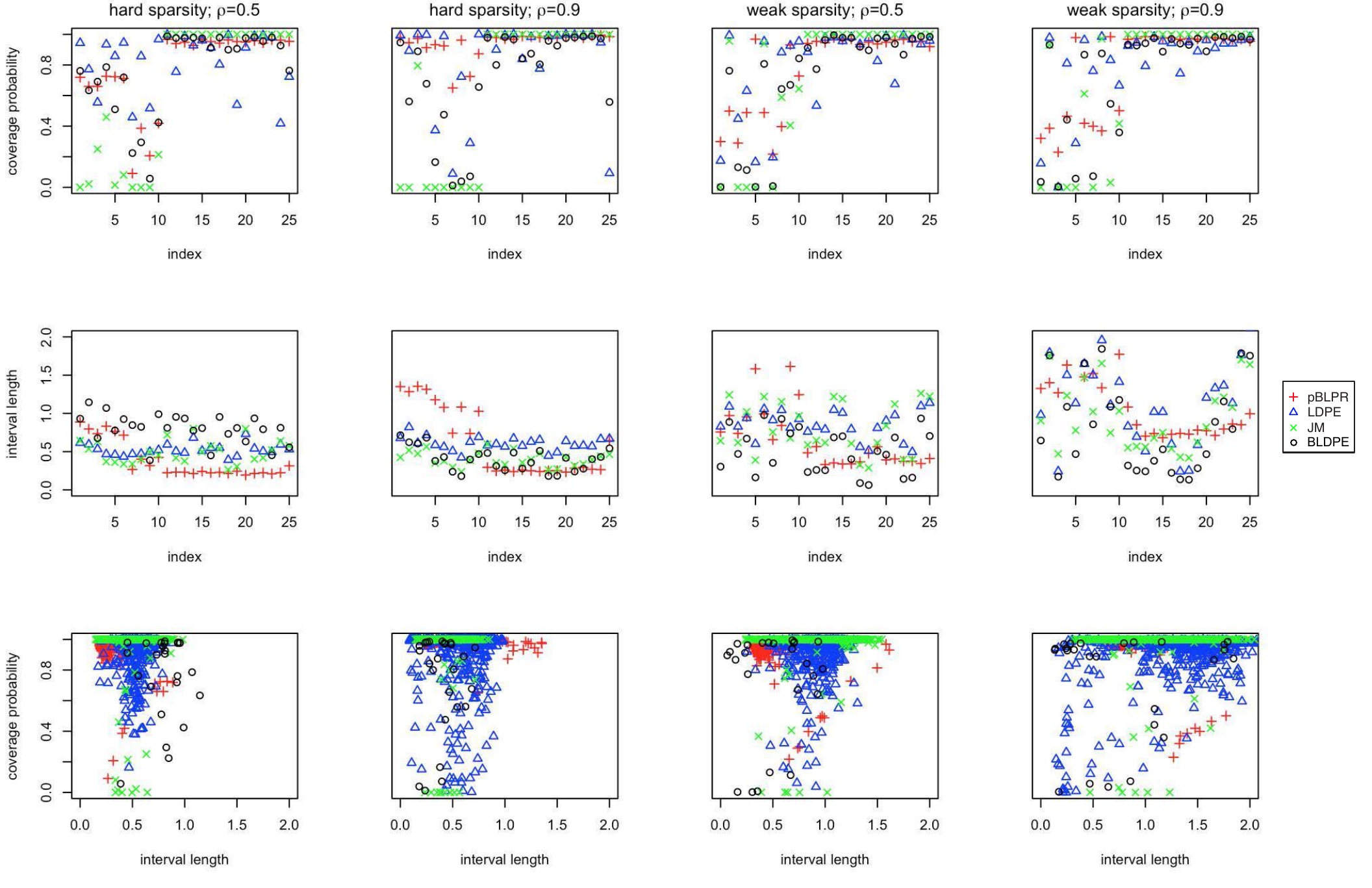}}
	\caption[Comparison of pBLPR, \ZZ{}, \JM{} and BLDPE -- $t_2$ design with a Toeplitz type covariance matrix]{See caption of Figure~\ref{fig:paired-prob-length-N1} with the only difference being the type of design matrix. In this plot, the design matrix is generated from $t_2$ distribution with a Toeplitz type covariance matrix.}\label{fig:paired-prob-length-T}
\end{figure}

\hanzhongg{Figures~\ref{fig:paired-prob-length-N1},~\ref{fig:paired-prob-length-N3}, and~\ref{fig:paired-prob-length-T} show the comparison results of \pBLPR{}, \JL{\ZZ{}, \JM{}, \finalrevise{and BLDPE}, under a Normal design matrix with a Toeplitz type covariance matrix, with an Equi.corr type covariance matrix, and a $t_2$ distributed design matrix with a Toeplitz type covariance matrix, respectively.} From Figure~\ref{fig:paired-prob-length-N1},} we see that the \pBLPR{} gives promising results. Overall, it has good performance for large and small components of $\beta^0$, and in some cases it outperforms \ZZ{} and \JM{}, \hanzhongg{\JL{by producing confidence intervals with, on average, $50\%$ shorter lengths} (see the comparison results in \JL{Tables}~\ref{tab:cplarge} -- \ref{tab:lengthsmall} in the Appendix). When the predictors have high correlations (see the results for \finalrevise{$\rho = 0.9$}), \pBLPR{} gives confidence intervals with higher coverage probabilities for large coefficients, and for small and zero coefficients, it gives shorter confidence interval lengths with good coverage probabilities.} \hanzhongg{Following the \JL{evaluation scheme} in \citep{Nicolai2014}, we also \JL{show more details of the comparison results} in Figures~\ref{fig:paired-hdi-N1-1} -- \ref{fig:paired-hdi-N1-4} in the Appendix, \JL{which clearly show the advantegeous performance of the \pBLPR{} in constructing confidence intervals for a broad range of coefficients}.}

\begin{figure}[ht]
	\centerline{\includegraphics[width=\textwidth]{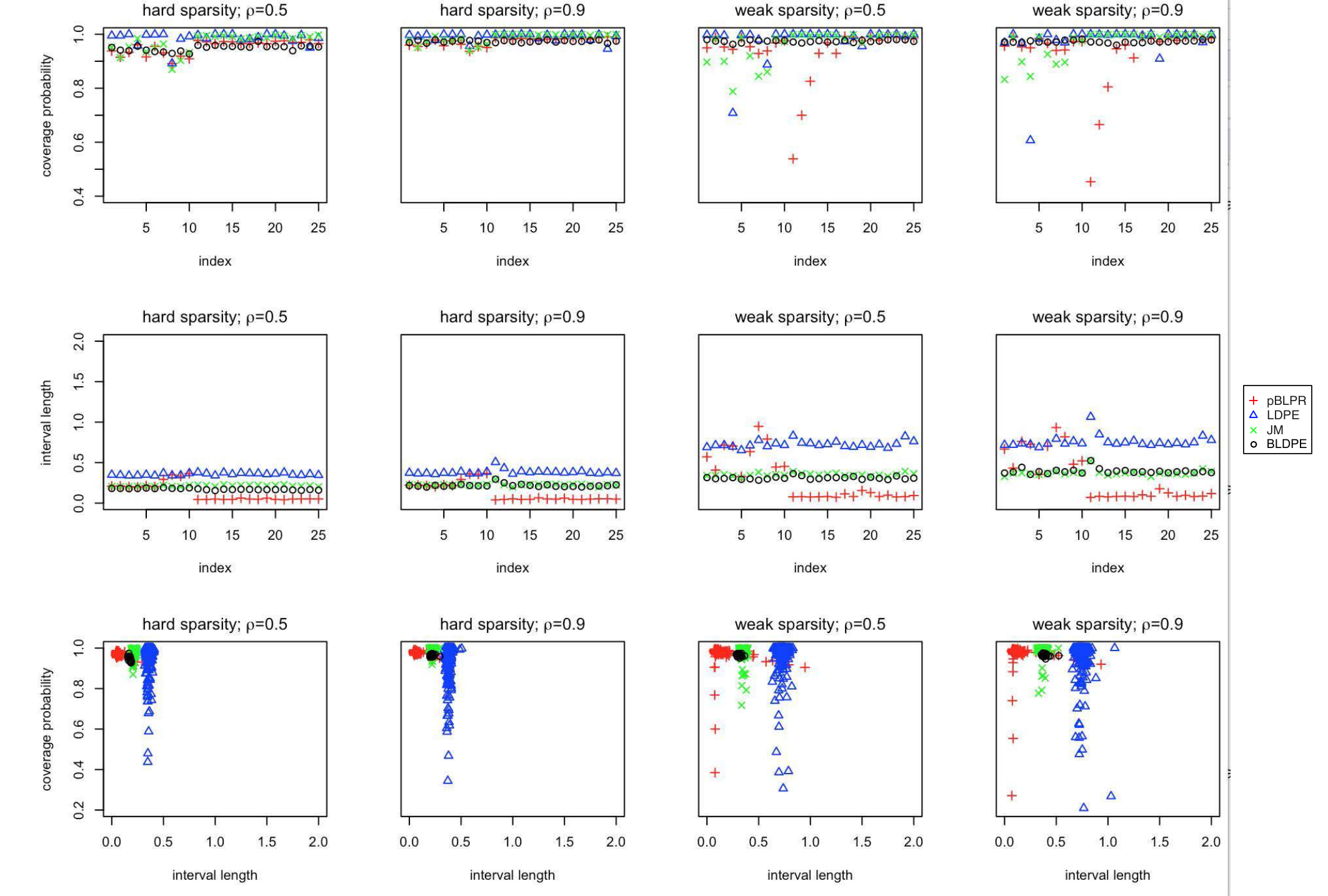}}
	\vspace*{0.05in}
	\caption[Comparison of pBLPR, \ZZ{}, \JM{} and BLDPE -- Normal design with an Exp.decay type covariance matrix]{See caption of Figure~\ref{fig:paired-prob-length-N1} with the only difference being that the covariance matrix is Exp.decay type.}\label{fig:paired-prob-length-N2}
\end{figure}

\begin{figure}[ht]
	\centerline{\includegraphics[width=\textwidth]{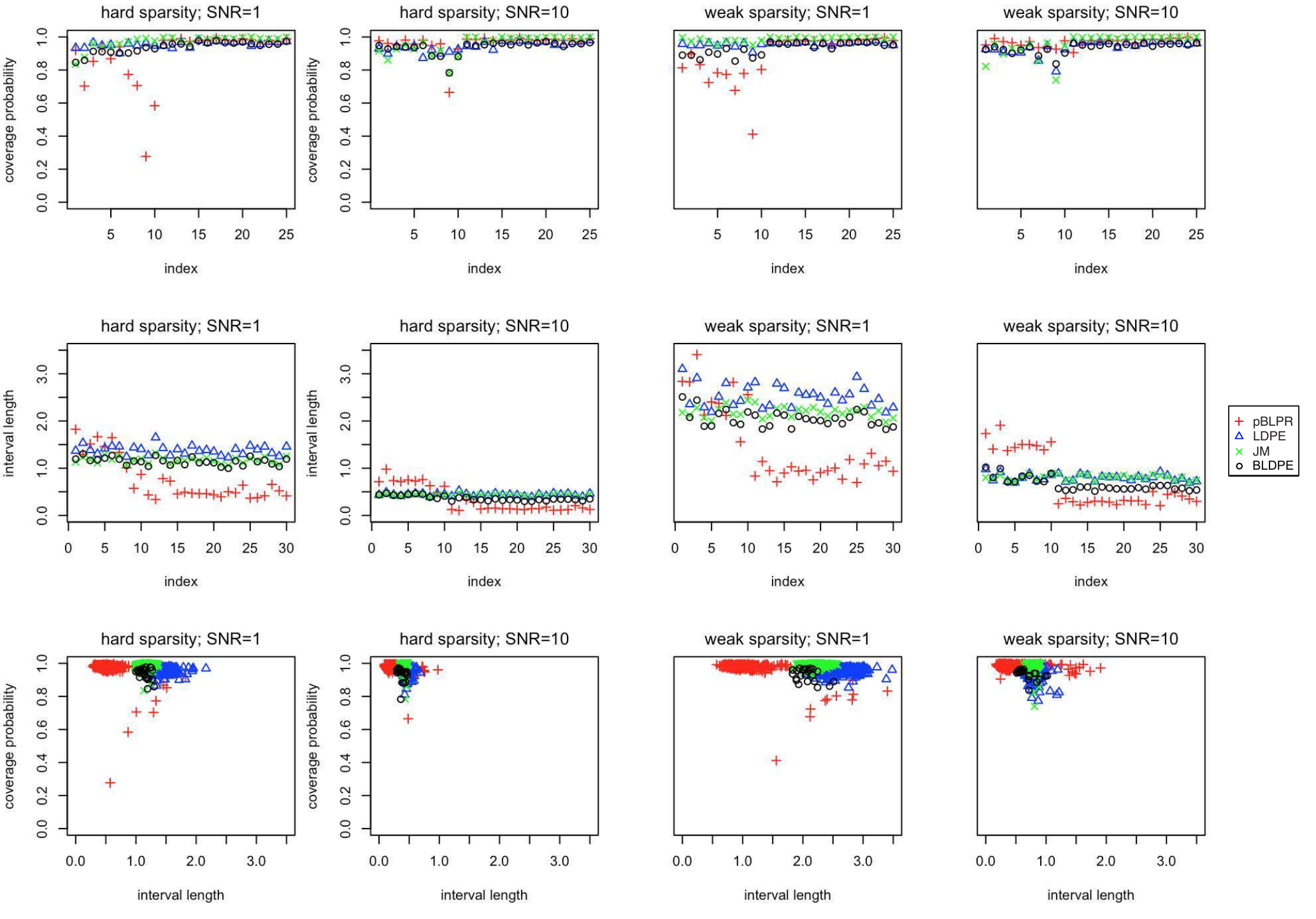}}
	\vspace*{0.05in}
	\caption[Comparison of pBLPR, \ZZ{}, \JM{} and BLDPE -- fMRI design]{See caption of Figure~\ref{fig:paired-prob-length-N1} with the only difference being that the design matrix is generated from the fMRI data.}\label{fig:paired-prob-length-real}
\end{figure}

\hanzhongg{\JL{Under a Normal design matrix with an Equi.corr type covariance matrix} (see Figure~\ref{fig:paired-prob-length-N3}), the \JM{} does not work well \JL{when $\rho=0.9$,} because it dramatically overestimates the noise variance. Our method \JL{also} has unsatisfactory performance \JL{in terms of} coverage probabilities \JL{for large coefficients}, because \JL{the} lasso cannot correctly select the large \JL{predictors due to the strong collinearity among the} predictors.} Under a $t_2$ design matrix, Figure~\ref{fig:paired-prob-length-T} shows that no methods perform well, leaving large space for improvement. 
For other covariance structures, \hanzhongsep{the comparison results are shown in Figures~\ref{fig:paired-prob-length-N2} and \ref{fig:paired-prob-length-real}.}

\revise{The bootstrap version LDPE method (BLDPE) does improve the performance of LDPE. It has the best coverage probabilities among the considered methods, but its confidence interval lengths are close to or slightly shorter than the better one of LDPE and JM and, hence, larger than the \pBLPR{} method.}

\begin{figure}[ht]
	\centering
	\includegraphics[width=0.8\textwidth]{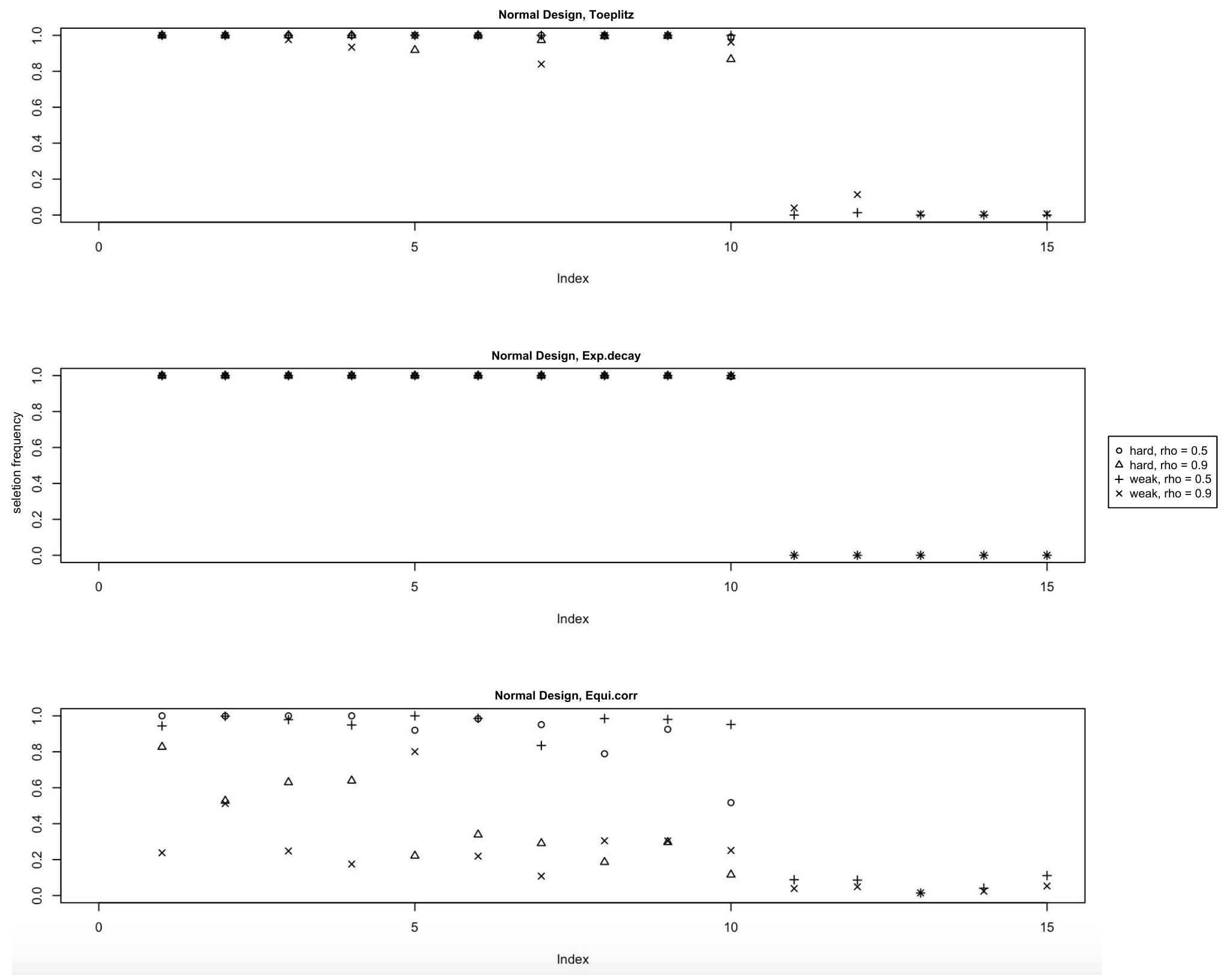}
	\caption[Selection frequency of each coefficient ]{The selection frequency of each coefficient in 1000 simulation runs by the lasso (the 10 nonzero coefficients in hard sparsity case and the first 15 largest coefficients in absolute values in weak sparsity case). }
	\label{fig:selection1}
\end{figure}

\begin{figure}[ht]
	\centering
	\includegraphics[width=0.8\textwidth]{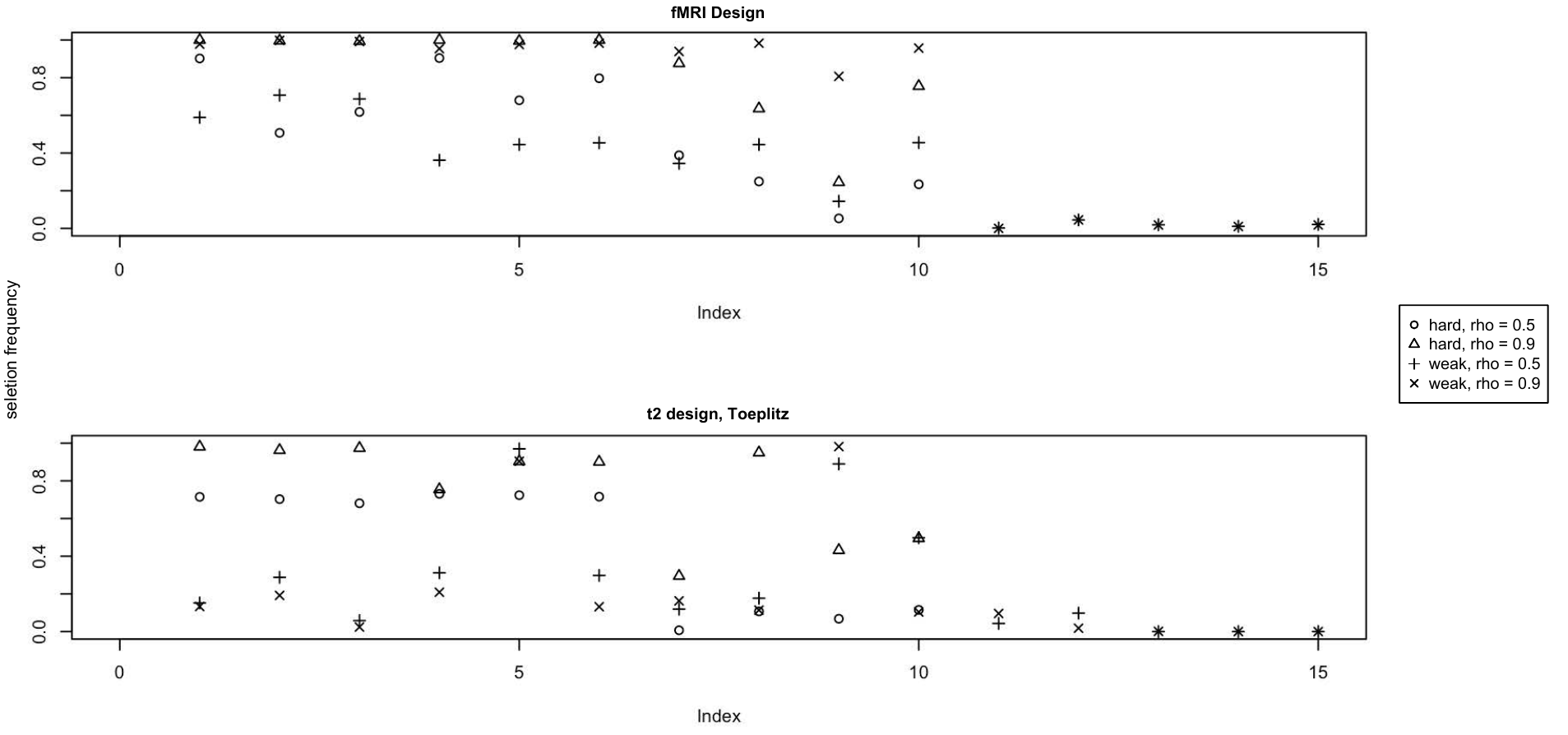}
	\caption[Selection frequency of each coefficient --  continue]{The selection frequency of each coefficient in 1000 simulation runs by the lasso (the 10 nonzero coefficients in hard sparsity case and the first 15 largest coefficients in absolute values in weak sparsity case).}
	\label{fig:selection2}
\end{figure}

\revise{The selection frequency of each coefficient in the 1000 simulation runs is shown in Figure~\ref{fig:selection1} and \ref{fig:selection2}. Although some important coefficients are missed by the lasso, their empirical coverage probabilities are still good. This maybe because the bootstrap runs help to correct the selection and the LPR estimator is \finalrevise{no longer} sparse due to the partial ridge penalty.}

\revise{The comparison results for rBLPR can be found in Figures~\ref{fig:paired-prob-length-N1-res} -- \ref{fig:paired-prob-length-real-res} in the Appendix.}

\begin{figure}[ht]
	\centerline{\includegraphics[width=\textwidth]{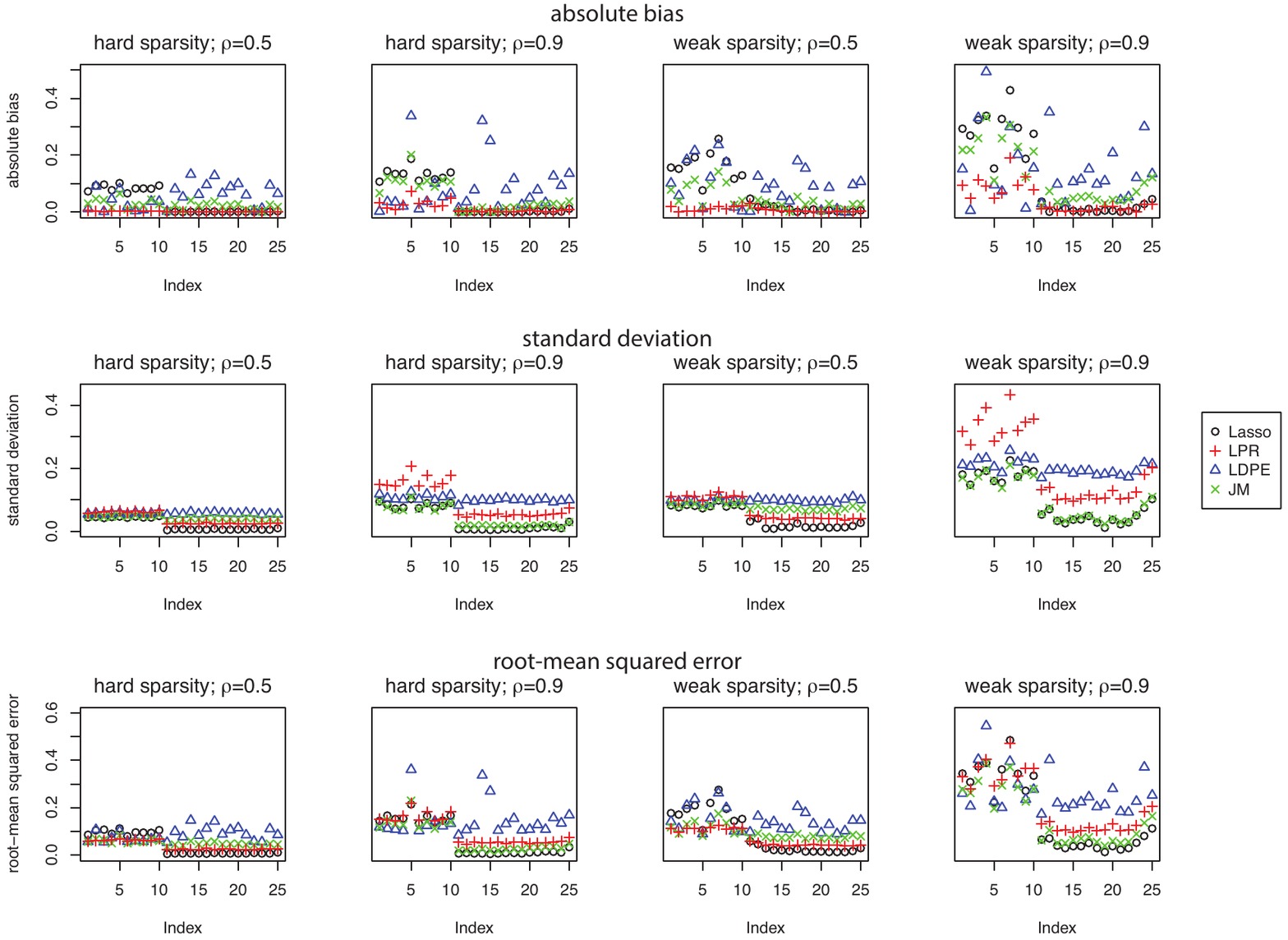}}
	\caption[Comparison of Bias, SD and MSE -- Normal design with a Toeplitz type covariance matrix]{Comparison of bias, standard deviation and root-mean squared error. The design matrix is generated from the Normal distribution with a Toeplitz type covariance matrix.}\label{fig:com-bias-sd-mse-N1}
\end{figure}

In addition, we also compare the bias, standard deviation (SD) and root-mean square error (RMSE) of the de-sparsified estimators and the \LPR{} estimator, in order to see to what extent these methods reduce the lasso bias. Figure~\ref{fig:com-bias-sd-mse-N1} shows the results. We found that, \hanzhongg{compared with \ZZ{} and \JM{}, \JL{the} \LPR{} estimator has smaller biases ($99\%$ and $72\%$ \JL{smaller}, on average, \JL{than that of \ZZ{} and \JM{}}, respectively) for almost all \JL{coefficients}, but the \LPR{} estimator has larger SDs ($30\%$ and $62\%$ \JL{larger}, on average, \JL{than that of \ZZ{} and \JM{}}, respectively) for large \JL{coefficients}. Overall, \LPR{} has $60\%$ smaller RMSE than \ZZ{}, but $42\%$ larger RMSE than \JM{}. Another interesting finding is that although de-sparsified estimators can dramatically decrease the biases \hanzhong{of the lasso} by more than $40\%$ for large  $\beta_j^*$'s, they can increase the biases more than twice for small, or zero $\beta_j^*$'s.}

\subsection{Robustness to \JL{signal-to-noise ratios}}
\label{sec:robust_to_snr}

\begin{figure}[ht]
	\centerline{\includegraphics[width=\textwidth]{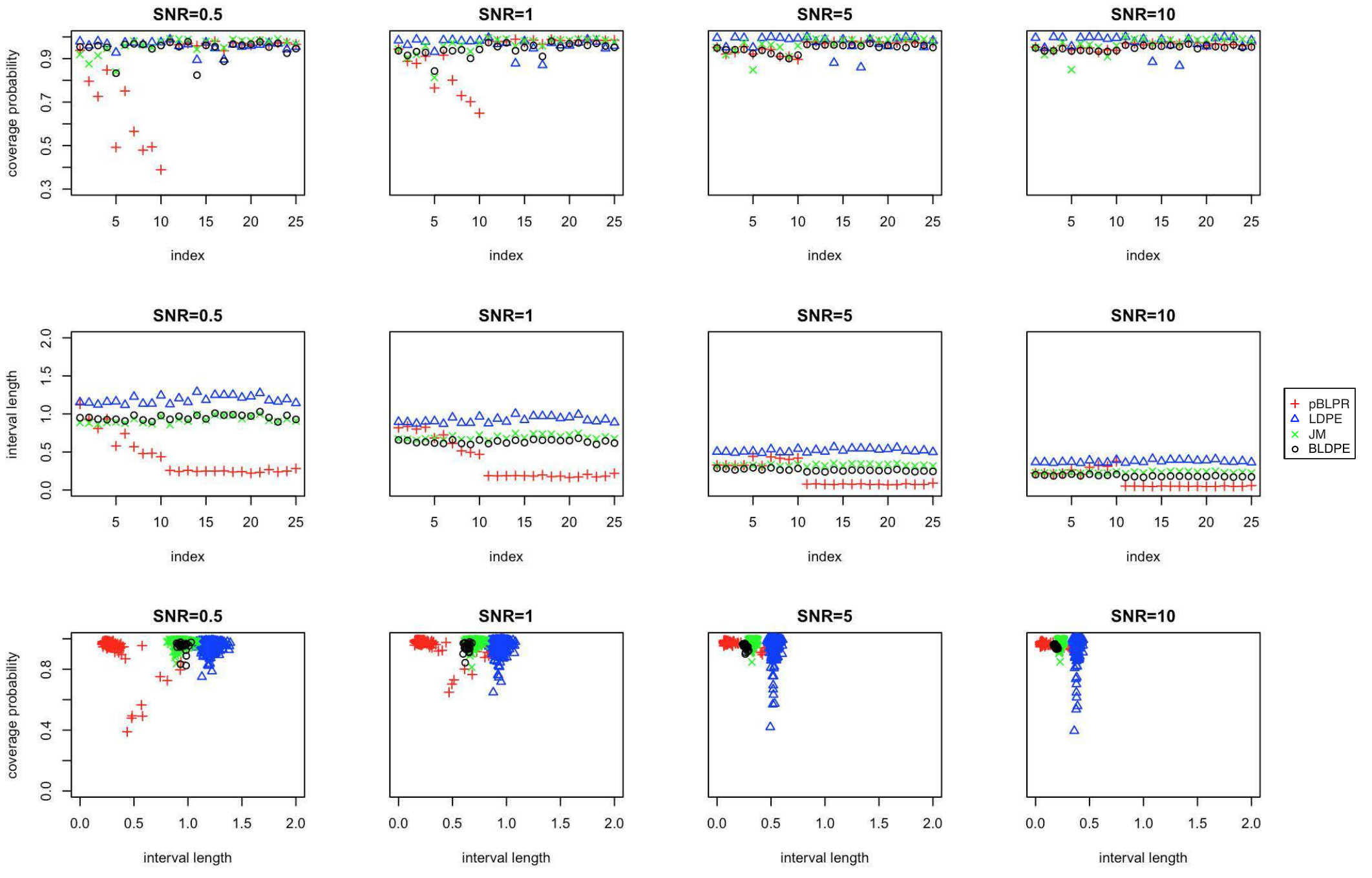}}
	\caption[Comparison of pBLPR, \ZZ{}, \JM{} and BLDPE -- changing SNR]{Comparison of coverage probabilities (first row) and mean interval lengths (second row) produced by pBLPR, \ZZ{}, \JM{}, and BLDPE, when SNR changes. The third row shows the coverage probabilities v.s. mean interval lengths. The design matrix is generated from a Normal distribution with a Toeplitz type covariance matrix, and $\rho=0.5$. }\label{fig:change-snr}
\end{figure}

Figure~\ref{fig:change-snr} shows the comparison results \JL{under varying signal-to-noise ratios (SNRs)}. We can see that the coverage performance of the de-sparsified methods is more robust to SNR changes. On the other hand, the \pBLPR{} method works well when SNR is high (say, larger \JL{than} 5), but it may have low coverage probabilities for nonzero \JL{coefficients} when SNR is low. This is reasonable because \JL{the} lasso cannot identify nonzero \JL{coefficients} with high probability when SNR is low. The \pBLPR{} method depends more on the model selection performance of \JL{the} lasso. However, it has much shorter (more than $20\%$, on average) confidence interval lengths for zero \JL{coefficients} even when SNR is low.

\subsection{Comparison of different methods under the misspecified model}
\label{sec:misspecified}

\begin{figure}[ht]
	\centerline{\includegraphics[width=.9\textwidth]{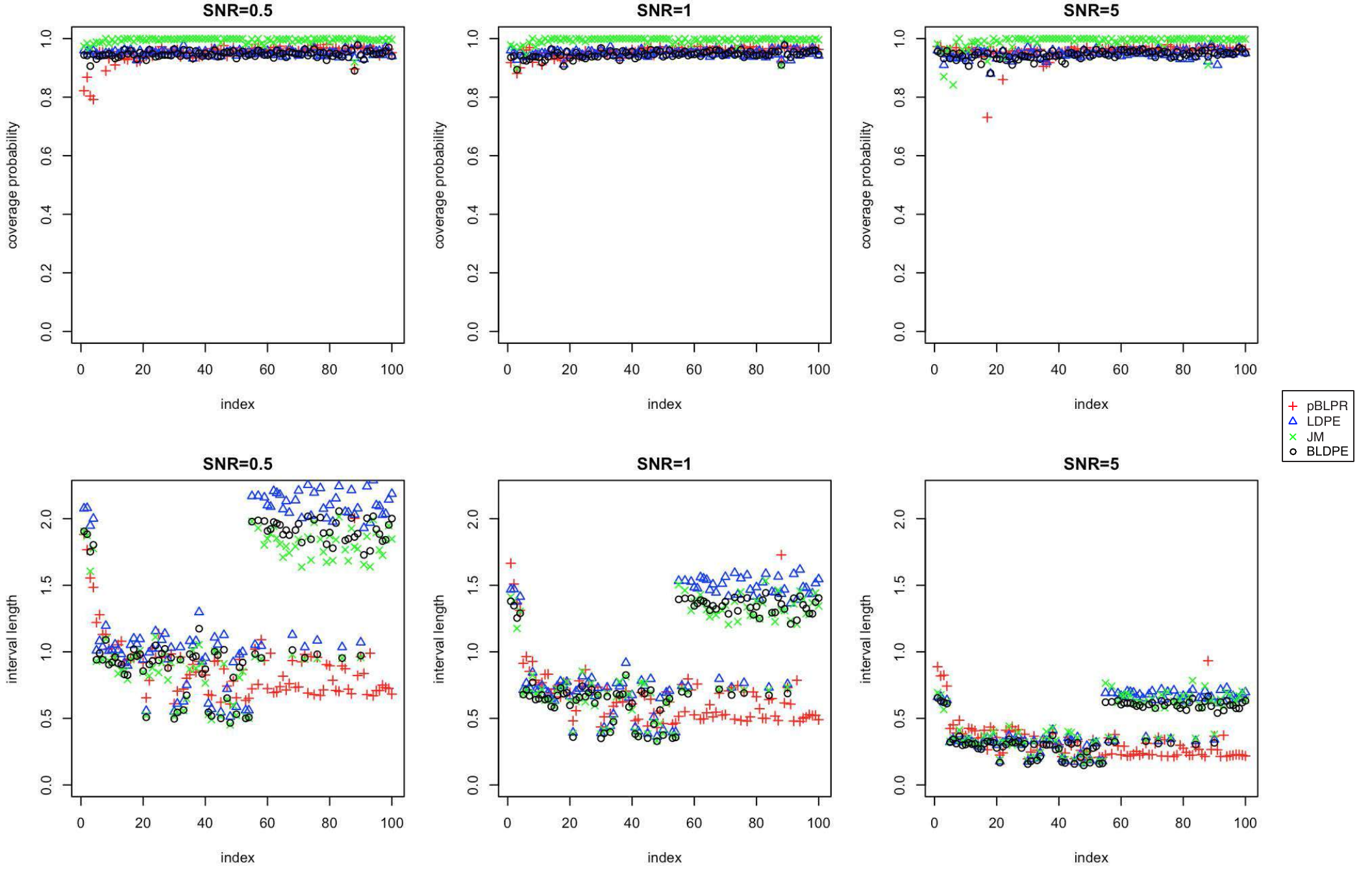}}
	\caption[Comparison of pBLPR, \ZZ{}, \JM{} and BLDPE -- misspecified model]{Comparison of coverage probabilities and mean interval lengths produced by pBLPR, \ZZ{}, \JM{}, and BLDPE. The results is based on data simulated from the misspecified linear model \eqref{eqn:yi-mis}.}\label{fig:prob-length-mis}
\end{figure}

Figure~\ref{fig:prob-length-mis} compares the performance of \pBLPR{}, \ZZ{}, \JM{}, \finalrevise{and BLDPE} under the misspecified linear model. The \pBLPR{} performs \revise{slightly worse than} the other three methods in terms of coverage probabilities, but it produces more than $50\%$, on average, shorter confidence intervals. 

\section{Real-data case study 1: fMRI data}\label{real_data}


In this section, we demonstrate the performance of our method pBLPR on a real fMRI data set and compare its performance with that of two de-sparsified methods. The fMRI data were provided by the Gallant Lab at UC Berkeley \cite{fMRI}. The fMRI measured blood oxygen level-dependent activity at 1331 discretized 3D brain volumes ($2 \times 2 \times 2.5$ millimeters): cube-like units called voxels. We use a sub-data set focusing on the responses in the ninth voxel, located in the brain region responsible for visual functions. A single human subject was shown pictures of everyday objects, such as trees, stars, and so on. Each picture was a 128 pixel by 128 pixel grayscale image, reduced to a vector of length 10921, as follows: (1) use a Gabor transform of the gray image to generate local contrast energy features $Z_j$; and (2) take the nonlinear transformation $X_j=\log(1+\sqrt{Z_j})$, for $j=1,\ldots,10921$. Training and validation data sets were collected during the experiment. There were 1750 natural images in the training data, consisting of a design matrix of dimensions $1750 \times 10921$. The validation data set contained responses to 120 natural images (we do not use the validation data in this study).

After reading the training data set into R, we calculate the variance of each feature (column) in $X$, and delete those columns with variances $\leq 1e^{-4}$. Then, we have a matrix of dimension $1750 \times 9076$. We further reduce the dimension of the design matrix using correlation screening, that is, sorting the correlations (Pearson correlation between every feature in $X$ and the response $Y$) in decreasing order of absolute value, and then selecting the top 500 features with the largest absolute correlations. We use the lasso+ols estimate of the feature coefficients, based on the $1750 \times 500$ design matrix, as the pseudo-true parameter values, denoted by $\beta^0$. We randomly choose a subset of $n=200$ rows to create a high-dimensional simulation setting, and then generate $Y$ from a linear regression model $y_i = x_i^\T \beta^0 + \epsilon_i$.  We \hanzhongsep{set $B=1000$ for the number of replications in the bootstrap}, and compare the performance of the \pBLPR{} method with that of \ZZ{} and \JM{}. 

\begin{figure}[ht]
	\centering
	\includegraphics[width=0.8\textwidth]{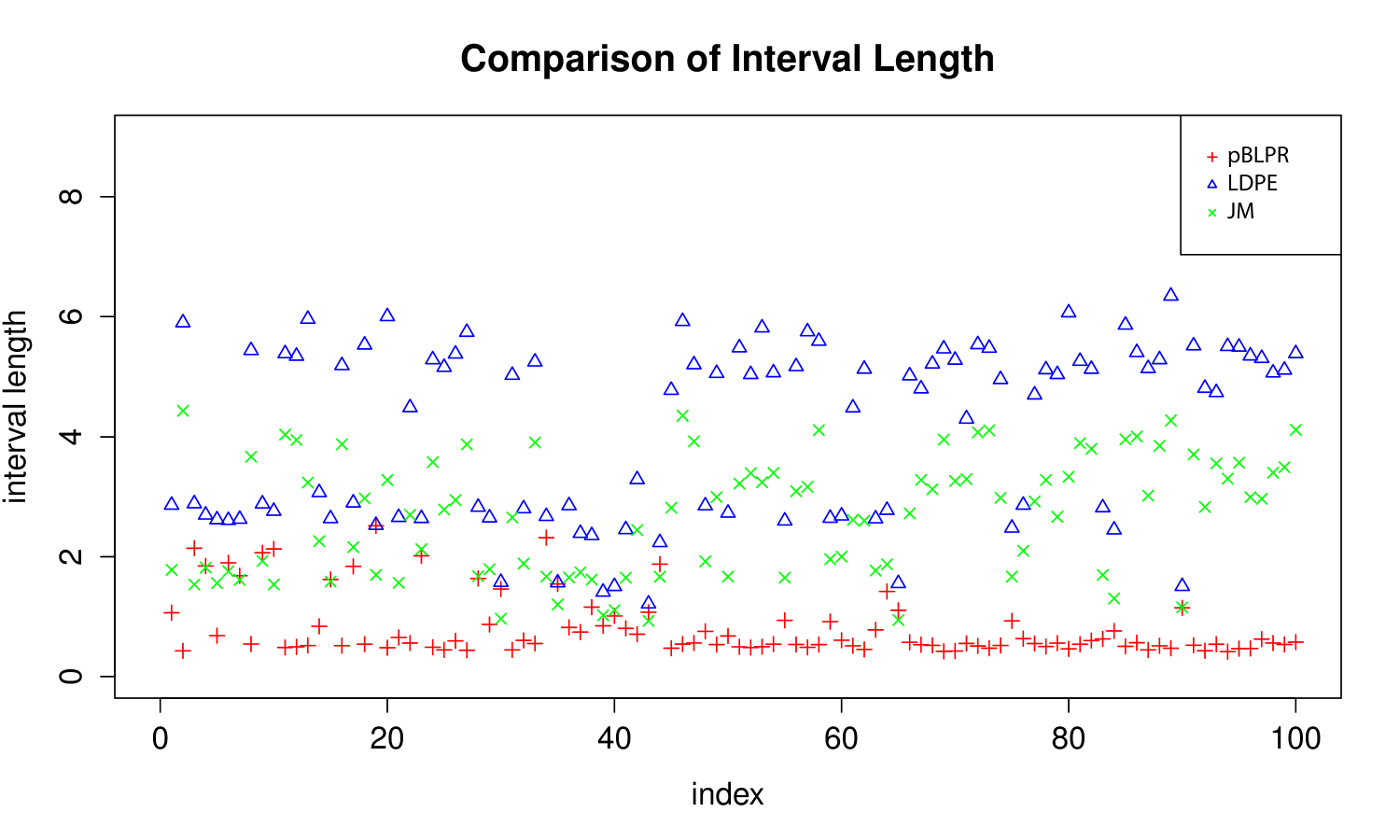}
	\caption[Comparison of pBLPR, \ZZ{}, \JM{} and BLDPE -- real fMRI data]{Comparison of interval lengths produced by pBLPR, \ZZ{}, \JM{}, and BLDPE. The plot is generated using the ninth voxel as the response in the fMRI data.}
	\label{fig:CI-fMRI}
\end{figure}

Based on the sub-data set with $n=200$ and $p=500$, we evaluate the performance of \pBLPR{}, \ZZ{}, and \JM{} in their construction of confidence intervals. The $95\%$ confidence intervals constructed by these three methods cover $95.8\%$, $97\%$, and $99.6\%$, respectively, of the $500$ components of $\beta^0$. All three methods cover more than $95\%$ of the pseudo-true values and, thus, have satisfactory performance in terms of coverage. In terms of interval lengths, however, our \pBLPR{} method produces much shorter confidence intervals than those of the other two methods for most of the coefficients, especially the small ones. Figure~\ref{fig:CI-fMRI} shows the confidence interval lengths of 100 coefficients (44 nonzero coefficients in $\beta^0$ and 56 randomly chosen zero coefficients) produced by the three methods. The satisfactory coverage and much shorter lengths of the confidence intervals produced by \pBLPR{}, demonstrate that it outperforms \ZZ{} and \JM{}, overall, in this real-data case study.

\revise{
\section{Real-data case study 2: neuroblastoma gene expression data}\label{real_data2}


In this section, we apply our \pBLPR{} and \rBLPR{} methods and three de-sparsified lasso methods, \ZZ{}, \finalrevise{\JM{}}, and BLDPE, to a data set containing $43,827$ gene expression measurements from the Illumina RNA sequencing of $498$ neuroblastoma samples (Gene Expression Omnibus accession number GSE62564, with the file name \texttt{GSE62564\_SEQC\_NB\_RNA-Seq\_log2RPM.txt.gz}) generated by the Sequencing Quality Control (SEQC) consortium \citep{wang2014concordance, su2014comprehensive, munro2014assessing, su2014investigation}. Each neuroblastoma sample was labeled as high-risk (HR) or non-HR, indicating whether the sample belonged to a HR patient based on clinical evidence. There were 176 HR samples and 322 non-HR samples.

Constructing gene-gene regulatory relationships is of primary interest for this data set. We encode the sample labels as a binary vector $Z \in R^{498}$, with \finalrevise{$Z_i = 1$} if the $i$th sample is HR, and $Z_i = 0$ otherwise. For the $j$th gene, we calculate the Pearson correlation between its gene expression vector $X_j \in R^{498}$ and $Z$, and we check the ten genes with the highest correlations. Among these ten genes, we find a gene \textit{CAMTA1}, which has been reported as a gene related to medulloblastoma \citep{wu2012clonal}, a type of cancer closely related to neuroblastoma, and included in the Candidate Cancer Gene Database (CCGD) \citep{abbott2015candidate}. We use the gene expression vector of \textit{CAMTA1} as the response vector $Y$, and we consider the gene expression matrix of the other $43,826$ genes as the design matrix of dimensions $498 \times 43,826$. Our goal is to find genes that have significant effects on predicting the expression of \textit{CAMTA1} in a multiple linear model. Given our lack of knowledge on the complex regulatory relationships between genes, the linear model is almost certainly a misspecified model. However, this case study would serve as a reasonable real-data example to demonstrate the ability of our \pBLPR{} and \rBLPR{} methods and three de-sparsified lasso methods (\ZZ{}, \JM{} and BLDPE) to identify significant predictors in a misspecified linear model.



Table~\ref{tab:SEQC} shows the numbers of significant genes found by the $95\%$ confidence intervals constructed by five methods. We find that  \ZZ{} and its bootstrap version, BLDPE, find the most significant genes; \pBLPR{} and \rBLPR{} find $91$ and $26$ significant genes, respectively; \JM{} finds only one significant genes. We investigate the biological functions of those significant genes by performing a Gene Ontology (GO) analysis using a bioinformatics online tool GOrilla \citep{eden2009tool}.  Specifically, between one of our methods (\pBLPR{} or \rBLPR{}) and one of the de-sparsified lasso methods (\ZZ{}, \JM{}, or BLDPE), we check the significant genes found by one method but not the other, and we obtain the functions (i.e., Biological Process GO terms) enriched in those genes by GOrilla. An interesting observation is that the functions related to natural and regulated cell deaths (e.g., apoptosis and autophagy), which are key processes used to prevent cancer, are only enriched in the significant genes found by \pBLPR{} or \rBLPR{}, but not in those found by any of the de-sparisied lasso methods. On the other hand, only general functions, such as basic processes in cells, are enriched in the significant genes found by a de-sparsified lasso method, but not by our methods. This suggests that \pBLPR{} and \rBLPR{} find significant features that are more reasonable and interpretable based on domain knowledge, implying that they are robust to model misspecification. Table \ref{tab:SEQC_GO} provides a summary of the numbers of the enriched GO terms and the specific terms related to apoptosis or autophagy. The detailed GO analysis results are provided in the Supplementary File.

\begin{table}
\revise{
	\centering
	\begin{tabular}{lccccc}
		\hline
		Method & pBLPR  &    rBLPR &   LDPE    &     JM   &   BLDPE \\
		\hline
       \# significant genes & 91     &    26     &     501    &      1   &     135 \\
       \hline
	\end{tabular}
	\caption{Numbers of significant genes found by the $95\%$ confidence intervals constructed by five methods.} \label{tab:SEQC}
	}
\end{table}

\begin{table}
\centering
\begin{tabular}{|c|ccc|}
\hline
$A \backslash B$ & LDPE & JM & BLDPE\\
\hline
pBLPR & $4/19$ & $6/15$ & $4/11$ \\
rBLPR & $2/14$ & $5/18$ & $2/11$ \\
\hline	
\end{tabular}
\begin{tabular}{|c|ccc|}
\hline
$B \backslash A$ & LDPE & JM & BLDPE\\
\hline
pBLPR & $0/67$ & $0/0$ & $0/18$ \\
rBLPR & $0/78$ & $0/0$ & $0/18$ \\
\hline	
\end{tabular}
\caption{The numbers of Biological Process GO terms enriched in the significant genes found by method $A$, but not by method $B$. The numerators are the numbers of GO terms related to apoptosis or autophagy, and the denominators are the total numbers of GO terms enriched in the significant genes. For example, $4/19$ in the left table indicates that there are $19$ GO terms enriched in the significant genes found by pBLPR, but not by LDPE, and among these $19$ terms, $4$ terms are related to apoptosis or autophagy.} \label{tab:SEQC_GO}
\end{table}
}

\section{Conclusion and future work}


Assigning p-values and constructing confidence intervals for parameters in high-dimensional sparse linear models are challenging tasks. The bootstrap, as a standard inference tool, has been shown useful in addressing this problem. However, previous works that extend the bootstrap technique to high-dimensional models rely on two key assumptions: the hard sparsity and beta-min condition. The beta-min condition is rather restrictive. In order to relax it, we propose two new bootstrap procedures based on a new two-stage estimator, called lasso+partial ridge. Our methods improve the performance of the bootstrap lasso+ols method proposed in \citep{Liu2013} when there exist a group of small, but nonzero regression coefficients. We conduct extensive simulation studies to compare our methods with \revise{three de-sparsified methods (\ZZ{}, \JM{}, and the bootstrap version of \ZZ{} (BLDPE))}. We find that our methods yield comparable coverage probabilities, but shorter (on average) intervals, and \JL{are more robust to} misspecified models than the other methods are \revise{under many scenarios}. We \JL{apply our methods to an} fMRI data set, finding that it gives reasonable coverage probabilities and shorter interval lengths than those of \ZZ{}, \JM{}, and BLDPE. \revise{In a second real-data application, we applied our methods to identify genes that have significant effects on predicting a cancer gene's expression levels in a (likely) misspecified linear model. Compared with three de-sparsified lasso methods, our methods find genes that are biologically more reasonable and interpretable, suggesting that our methods are robust to model misspecification \secrevise{in certain applications, despite the lack of rigorous theoretical analysis in this work. Future work is needed to investigate the robustness of various inference methods to different types of model misspecification, from both theoretical and empirical perspectives.}} 


\secrevise{A disadvantage of our method is that its resulting inference is not uniformly valid over the class of sparse models, owing to the cliff-weak-sparsity assumption. It is possible that our methods are uniformly valid for some pseudo-true parameter, that is, the parameters of the nearest model that satisfies the cliff-weak-sparsity; we leave this to future work. Moreover, compared with uniformly valid inference procedure such as the de-sparsified lasso methods, our empirical studies show that our methods are more likely to identify small, but nonzero coefficients, owing to the shorter confidence interval lengths returned by our methods. In many real-world applications, the covariates (or features) with small effects are not negligible, but may be important. For example, in genomic applications, where complex gene-gene regulatory relationships are of primary interest, researchers searching for regulators of a target gene are not only interested in the genes with large effects, but also in other genes with small effects. This is because many small effects have been discovered to play important functional roles in biological mechanisms. In this application, our methods provide a means to identify genes with small effects. However, note that subsequently experiments are still required to validate the identified genes. Furthermore, when an individual coefficient is too small, no method can successfully identify it; then, a statistical procedure should instead aim to detect the joint significance of a set of covariates.

}

Overall, the bootstrap lasso+ols method has the shortest confidence interval lengths, with good coverage probabilities, for large coefficients. However, for small, but nonzero coefficients, the bootstrap \LPR{} method (\rBLPR{} and \pBLPR{}) has the shortest confidence interval lengths, with good coverage probabilities. \secrevise{Therefore, if practitioners focus on the confidence intervals for large coefficients, we recommend the bootstrap lasso+ols method; however, if they are also interested in identifying small, but significant coefficients in a possibly misspecified linear model, we recommend our bootstrap \LPR{} methods. Nevertheless, note that the confidence intervals of the coefficients, with magnitudes of order $1/\sqrt{n}$, may be invalid. If practitioners' major concern is the coverage probabilities of confidence intervals, they should use the de-sparsified lasso methods, which are uniformly valid over the class of sparse models. Moreover, from an application perspective, our bootstrap \LPR{} methods have the advantages of being technically simple, interpretable, and easy to implement and parallelize.}


\secrevise{Finally,} multiple testing is another important task in hypothesis testing, and is closely related to confidence interval construction. Several procedures, such as the Bonferroni correction, Benjamini--Hochberg procedure and FDR control, have been proposed to correct multiple testing in low-dimensional settings. However, these procedures are based on accurate estimations of the $p$-values of each test, where small $p$-values can only be obtained using large numbers of bootstrap runs (e.g., a $p$-value of $0.001$ requires at least $1000$ runs), thus imposing too much computational complexity. We leave the correction for multiple testing in high-dimensional models as future work.


\section*{Acknowledgments}

The authors would like to thank the Gallant Lab at UC Berkeley for providing the fMRI data, Simon Walter (UC Berkeley) and Dr. Chad Hazlett (UCLA) for their edits and suggestions that have helped clarify the text, and Prof. Bin Yu at UC Berkeley for her helpful discussions and comments that have helped improve the quality of the paper. Dr. Hanzhong Liu's research is partially supported by NSF grants DMS-1613002, DMS-1228246, AFOSR grant FA9550-14-1-0016, and the National Natural Science Foundation of China 11701316. Dr. Jingyi Jessica Li's research is supported by the Hellman Fellowship, the PhRMA Foundation Research Starter Grant in Informatics, NIH/NIGMS grant R01GM120507, and NSF grant DMS-1613338.


\appendix

\section{Organization of the Appendix}

The Appendix is organized as follows. Section A1 contains proofs of the results in the main text. Section A2 provides examples that satisfy (or do not satisfy) Condition 8. Section A3 gives an example that satisfies Condition 11. Section A4  provides additional figures and tables of simulation results.  The details of the algorithm for cv(lasso+ols) is provided in Section A5.

\setcounter{section}{0}
\setcounter{equation}{0}
\def\theequation{A\arabic{section}.\arabic{equation}}
\def\thesection{A\arabic{section}}

\section{Proof of Theorems}

\subsection{Proof of Theorem~\ref{thm:selection-consistency-lasso}}
We will follow the proof of the sign-consistency of the lasso in \citet{ZhaoYu2006} with modifications when necessary. Before proving Theorem~\ref{thm:selection-consistency-lasso}, we first state the following Proposition 1 which is similar to the Proposition 1 in \cite{ZhaoYu2006}. 


\begin{proposition}
	\label{prop1}
	Assume Condition \ref{cond:irrepresentable} holds with a constant $\eta > 0$, then
	\begin{equation}
	\pr \left((\hat{\beta}_{\textnormal{lasso}})_S=_s \beta^0_S,  ~(\hat{\beta}_{\textnormal{lasso}})_{S^c}=0\right)\geq \pr (A_n \cap B_n)
	\end{equation}
	for
	$$A_n=\left\{|C_{11}^{-1}W_S| < n^{\frac{1}{2}} \left(|\beta^0_S|-\lambda_{1}|C_{11}^{-1}\sign(\beta^0_S)|-|C_{11}^{-1}C_{12}\beta^0_{S^c}| \right) \right\},$$
	$$B_n=\left\{|C_{21}C_{11}^{-1}W_S-W_{S^c}|\leq n^{\frac{1}{2}}\lambda_{1}\eta-|n^{\frac{1}{2}}(C_{21}C_{11}^{-1}C_{12}-C_{22})\beta^0_{S^c}| \right\},$$
	where
	$$W_S=n^{-\frac{1}{2}}X_S^{\T}\epsilon,\quad \ W_{S^c}=n^{-\frac{1}{2}}X_{S^c}^{\T}\epsilon.$$
\end{proposition}
Setting $\beta^0_{S^c}=0$, then Proposition 1 gives back to the same proposition in \citet{ZhaoYu2006}.

\begin{proof}
	
	By Karush-Kuhn-Tucker condition for convex optimization, we obtain the following Lemma \ref{lemma:KKT} without giving the proof.
	
	\begin{lemma}
		\label{lemma:KKT}
		$\hat{\beta}_{\textnormal{lasso}}$ is the lasso estimator defined in \eqref{eqn:lasso} if and only if 
		$$\frac{1}{2n} \frac{d\left\|Y-X\beta\right\|_2^2}{d\beta_j}|_{\beta_j=(\hat{\beta}_{\textnormal{lasso}})_j}= - \lambda_1\sign\left((\hat{\beta}_{\textnormal{lasso}})_j\right)\ \ \  \textnormal{for} \ j, \ \textnormal{such that} \ (\hat{\beta}_{\textnormal{lasso}})_j \neq 0, $$
		$$\frac{1}{2n} \left | \frac{d\left\|Y-X\beta\right\|_2^2}{d\beta_j}|_{\beta_j=(\hat{\beta}_{\textnormal{lasso}})_j} \right | \leq \lambda_1 \ \ \ \textnormal{for} \ j, \ \textnormal{such that} \ (\hat{\beta}_{\textnormal{lasso}})_j = 0.$$
	\end{lemma}

	It is easy to obtain
	$$\frac{1}{2n} \frac{d\left\|Y-X\beta\right\|_2^2}{d\beta} = -\frac{1}{n} X^{\T}(Y-X\beta) = C (\beta - \beta^0) -\frac{1}{n} X^{\T} \epsilon, $$
	where $C=X^{\T}X/n.$ Then by definition of the lasso \eqref{eqn:lasso} and Lemma~\ref{lemma:KKT}, if there exist $\hat \beta = (\hat \beta_S^{\T}, 0_{S^c}^{\T})^{\T}$, such that the following holds: 
	\begin{equation}
	\label{eqn:KKT-1}
	n^{\frac{1}{2}} C_{11}(\hat \beta -\beta^0)_S- n^{\frac{1}{2}} C_{12}\beta^0_{S^c}- X_{S}^{\T}\epsilon/n^{\frac{1}{2}}=-n^{\frac{1}{2}}\lambda_{1}\sign(\beta^0_S),
	\end{equation}	
	\begin{equation}
	\label{eqn:KKT-2}
	-n^{\frac{1}{2}}\lambda_{1} {\bf 1} \leq n^{\frac{1}{2}} C_{21} (\hat \beta -\beta^0)_S -X_{S^c}^{\T}\epsilon/n^{\frac{1}{2}}- n^{\frac{1}{2}} C_{22}\beta^0_{S^c} \leq n^{\frac{1}{2}}\lambda_{1}  {\bf 1} ,
	\end{equation}
	\begin{equation}
	|(\hat{\beta} - \beta^0)_S|<|\beta^0_S|,
	\end{equation}
	then, $\hat \beta $ is the lasso solution, that is, $\hat \beta = \hat \beta_{\textnormal{lasso}}$ and, hence, $(\hat \beta_{\textnormal{lasso}})_{S^c}= \hat \beta_{S^c} = 0$ and $\sign((\hat{\beta}_{\textnormal{lasso}})_S)= \sign(\hat \beta_S)= \sign(\beta^0_S)$. Let $W=X^{\T}\epsilon/n^{\frac{1}{2}}$, then, $W_S= X_{S}^{\T}\epsilon/n^{\frac{1}{2}}$ and $W_{S^c}= X_{S^c}^{\T}\epsilon/n^{\frac{1}{2}}$.
	
	Substitute $(\hat \beta -\beta^0)_S$ and bound the absolute values, the existence of such $\hat \beta$ is implied by 
	\begin{equation}
	\label{eqn:An}
	|C_{11}^{-1}W_S| < n^{\frac{1}{2}}\left( |\beta^0_S|-\lambda_{1}|C_{11}^{-1}\sign(\beta^0_S)|-|C_{11}^{-1}C_{12}\beta^0_{S^c}| \right),
	\end{equation}
	\begin{equation}
	\label{eqn:Bn}
	|C_{21}C_{11}^{-1}W_S-W_{S^c}|\leq n^{\frac{1}{2}}\lambda_{1}\left( {\bf 1} -|C_{21}C_{11}^{-1}\sign(\beta^0_S)|\right)-|n^{\frac{1}{2}}\left( C_{21}C_{11}^{-1}C_{12}-C_{22} \right) \beta^0_{S^c}|.
	\end{equation}
	$\left\{ \eqref{eqn:An} \right\}$ coincides with $A_n$ and $\left\{ \eqref{eqn:Bn} \right\}  \subset B_n$. This implies Proposition 1. 

	
	To prove Theorem \ref{thm:selection-consistency-lasso}, we can follow the proof of Theorem 4 in \citet{ZhaoYu2006}, using our new Proposition \ref{prop1}.

	First, by Proposition \ref{prop1}, we have 
	$$\pr \left((\hat{\beta}_{\textnormal{lasso}})_S=_s \beta_S,  ~(\hat{\beta}_{\textnormal{lasso}})_{S^c}=0\right)\geq \pr (A_n \cap B_n).$$
	On the other hand, 
	\begin{equation}
	\begin{split}
	& 1-\pr (A_n\cap B_n) \leq \pr (A_n^c)+\pr (B_n^c)\\
	&\leq \sum_{i=1}^s\pr \left( |z_i|\geq n^{\frac{1}{2}}(|\beta^0_i|-\lambda_{1}b_i-h_i) \right)+\sum_{i=1}^{p-s}\pr \left(|\zeta_i|\geq n^{\frac{1}{2}}\lambda_{1}\eta_i-m_i\right), 
	\end{split}
	\end{equation}
	where 
	$$ z=(z_1,\dots,z_s)^{\T}=C_{11}^{-1}W_S, $$
	$$
	\zeta=(\zeta_1,\dots,\zeta_{p-s})^{\T}=C_{21}C_{11}^{-1}W_S-W_{S^c}, $$
	$$
	b=(b_1,\dots,b_s)=|C_{11}^{-1}\sign(\beta^0_S))|, $$
	$$
	h=(h_1,\dots,h_s)=|C_{11}^{-1}C_{12}\beta^0_{S^c}|, $$
	$$
	m=(m_1,\dots,m_{p-s})=|n^{\frac{1}{2}}\left(C_{21}C_{11}^{-1}C_{12}-C_{22}\right)\beta^0_{S^c}|.
	$$
	Due to Condition \ref{cond:subgaussian}, $\epsilon_i$ are independent and identically distributed sub-Gaussian random variables, with mean 0 and variance $\sigma^2$. Therefore, $z_i$'s and $\zeta_i$'s are all sub-Gaussian random variables, with mean 0. By simple algebra, we have
	$$E(zz^{\T})= \sigma^2 C_{11}^{-1}; \quad  E(\zeta \zeta^{\T})= \sigma^2 (C_{22}-C_{21}C_{11}^{-1}C_{12}).$$
	Therefore,
	$$ Ez_i^2 = \sigma^2 (C_{11}^{-1})_{ii} \leq \sigma^2  \Lambda_{\max}(C_{11}^{-1}) \leq \sigma^2  /\Lambda,$$
	where the last inequality is due to Condition \ref{cond:smallesteigen}. Moreover,
	$$E\zeta_i^2 = \sigma^2 (C_{22}-C_{21}C_{11}^{-1}C_{12})_{ii} \leq \sigma^2 (C_{22})_{ii} = \sigma^2 ,$$
	where the last equality is because of Condition \ref{cond:standard}.
	Therefore, $z_i$'s and $\zeta_i$'s are sub-Gaussian random variables, with mean 0 and finite variance. Hence, there exits a constant $c>0$, such that,  for all $  t > 0$, 
	$$\pr (|z_i| \geq t) \leq 2e^{-ct^2}; \quad  \pr (|\zeta_i| \geq t) \leq 2e^{-ct^2}.$$
	
	For $ i=1,\ldots,s$, using Cauchy-Schwarz inequality and Conditions \ref{cond:smallesteigen}, \ref{cond:scaling}, and \ref{cond:lambda1}, we have
	\begin{eqnarray}
	n^{\frac{1}{2}} \lambda_1 |b_i| & \leq & n^{\frac{1}{2}} \lambda_1 \Lambda_{\max}(C_{11}^{-1}) || \sign(\beta^0_S)||_2 \leq s^{\frac{1}{2}} n^{\frac{1}{2}} \lambda_1 /\Lambda  \nonumber \\
	& = & O(n^{\frac{1}{2}} n^{\frac{c_1+c_4-1}{2}}) = o(n^{\frac{1}{2}} n^{\frac{c_3-1}{2}}), \nonumber
	\end{eqnarray}
	where the last inequality holds because $c_4<c_3-c_1$ (see Condition \ref{cond:lambda1}).
	
	Condition \ref{cond:extra} implies that $n^{\frac{1}{2}}h_i = O(1),$ for $ i=1,\ldots,s $. Combining with Condition \ref{cond:cliff-weak}, we have
	$$ n^{\frac{1}{2}} \lambda_1 |b_i| +  n^{\frac{1}{2}}h_i = o(1) n^{\frac{1}{2}} |\beta_i^0|,  \ \textnormal{for} \ i = 1,\cdots,s. $$
	Therefore,
	\begin{equation}
	\begin{split}
	\sum_{i=1}^s \pr \left(|z_i|\geq n^{\frac{1}{2}}(|\beta_i^0|-  \finalrevise{ \lambda_{1} } b_i-h_i) \right) 
	&\leq \sum_{i=1}^s \pr \left(|z_i|\geq (1+o(1))n^{\frac{1}{2}}|\beta_i^0| \right)\\
	&\leq \sum_{i=1}^s \pr (|z_i|\geq n^{\frac{c_3}{2}})\\
	& = o(e^{-n^{c_2}}).
	\end{split}
	\end{equation}
	Due to Conditions \ref{cond:extra} and \ref{cond:lambda1}, $m_i = o(n^{c_4/2})$, and $n^{\frac{1}{2}} \lambda_1 = O(n^{c_4/2})$. Then,
	\begin{equation}
	\begin{split}
	\sum_{i=1}^{p-s}\pr \left( |\zeta_i|\geq n^{\frac{1}{2}}\lambda_{1}\eta_i-m_i \right)
	&\leq \sum_{i=1}^{p-s}\pr \left(|\zeta_i|\geq O(n^{\frac{c_4}{2}}) \right) = o(e^{-n^{c_2}} ).
	\end{split}
	\end{equation}
	Theorem 1 follows immediately.	
\end{proof}

\subsection{Proof of Theorem~\ref{thm:selection-consistency-bootlasso-resid}}

\begin{proof}
	We have to check that the residual bootstrap version\footnote{Replacing $(\beta^0, \epsilon, Y)$ with $(\hat \beta_{\textnormal{lasso+ols}}, \epsilon^*, Y^*_{\textnormal{rboot}})$.} of Conditions \ref{cond:subgaussian} -- \ref{cond:lambda1} hold, with conditional probability, given $\epsilon$, going to one. For the residual bootstrap sample, we have
	\[  Y^*_{\textnormal{rboot}} = X \hat \beta_{\textnormal{lasso+ols}} +  \epsilon^*.\]
	Conditions \ref{cond:standard}, \ref{cond:smallesteigen} and \ref{cond:lambda1} depend only on $X$ and $\lambda_{1}$ which are the same for the original sample $(X,Y)$ and bootstrap sample $(X,Y^*_{\textnormal{rboot}})$, therefore, they hold obviously. We next show, one by one, the bootstrap version of Conditions \ref{cond:subgaussian}, \ref{cond:scaling} -- \ref{cond:extra} hold, with probability going to one. We need the following Lemma.
	\begin{lemma}
		\label{lemma:lasso+ols}
		Under Conditions \ref{cond:subgaussian} -- \ref{cond:lambda1}, and for the constant $M$ in Condition \ref{cond:cliff-weak}, we have
		\begin{equation}
		\label{eqn:infinity-norm-lassoOLS}
		\pr \left( ||\hat \beta_{\textnormal{lasso+ols}} - \beta^0||_{\infty} \leq 2M n^{\frac{c_1-1}{2}} \right) \rightarrow 1.
		\end{equation}
	\end{lemma}
	Lemma \ref{lemma:lasso+ols} bounds element-wise estimation error of the lasso+ols estimator, the proof of which can be founded in the following subsection~\ref{proof:lemma}.
	
	Now, we can show that residual bootstrap version of Conditions \ref{cond:subgaussian}, \ref{cond:scaling} -- \ref{cond:extra} hold, with probability going to one. Under Conditions \ref{cond:subgaussian} -- \ref{cond:lambda1} and using Theorem \ref{thm:selection-consistency-lasso}, the lasso $\hat \beta_{\textnormal{lasso}}$ has sign-consistency, that is,
	\[ \pr (\hat S = S ) = 1 - o(e^{-n^{c_2}}) \rightarrow 1.\]
	
	
	In what follows, we always condition on $\{ \hat S = S \}$. By Lemma \ref{lemma:lasso+ols}, it is easy to show that
	\[ \pr \left( (\hat \beta_{\textnormal{lasso+ols}})_S =_s \beta^0_S  \right) \rightarrow 1,\]
	which guarantees that bootstrap version of Conditions \ref{cond:scaling} -- \ref{cond:extra} hold, with probability going to one. Therefore, we only need to show the bootstrap version of Condition \ref{cond:subgaussian} holds, with probability going to 1, that is, 	
	\begin{condition}
		\label{cond:subgaussian*}
		$\epsilon_i^*$ are conditionally independent and identically distributed sub-Gaussian random variables, with mean 0. That is, there exists constant $C^*>0$ and $c^*>0$, such that
		\begin{equation}
		\label{eqn:subguassionbootstrap}
		\pr \left(|\epsilon_i^*| \geq t  \mid \epsilon \right) \leq C^* e^{-c^*t^2}, \ \forall t\geq 0,
		\end{equation}
		holds in probability.
	\end{condition}

\begin{lemma}
\label{lemma:subgaussian*}	
Conditions \ref{cond:subgaussian} -- \ref{cond:s-square-1} imply Condition \ref{cond:subgaussian*}.
\end{lemma} 	
	
	The proof is similar to that in \citet{Liu2013} with modifications accounting for cliff-weak-sparsity. Let $\mathbb{I}_\cdot$ denote the indicator function. Note that $\pr(|\epsilon_i^*| \geq t \mid \epsilon) = (\sum_{i=1}^{n} \mathbb{I}_{|\hat \epsilon_i - \tilde{\epsilon}| \geq t})/n$, hence, \ref{eqn:subguassionbootstrap} is equivalent to
	\begin{equation}
	\label{eqn:subguassionbootstrap1}
	\mathop {\sup}\limits_{t\geq 0} \left\{ \frac{1}{n} \sum\limits_{i=1}^{n} e^{c^*t^2} \mathbb{I}_{|\hat \epsilon_i - \tilde{\epsilon}| \geq t} \right\} \leq C^*.
	\end{equation}
	We know that
	\begin{eqnarray}
	\hat \epsilon_i - \tilde{\epsilon} & = & y_i - x_i^{\T} \hat \beta_{\textnormal{lasso+ols}} - (\overline{y} - \overline{x}^{\T} \hat \beta_{\textnormal{lasso+ols}} ) \nonumber \\
	& = & x_i^{\T} \beta^0 + \epsilon_i - x_i^{\T} \hat \beta_{\textnormal{lasso+ols}}  - (\overline{x}^{\T} \beta^0 + \overline{\epsilon} - \overline{x}^{\T} \hat \beta_{\textnormal{lasso+ols}} ) \nonumber \\
	& = & x_i ^{\T} (\beta^0 - \hat \beta_{\textnormal{lasso+ols}} ) + \epsilon_i - \bar{\epsilon},
	\label{eqn:1121}
	\end{eqnarray}
	where $x_i^{\T}$ is the $i$th row of $X$, $\bar{y}=\sum_{i=1}^{n} {y_i}/n$, $\bar{\epsilon}= \sum_{i=1}^{n} {\epsilon_i}/n$, and $\bar{x}=  \sum_{i=1}^{n} {x_i}/n = 0$.
	It is easy to see that $\mathop {\sup}_{t\geq 0} \left\{  ( \sum_{i=1}^{n} e^{c^*t^2} \mathbb{I}_{|\hat \epsilon_i -  \tilde{\epsilon} | \geq t})/n \right\}$ can be bounded by
	\begin{eqnarray}
	\frac{1}{n} \sum\limits_{i=1}^{n} \left \{ \mathop {\sup}\limits_{t\geq 0} \left\{ e^{c^*t^2} \mathbb{I}_{|x_i^{\T} (\beta^0 - \hat \beta_{\textnormal{lasso+ols}})| \geq t/3} \right\} + \mathop {\sup}\limits_{t\geq 0} \left\{ e^{c^*t^2} \mathbb{I}_{|\overline{\epsilon}| \geq t/3} \right\} + \mathop {\sup}\limits_{t\geq 0} \left\{ e^{c^*t^2} \mathbb{I}_{|\epsilon_i| \geq t/3} \right\} \right \}. \nonumber \\
	\label{eqn:11211}
	\end{eqnarray}
	We can bound the second and third terms exactly the same as those in \citet{Liu2013}, that is, there exist a constant $C^*_1>0$, such that for $c^* = 1/(36\sigma^2)$,
	\begin{equation}
	\label{eqn:part2}
	\pr \left( \frac{1}{n} \sum\limits_{i=1}^{n} \mathop {\sup}\limits_{t\geq 0} \left\{ e^{c^*t^2} \mathbb{I}_{|\overline{\epsilon}| \geq t/3} \right\} \leq C_1^* \right) \rightarrow 1.
	\end{equation}
	\begin{equation}
	\label{eqn:part3}
	\pr \left( \frac{1}{n} \sum\limits_{i=1}^{n}  \mathop {\sup}\limits_{t\geq 0} \left\{ e^{c^*t^2} \mathbb{I}_{|\epsilon_i| \geq t/3} \right\} \leq C_1^*   \right) \rightarrow 1.
	\end{equation}
	Since the proof is exactly the same, we omit it here. Next, we bound the first term, which is different from that in \citet{Liu2013}, because of the weaker cliff-weak-sparsity assumption. For the constant $D>0$ appearing in Condition \ref{cond:s-square-1},
	\begin{eqnarray}
	& & \pr \left(\mathop {\max}\limits_{1\leq i \leq n} |x_i^{\T} (\beta^0 - \hat \beta_{\textnormal{lasso+ols}})| \geq 2D \right) \nonumber \\
	& = & \pr \left(\mathop {\max}\limits_{1\leq i \leq n} |x_i^{\T} (\beta^0 - \hat \beta_{\textnormal{lasso+ols}})| \geq 2D,\hat S=S \right) \nonumber \\
	&& + \pr \left(\mathop {\max}\limits_{1\leq i \leq n} |x_i^{\T} (\beta - \hat \beta_{\textnormal{lasso+ols}})| \geq 2D,\hat S \neq S \right)  \nonumber \\
	& \leq & \pr \left(\mathop {\max}\limits_{1\leq i \leq n} | x_{i,S}^{\T} (\beta^0_S - (\hat \beta_{\textnormal{lasso+ols}})_S) + x_{i,S^c}^{\T} \beta^0_{S^c}  | \geq 2D \right)  + \pr (\hat S \neq S) \nonumber \\
	& \leq & \pr \left( \mathop {\max}\limits_{1\leq i \leq n} | x_{i,S}^{\T} (\beta^0_S - (\hat \beta_{\textnormal{lasso+ols}})_S)  | \geq D \right) + \pr \left( \mathop {\max}\limits_{1\leq i \leq n} | x_{i,S^c}^{\T} \beta^0_{S^c}  | \geq D \right) \nonumber \\
	&& + \pr (\hat S \neq S) \nonumber \\
	& = & \pr \left( \mathop {\max}\limits_{1\leq i \leq n} | x_{i,S}^{\T} (\beta^0_S - (\hat \beta_{\textnormal{lasso+ols}})_S   | \geq D \right) + \pr (\hat S \neq S),
	\end{eqnarray}
	where the last equality holds because of Condition \ref{cond:s-square-1}. Using Cauchy-Schwarz inequality and Lemma \ref{lemma:lasso+ols}, we have
	\begin{eqnarray}
	& &  \mathop {\max}\limits_{1\leq i \leq n}  | x_{i,S}^{\T} (\beta^0_S - (\hat \beta_{\textnormal{lasso+ols}})_S)|  \leq \mathop {\max}\limits_{1\leq i \leq n} || x_{i,S}||_2 ||\beta^0_S - (\hat \beta_{\textnormal{lasso+ols}})_S||_2. \nonumber 
	\end{eqnarray}
	Conditional on $\{ \hat S = S \}$, the lasso+ols estimator has the following form:
	\begin{eqnarray}
	(\hat \beta_{\textnormal{lasso+ols}})_S  = (X_S^{\T}X_S)^{-1}X_S^{\T}Y = \beta^0_S + C_{11}^{-1} C_{12} \beta^0_{S^c} + (X_S^{\T}X_S)^{-1}X_S^{\T} \epsilon;
	\end{eqnarray}
	\[ \  (\hat \beta_{\textnormal{lasso+ols}})_{S^c} = 0. \]	
	Therefore, together with Condition \ref{cond:extra},
	\begin{eqnarray}
	 ||\beta^0_S - (\hat \beta_{\textnormal{lasso+ols}})_S||_2 & \leq || C_{11}^{-1} C_{12} \beta^0_{S^c} ||_2 + || (X_S^{\T}X_S)^{-1}X_S^{\T} \epsilon ||_2 \nonumber  \\
	 &= O\left( (s/n)^{1/2} \right) + || (X_S^{\T}X_S)^{-1}X_S^{\T} \epsilon ||_2 . 
	\end{eqnarray}
	Hence, by Condition \ref{cond:s-square-1},
	\begin{eqnarray}
	\label{eqn:x-S-beta1}
	&& \mathop {\max}\limits_{1\leq i \leq n}  | x_{i,S}^{\T} (\beta^0_S - (\hat \beta_{\textnormal{lasso+ols}})_S)| \nonumber \\
	& \leq&  o(n^{1/4}(s/n)^{1/2} ) +  \mathop {\max}\limits_{1\leq i \leq n} || x_{i,S}||_2 || (X_S^{\T}X_S)^{-1}X_S^{\T} \epsilon ||_2 \nonumber \\
	& =& o(1) + \mathop {\max}\limits_{1\leq i \leq n} || x_{i,S}||_2 || (X_S^{\T}X_S)^{-1}X_S^{\T} \epsilon ||_2.
	\end{eqnarray}
	It is easy to show that
	\begin{equation}
	\label{eqn:x-S-beta2}
	\mathop {\max}\limits_{1\leq i \leq n} || x_{i,S}||_2 || (X_S^{\T}X_S)^{-1}X_S^{\T} \epsilon ||_2 =o_p (1), 
	\end{equation}
	therefore,
	\begin{eqnarray}
	 \pr \left( \mathop {\max}\limits_{1\leq i \leq n} || x_{i,S}||_2 || (X_S^{\T}X_S)^{-1}X_S^{\T} \epsilon ||_2  
	 \geq  D \right)\rightarrow 0. 
	\end{eqnarray}
	Hence, 
	\begin{equation}
	\label{eqn:max_xi_beta}
	\pr \left(\mathop {\max}\limits_{1\leq i \leq n} |x_i^{\T} (\beta^0 - \hat \beta_{\textnormal{lasso+ols}})| \geq 2D \right) \rightarrow 0. 
	\end{equation}
	Therefore,
	\begin{eqnarray}
	& & \pr \left(\frac{1}{n} \sum\limits_{i=1}^{n} \mathop {\sup}\limits_{t\geq 1} \left\{ e^{c^*t^2} \mathbb{I}_{|x_i^{\T} (\beta^0 - \hat \beta_{\textnormal{lasso+ols}})| \geq t/3} \right\} \leq  e^{36D^2c^*} \right) \nonumber \\
	&\geq &  \pr \left(\mathop {\max}\limits_{1\leq i \leq n} |x_i^{\T} (\beta^0 - \hat \beta_{\textnormal{lasso+ols}})| < 2D \right) \rightarrow 1.
	\end{eqnarray}
	The above inequality holds, because it is easy to show that
	\begin{eqnarray}
	&& \left\{\frac{1}{n} \sum\limits_{i=1}^{n} \mathop {\sup}\limits_{t\geq 1} \left\{ e^{c^*t^2} \mathbb{I}_{|x_i^{\T} (\beta^0 - \hat \beta_{\textnormal{lasso+ols}})| \geq t/3} \right\} \leq  e^{36D^2c^*} \right \}  \nonumber \\
	& \supseteq & \left\{ \mathop {\max}\limits_{1\leq i \leq n} |x_i^{\T} (\beta^0 -  \hat \beta_{\textnormal{lasso+ols}})| < 2D \right\}. 
	\end{eqnarray}
	It is clear that
	\[ \frac{1}{n} \sum\limits_{i=1}^{n} \mathop {\sup}\limits_{0\leq t \leq 1} \left\{ e^{c^*t^2} \mathbb{I}_{|x_i^{\T} (\beta^0 - \hat \beta_{\textnormal{lasso+ols}})| \geq t/3} \right\} \leq e^{c^*} . \]
	Therefore, with probability going to $1$, we have
	\begin{eqnarray}
	\label{eqn:part1}
	&& \frac{1}{n} \sum\limits_{i=1}^{n} \mathop {\sup}\limits_{t\geq 0} \left\{ e^{c^*t^2} \mathbb{I}_{|x_i^{\T} (\beta^0 - \hat \beta_{\textnormal{lasso+ols}})| \geq t/3} \right\} \nonumber \\
	&= &  \max\left\{ \frac{1}{n} \sum\limits_{i=1}^{n} \mathop {\sup}\limits_{0 \leq t\leq 1} \left\{ e^{c^*t^2} \mathbb{I}_{|x_i^{\T} (\beta^0 - \hat \beta_{\textnormal{lasso+ols}})| \geq t/3} \right\}, \right. \nonumber\\
	&& \left. \frac{1}{n} \sum\limits_{i=1}^{n} \mathop {\sup}\limits_{t\geq 1} \left\{ e^{c^*t^2} \mathbb{I}_{|x_i^{\T} (\beta^0 - \hat \beta_{\textnormal{lasso+ols}})| \geq t/3} \right\}  \right\} \nonumber \\
	& \leq & \max \left\{ e^{c^*}, e^{36D^2c^*} \right\}. 
	\end{eqnarray}
	Let $C^*=2C_1^* + \max\left\{   e^{c^*}, e^{36D^2c^* } \right\}$, and combine \eqref{eqn:part1}, \eqref{eqn:part2}, and \eqref{eqn:part3}, 
	\[ \pr \left( \mathop {\sup}\limits_{t\geq 0} \left\{ \frac{1}{n} \sum\limits_{i=1}^{n} e^{c^*t^2} \mathbb{I}_{|\hat \epsilon_i - \tilde{\epsilon}| \geq t} \right\} \leq C^* \right) \rightarrow 1. \]
\end{proof}

\subsection{Proof of Theorem~\ref{thm:valid-resid-boot}}

\begin{proof}
First, by Theorem~\ref{thm:selection-consistency-lasso} and Theorem~\ref{thm:selection-consistency-bootlasso-resid}, both the lasso, $\hat \beta_{\textnormal{lasso}}$, and the residual bootstrap lasso, $\hat \beta_{\textnormal{rBlasso}}$, have model selection consistency. We can continue our argument by conditioning on $\{\hat S =S\}$ and $\{\hat S^*_{\textnormal{rBlasso}} =S\}$.

Second, we next show that
\begin{equation}\label{dist-lpr}
	n^{\frac{1}{2}}u^{\T}(\hat \beta_{\textnormal{\LPR{}}}-\beta^0) = n^{-\frac{1}{2}}u^{\T} C_{\lambda_2}^{-1} X^{\T}\epsilon + o_p (1);
	\end{equation}
\begin{equation}\label{dist-boot-lpr}
	n^{\frac{1}{2}}u^{\T}(\hat \beta^*_{\textnormal{\rBLPR{}}}-\hat \beta_{\textnormal{lasso+ols}}) = n^{-\frac{1}{2}}u^{\T} C_{\lambda_2}^{-1} X^{\T}\epsilon^* + o_p (1).
	\end{equation}
By definition, $\hat \beta_{\textnormal{\LPR{}}}$ is the solution of the following equation:
\[  -\frac{1}{n} X^{\T} ( Y - X \hat \beta_{\textnormal{\LPR{}}} ) + \lambda_2 \left({\bf 0}^{\T}, \left( \hat \beta_{\textnormal{\LPR{}}, S^c} \right)^{\T} \right)^{\T} = 0.   \]
Since $Y = X \beta^0 + \epsilon$, we have
\[  \frac{1}{n} X^{\T}   X ( \hat \beta_{\textnormal{\LPR{}}} - \beta^0 )  -  \frac{1}{n} X^{\T}     \epsilon + \lambda_2 \left({\bf 0}^{\T}, \left( \hat \beta_{\textnormal{\LPR{}}, S^c} \right)^{\T} \right)^{\T} = 0.   \]
Simple linear algebra gives 
\[  C_{\lambda_2}  ( \hat \beta_{\textnormal{\LPR{}}} - \beta^0 )  =  \frac{1}{n} X^{\T}     \epsilon  -    \lambda_2 \left({\bf 0}^{\T}, \left( \beta^0_{S^c} \right)^{\T} \right)^{\T}  .   \]
Therefore,
\begin{eqnarray}\label{dist-lpr}
	n^{\frac{1}{2}}u^{\T}(\hat \beta_{\textnormal{\LPR{}}}-\beta^0) & = & n^{-\frac{1}{2}}u^{\T} C_{\lambda_2}^{-1} X^{\T}\epsilon - \lambda_2 n^{\frac{1}{2}} u^{\T} C_{\lambda_2}^{-1} \left({\bf 0}^{\T}, \left( \beta^0_{ S^c} \right)^{\T} \right)^{\T} \nonumber \\
	&  = & n^{-\frac{1}{2}}u^{\T} C_{\lambda_2}^{-1} X^{\T}\epsilon + o_p (1),
	\end{eqnarray}
where the second equality is due to Condition~\ref{cond:orthogonal} and $\lambda_2 \propto n^{-1}$ in Condition \ref{cond:lambda1}.

Similarly, note that $Y^*_{\textnormal{rboot}} = X \hat \beta_{\textnormal{lasso+ols}} + \epsilon^*$, and with probability going to 1, $  \hat \beta_{\textnormal{lasso+ols}, S^c} = 0 $, we have
\begin{equation}\label{dist-boot-lpr}
	n^{\frac{1}{2}}u^{\T}(\hat \beta^*_{\textnormal{\rBLPR{}}}-\hat \beta_{\textnormal{lasso+ols}}) = n^{-\frac{1}{2}}u^{\T} C_{\lambda_2}^{-1} X^{\T}\epsilon^* + o_p (1).
	\end{equation}

Third, let $U =  n^{-1/2}u^{\T} C_{\lambda_2}^{-1} X^{\T}\epsilon$, and $U^* =  n^{-1/2}u^{\T} C_{\lambda_2}^{-1} X^{\T}\epsilon^*$. We can show that both $U$ and $(U^* \mid \epsilon)$ converge in distribution to $N(0,  \sigma_1^2)$, where $C  = X^{\T}X/n$ and 
\[ \sigma_1^2  = \lim_{n \rightarrow \infty} \left(  u^{\T}  C_{\lambda_2}^{-1}   C  (C_{\lambda_2}^{-1})^{\T} u \right) \sigma^2. \]

\noindent For simplicity, denote
\[  \sigma_2^2 = \left(  u^{\T}  C_{\lambda_2}^{-1}   C  (C_{\lambda_2}^{-1})^{\T} u \right) \sigma^2 .\]
We need to check Linderberg condition for the asymptotic normality. 
	For deriving the asymptotic normality of $U = n^{-1/2} u^{\T}  C_{\lambda_2}^{-1}  X^{\T}\epsilon$, denote 
	$$v= n^{-1/2} X \left( C_{\lambda_2}^{-1} \right)^{\T} u = (v_1,\ldots,v_n)^{\T},$$ 
	where $v_k=n^{-1/2} x_k^{\T}  \left( C_{\lambda_2}^{-1} \right)^{\T} u $. It is easy to show that
	\begin{equation} \label{eqn:Linderberg}
	 \sum_{k=1}^{n} E(v_k \epsilon_k)^2 = \left( \sum_{k=1}^{n} v_k^2 \right) \sigma^2 =  \left( u^{\T}  C_{\lambda_2}^{-1}   C  (C_{\lambda_2}^{-1})^{\T} u \right) \sigma^2  = \sigma_2^2.
	\end{equation}	
The Linderberg condition holds if for any $\delta>0$,
	\begin{equation} \label{eqn:Linderberg}
	\frac{1}{\sigma_2^2} \sum_{k=1}^{n} v_k^2 E \left\{ \epsilon_k^2 I_{|v_k \epsilon_k| > \delta \sigma_2 }  \right\} \rightarrow 0.
	\end{equation}
Since the errors $\epsilon_i$ are independent and identically distributed sub-Gaussian random variables, it is easy to see that,
\begin{eqnarray}
 \frac{1}{\sigma_2^2} \sum_{k=1}^{n} v_k^2 E \left\{ \epsilon_k^2 I_{|v_k \epsilon_k| > \delta \sigma_2 } \right\}  &\leq & \frac{1}{\sigma^2} \mathop {\max}\limits_{1\leq k \leq n} E \left\{ \epsilon_k^2 I_{| v_k \epsilon_k| > \delta \sigma_2} \right\} \nonumber \\
 &\leq &  \frac{1}{\sigma^2} E \left\{ \epsilon_1^2 I_{ | \epsilon_1| > \frac{ \delta  } { \mathop {\max}_{1\leq k \leq n}|v_k| / \sigma_2} } \right\} , \nonumber \\
 & = & o(1),
	\end{eqnarray}
where the last equality is because Condition \ref{cond:orthogonal}. That is,
\[  \mathop {\max}\limits_{1\leq k \leq n}|v_k| / \sigma_2 = n^{-\frac{1}{2}} \max_{1 \leq k \leq n} \left|  u^{\T} C_{\lambda_2}^{-1} x_k  \right| / \left\{  \sigma^2 u^{\T}  C_{\lambda_2}^{-1}   C  (C_{\lambda_2}^{-1})^{\T} u     \right\}^{1/2} =  o(1).  \]	

Finally, we prove the asymptotic normality of  $ U^* =  n^{-1/2} u^{\T} C_{\lambda_2}^{-1} X^{\T}\epsilon^*$, given $\epsilon$. By Lemma~\ref{lemma:subgaussian*},  $\epsilon^*_i$ are conditionally (given $\epsilon$) independent and identically distributed sub-Gaussian random variables, with mean 0 and variance $(\sigma^*)^2$. Similar arguments lead to the same asymptotic normality of $ U^*$, given $ \epsilon$,  as those for $U$, as long as $\sigma^* \rightarrow_p \sigma$. The reminder of the proof is to show that $\sigma^* \rightarrow_p \sigma$.

Note that
\[  (\sigma^*)^2 = \frac{1}{n-1} \sum_{i=1}^{n} (\hat \epsilon_i - \tilde{\epsilon})^2 = \frac{1}{n-1} \sum_{i=1}^{n} \left[x_i ^{\T} (\beta^0 - \hat \beta_{\textnormal{lasso+ols}}) + \epsilon_i - \overline{\epsilon}\right]^2.  \]	
	By Strong Law of Large Number, we have
	\begin{equation}\label{SLLN:sigma}
	\frac{1}{n-1} \sum_{i=1}^{n} (\epsilon_i - \overline{\epsilon})^2 \rightarrow \sigma^2, \ \textnormal{almost surely}.
	\end{equation}
	Since
	\begin{eqnarray} \label{eqn:x-beta}
	& & \frac{1}{n-1} \sum_{i=1}^{n} \left[x_i ^{\T} (\beta^0 - \hat \beta_{\textnormal{lasso+ols}})\right]^2 \nonumber \\
	& = & \frac{1}{n-1} \sum_{i=1}^{n} \left[x_{i,S} ^{\T} (\beta^0_S - (\hat \beta_{\textnormal{lasso+ols}})_{S} ) + x_{i,S^c}^{\T}\beta^0_{S^c} \right]^2 \nonumber \\
	& \leq &  2\left\{ \frac{1}{n-1} \sum_{i=1}^{n} \left[x_{i,S} ^{\T} (\beta^0_S - (\hat \beta_{\textnormal{lasso+ols}})_S )\right]^2 +  \frac{1}{n-1} \sum_{i=1}^{n} (x_{i,S^c}^{\T}\beta^0_{S^c})^2  \right\} \nonumber \\
	& \leq & \frac{2n}{n-1} \left\{  \mathop {\max}\limits_{1\leq i \leq n}  \left[ x_{i,S}^{\T} (\beta^0_S - (\hat \beta_{\textnormal{lasso+ols}})_S) \right]^2 +  \left( \beta^0_{ S^c} \right)^{\T} C_{22}  \left( \beta^0_{ S^c} \right)  \right\}   \nonumber \\
	& = & o_p (1),
	\end{eqnarray}
	where the last equality holds because of \eqref{eqn:x-S-beta1}, \eqref{eqn:x-S-beta2}, and Condition \ref{cond:orthogonal}. Combining~\eqref{SLLN:sigma} and~\eqref{eqn:x-beta}, we have
	\[ (\sigma^*)^2 \rightarrow_p \sigma^2. \]
	
\end{proof}

\subsection{Proof of Lemma \ref{lemma:lasso+ols}}
\label{proof:lemma}

\begin{proof}

	Under Conditions \ref{cond:subgaussian} -- \ref{cond:lambda1}, and using Theorem \ref{thm:selection-consistency-lasso}, the lasso, $\hat \beta_{\textnormal{lasso}}$, has model selection consistency, that is,
	\[ \pr (\hat S = S ) = 1 - o(e^{-n^{c_2}}) \rightarrow 1.\]
	Conditional on $\{ \hat S = S \}$, the lasso+ols estimator has the following form:
	$$
	(\hat \beta_{\textnormal{lasso+ols}})_S  = (X_S^{\T}X_S)^{-1}X_S^{\T}Y = \beta^0_S + C_{11}^{-1} C_{12} \beta_{S^c} + (X_S^{\T}X_S)^{-1}X_S^{\T} \epsilon;
	$$
	$$  (\hat \beta_{\textnormal{lasso+ols}})_{S^c} = 0. $$
	Therefore,
	\begin{equation}
	||\hat \beta_{\textnormal{lasso+ols}} - \beta^0||_{\infty} \leq ||C_{11}^{-1} C_{12} \beta^0_{S^c}||_{\infty} + ||(X_S^{\T}X_S)^{-1}X_S^{\T} \epsilon||_{\infty} + ||\beta^0_{S^c}||_{\infty}.
	\end{equation}
	By Condition \ref{cond:extra}, we have $||C_{11}^{-1} C_{12} \beta^0_{S^c}||_{\infty} = o(n^{(c_1-1)/2})$. Condition \ref{cond:cliff-weak} gives $||\beta^0_{S^c}||_{\infty} \leq M n^{-(1+c_1)/2}$. Since $(X_S^{\T}X_S)^{-1}X_S^{\T} \epsilon$ are sub-Gaussian random variables, with covariance matrix $\sigma^2 C_{11}^{-1}/n$, it is not hard to show that
	\[ \pr \left( ||(X_S^{\T}X_S)^{-1}X_S^{\T} \epsilon||_{\infty} \leq M n^{\frac{c_1-1}{2}} \right) \rightarrow 1. \]
	Therefore,
	$$\pr \left( ||\hat \beta_{\textnormal{lasso+ols}} - \beta^0||_{\infty} \leq 2M n^{\frac{c_1-1}{2}} \right) \rightarrow 1.$$

\end{proof}

\section{Examples related to Condition 8}
\setcounter{equation}{0}

We provide three examples of design matrices, which satisfy or do not satisfy  Condition \ref{cond:extra}.

\noindent \textbf{Example 1. Orthogonal design.} $X$ is orthogonal such that $X^{\T}X/n$ is an identity matrix. In this case, $C_{11}$ and $C_{22}$ are identity matrices, and $C_{12}$ is a zero matrix. As $\left\| \beta^0_{S^c}\right\|_{\infty} = o\left( n^{-1/2} \right)$, we have
$$
\left\|C_{11}^{-1}C_{12}\beta^0_{S^c}\right\|_{\infty} = 0,
$$
$$
\left\|n^{\frac{1}{2}}(C_{21}C_{11}^{-1}C_{12}-C_{22})\beta^0_{S^c}\right\|_{\infty}= \left\|n^{\frac{1}{2}}\beta^0_{S^c}\right\|_{\infty}=O(1).
$$
Thus, this example satisfies  Condition \ref{cond:extra}.

\noindent \textbf{Example 2. Exponential decay.} In this example, $X$ has the following pattern:
\begin{equation}       
\frac{1}{n}X^{\T} X= \left(                 
\begin{array}{ccccc}   
1 & \rho & \rho^2 & \cdots & \rho^{p-1}\\  
\rho & 1 & \rho &  \cdots & \rho^{p-2}\\  
\rho^2 & \rho & 1 & \cdots & \rho^{p-3}\\  
\vdots & \vdots & \vdots & \ddots & \vdots\\
\rho^{p-1}& \rho^{p-2} & \rho^{p-3} & \cdots & 1\\
\end{array}
\right) .   \nonumber             
\end{equation}
In this case,
\begin{equation}       
C_{11}= \left(            
\begin{array}{ccccc}   …  1 & \rho & \rho^2 & \cdots & \rho^{s-1}\\  …ƒ    \rho & 1 & \rho &  \cdots & \rho^{s-2}\\…ƒ    \rho^2 & \rho & 1 & \cdots & \rho^{s-3}\\  
\vdots & \vdots & \vdots & \ddots & \vdots\\
\rho^{s-1}& \rho^{s-2} & \rho^{s-3} & \cdots & 1\\
\end{array}
\right) .   \nonumber             
\end{equation}
Using mathematical induction, we can prove that
\[  
(1-\rho^2)C_{11}^{-1}= \left(               
\begin{array}{ccccc}   …    1 & -\rho & 0 & \cdots & 0\\  
-\rho & 1+\rho^2 & -\rho &  \cdots & 0\\  
0 & -\rho & 1+\rho^2 & \cdots & 0\\  
\vdots & \vdots & \vdots & \ddots & \vdots\\
0& \cdots& -\rho & 1+\rho^2 & -\rho\\
0& 0 & \cdots& -\rho & 1\\
\end{array}
\right), \quad C_{12}= \left(               
\begin{array}{cccc}  …   \rho^s & \rho^{s+1}  & \cdots & \rho^{p-1}\\ …    \rho^{s-1} & \rho^{s}  & \cdots & \rho^{p-2}\\ …     \vdots & \vdots & \ddots & \vdots\\
\rho & \rho^2  & \cdots & \rho^{p-s}\\
\end{array}
\right). \nonumber 
\] 
Then, 
\begin{equation}      
C_{11}^{-1}C_{12}= \left(                
\begin{array}{cccc}   
0& 0  & \cdots & 0\\  …ƒ    \vdots & \vdots & \ddots & \vdots\\
0& 0  & \cdots & 0\\ 
\rho-\rho^3 & \rho^2-\rho^4  & \cdots & \rho^{p-s}-\rho^{p-s+2}\\
\end{array}
\right) . \nonumber                
\end{equation}
As $\left\| \beta^0_{S^c}\right\|_{\infty} = o\left( n^{-1/2} \right)$, we have
$$ \left\|n^{\frac{1}{2}}C_{11}^{-1}C_{12}\beta^0_{S^c}\right\|_{\infty}\leq n^{\frac{1}{2}}(\rho+\rho^2-\rho^{p-s+1}-\rho^{p-s+2})\left\| \beta^0_{S^c}\right\|_{\infty}=O(1),$$
$$ \left\|n^{\frac{1}{2}}(C_{21}C_{11}^{-1}C_{12}-C_{22})\beta^0_{S^c}\right\|_{\infty} \leq \left\|n^{\frac{1}{2}}C_{21}C_{11}^{-1}C_{12}\beta^0_{S^c}\right\|_{\infty}+\left\|n^{\frac{1}{2}}C_{22}\beta^0_{S^c}\right\|_{\infty}, $$
$$ \left\|n^{\frac{1}{2}}C_{21}C_{11}^{-1}C_{12}\beta^0_{S^c}\right\|_{\infty} \leq n^{\frac{1}{2}}(\rho^2+\rho^3-\rho^{p-s+2}-\rho^{p-s+3})\left\| \beta^0_{S^c}\right\|_{\infty}=O(1), $$
$$ \left\|n^{\frac{1}{2}}C_{22}\beta^0_{S^c}\right\|_{\infty} < \frac{n^{\frac{1}{2}}}{1-\rho}\left\| \beta^0_{S^c}\right\|_{\infty}=O(1).
$$
Thus, this example satisfies Condition \ref{cond:extra}.

\noindent \textbf{Example 3. Equal correlation.} The design matrix $X$ satisfies
\begin{equation}     
\frac{1}{n}X^{\T} X= \left(               
\begin{array}{ccccc}   
1 & \rho & \rho & \cdots & \rho\\  
\rho & 1 & \rho &  \cdots & \rho\\   
\rho & \rho & 1 & \cdots & \rho\\  
\vdots & \vdots & \vdots & \ddots & \vdots\\
\rho& \rho & \rho & \cdots & 1\\
\end{array}
\right).  \nonumber             
\end{equation}
In this case, 
\[  
C_{11}= \left(             
\begin{array}{ccccc}   …    1 & \rho & \rho & \cdots & \rho\\  …     \rho & 1 & \rho &  \cdots & \rho\\ …ƒ     \rho & \rho & 1 & \cdots & \rho\\  
\vdots & \vdots & \vdots & \ddots & \vdots\\
\rho& \rho & \rho & \cdots & 1\\
\end{array}
\right), \quad 
C_{12}=\left(             
\begin{array}{cccc}   …   \rho & \rho & \cdots & \rho\\  
\rho  & \rho &  \cdots & \rho\\
\rho & \rho  & \cdots & \rho\\  
\vdots & \vdots & \ddots & \vdots\\
\rho & \rho & \cdots & \rho\\
\end{array}
\right)  .           
\]
It is easy to show that $\textbf{1}=(1,1,\cdots,1)^{\T}$ is an eigenvector of $C_{11}$, hence, it is also an eigenvector of $C_{11}^{-1}$. Let $Sum_{S^c}$ be the sum of elements in $\beta^0_{S^c}$. Then,
$$\left\|C_{11}^{-1}C_{12}\beta^0_{S^c}\right\| = \frac{\rho}{1+\rho(s-1)} \left| Sum_{S^c} \right|.$$
As we do not assume a bound for $Sum_{S^c}$, this example does not always satisfy Condition \ref{cond:extra}.

\revise{\section{Examples related to Condition 11}
When the correlation between covariates satisfies cor$(X_i, X_j) = \rho^{| i - j | }$, with $\rho < 1/5$, condition 11 holds.} \revise{In this case, $p\leq n$, and
\begin{equation*}
\begin{split}
(\beta_{S^c}^0)^TC_{22}(\beta_{S^c}^0) &=\sum_{\substack{s<i\leq p\\  s<j\leq p}}(C_{22})_{ij}\beta_i^0\beta_j^0 =\sum_{\substack{s<i\leq p\\  s<j\leq p}}\rho^{|i-j|}\beta_i^0\beta_j^0  =o(1).
\end{split}
\end{equation*}

\begin{lemma}
For any $p \times 1$ vector $u, v$, and $p\times p$ symmetric matrix $A$, we have $\mu_p(A) \leq \frac{u^TAv}{u^Tv}\leq \mu_1(A)$, where $\mu_p(A)$ and $\mu_1(A)$ are the smallest and largest eigenvalues of $A$, respectively.
\end{lemma}
From the above lemma, 
\begin{center}
$\mu_p(C_{\lambda_2}^{-1}) \leq \frac{u^TC_{\lambda_2}^{-1}x_k}{u^Tx_k}\leq \mu_1(C_{\lambda_2}^{-1})$.
\end{center}
Assume that $\rho < \frac{1}{5}$, by Gershgorin circle theorem, there exists a $\delta>0$, such that 
\begin{center}
$3>\mu_1(C_{\lambda_2})\geq \mu_2(C_{\lambda_2}) \geq \dots \mu_p(C_{\lambda_2}) > \delta >0$.
\end{center}
Then we have 
\begin{center}
$\frac{1}{\delta}>\mu_1(C_{\lambda_2}^{-1})\geq \mu_2(C_{\lambda_2}^{-1}) \geq \dots \mu_p(C_{\lambda_2}^{-1}) > \frac{1}{3}$.
\end{center}
Thus,
\begin{center}
$|\frac{u^TC_{\lambda_2}^{-1}x_k}{u^Tx_k}| \leq \frac{1}{\delta}, \quad \max_{1\leq k \leq n}|u^TC_{\lambda_2}^{-1}x_k| \leq \frac{1}{\delta}\max_{1\leq k \leq n}|u^Tx_k|$.
\end{center}
Therefore, Condition 11 is guaranteed by assuming
$$
\max_{1\leq k \leq n}|u^Tx_k|=o(\sqrt{n}), \quad u_{S^c}^{\T}  \beta^0_{ S^c}  = o(\sqrt{n}).
$$
}


\section{Figures and Tables}

\begin{figure}[ht]
	\centerline{\includegraphics[width=0.8\textwidth]{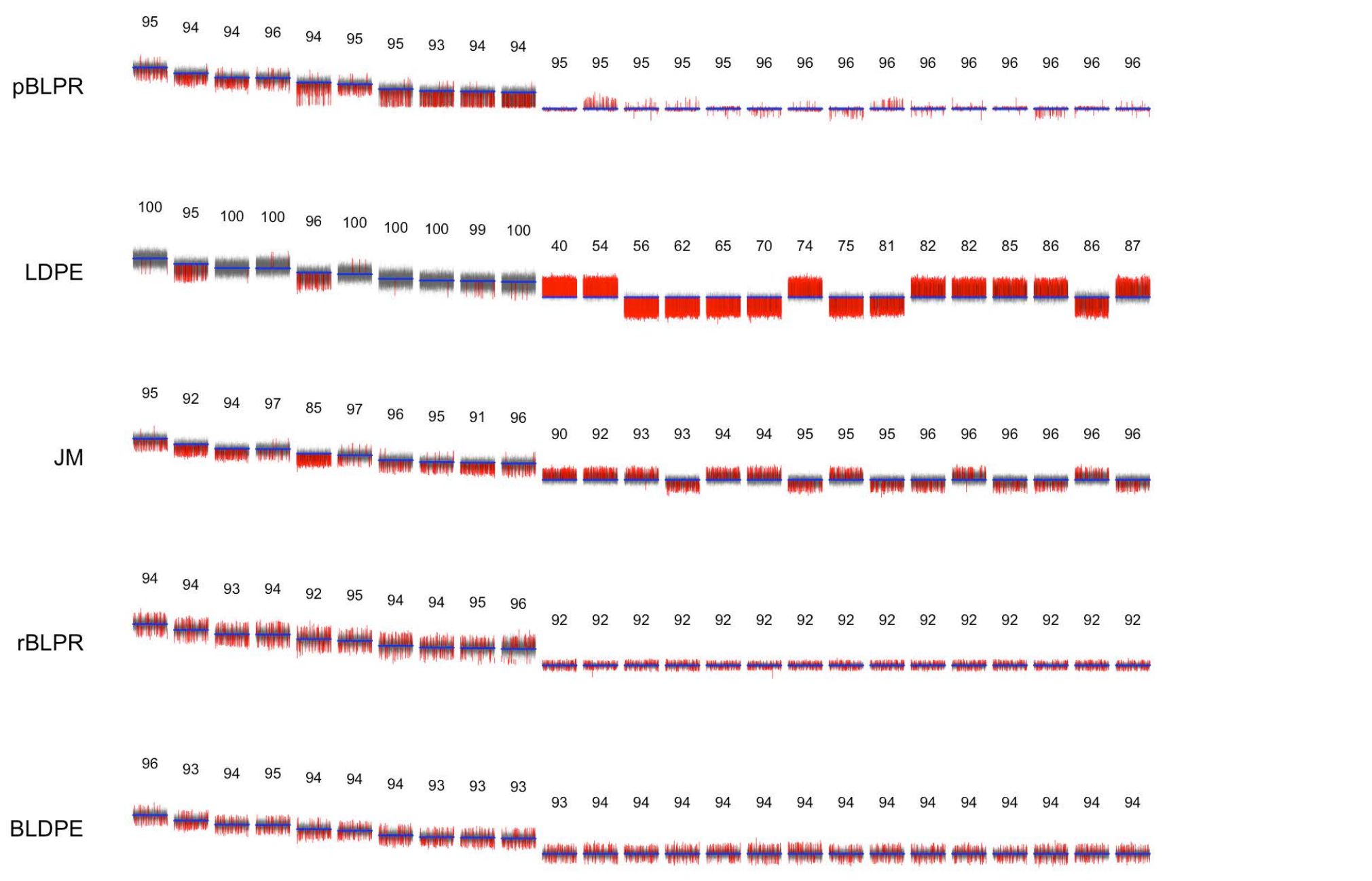}}
	\caption[Confidence intervals of pBLPR, LDPE, JM and BLDPE -- hard sparsity, Normal design, Toeplitz, $\rho=0.5$]{\JL{$1,000$} confidence intervals and their empirical coverage of the true coefficients (blue line). Black confidence intervals cover the truth, whereas red confidence intervals do not. The first 10 coefficients are the largest 10 (non-zero). The remaining 15 coefficients shown are those with the worst coverage for that method. The numbers above the intervals are the empirical coverage probabilities in percentages. This plot is for hard sparsity and a Normal design matrix with a Toeplitz type covariance matrix, and $\rho=0.5$. }\label{fig:paired-hdi-N1-1}
\end{figure}

\begin{figure}[ht]
	\centerline{\includegraphics[width=0.9\textwidth]{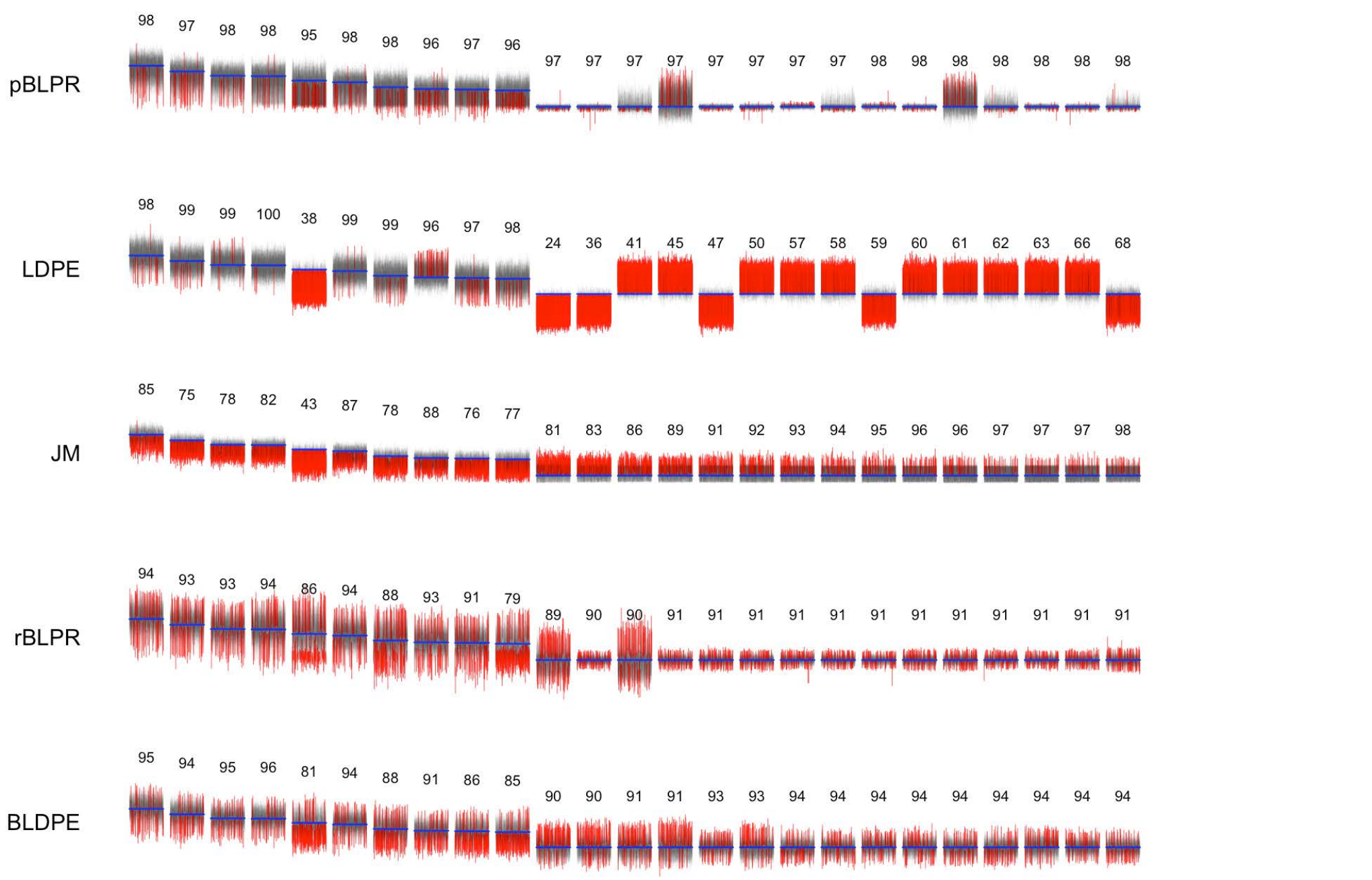}}
	\caption[Confidence intervals of pBLPR, LDPE, JM and BLDPE -- hard sparsity, Normal design, Toeplitz, $\rho=0.9$]{See caption of Figure~\ref{fig:paired-hdi-N1-1} with the only difference being $\rho=0.9$.}\label{fig:paired-hdi-N1-2}
\end{figure}

\begin{figure}[ht]
	\centerline{\includegraphics[width=0.9\textwidth]{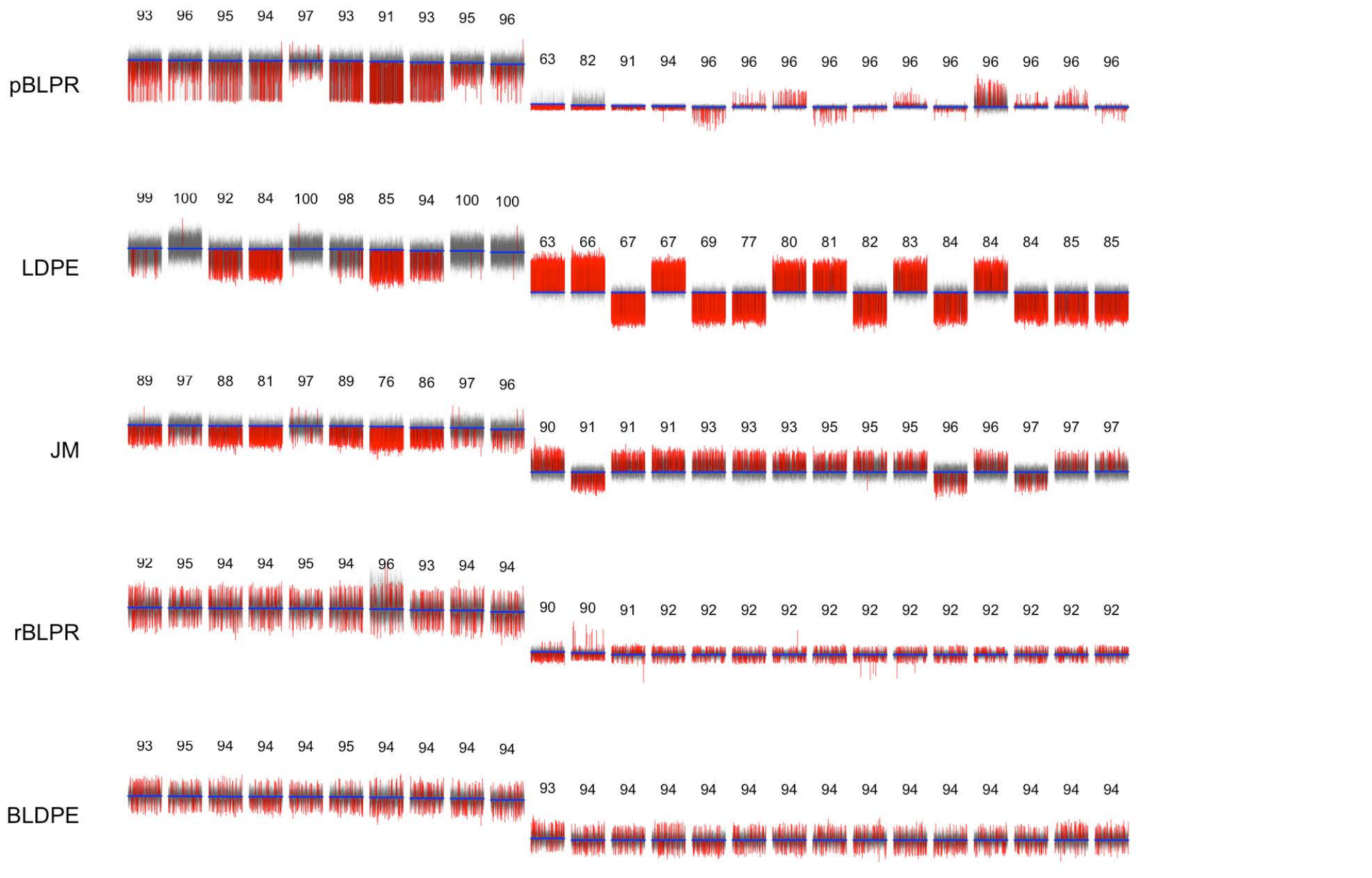}}
	\caption[Confidence intervals of pBLPR, LDPE, JM and BLDPE -- weak sparsity, Normal design, Toeplitz, $\rho=0.5$]{See caption of Figure~\ref{fig:paired-hdi-N1-1} with the only difference being weak sparsity.}\label{fig:paired-hdi-N1-3}
\end{figure}

\begin{figure}[ht]
	\centerline{\includegraphics[width=0.75\textwidth]{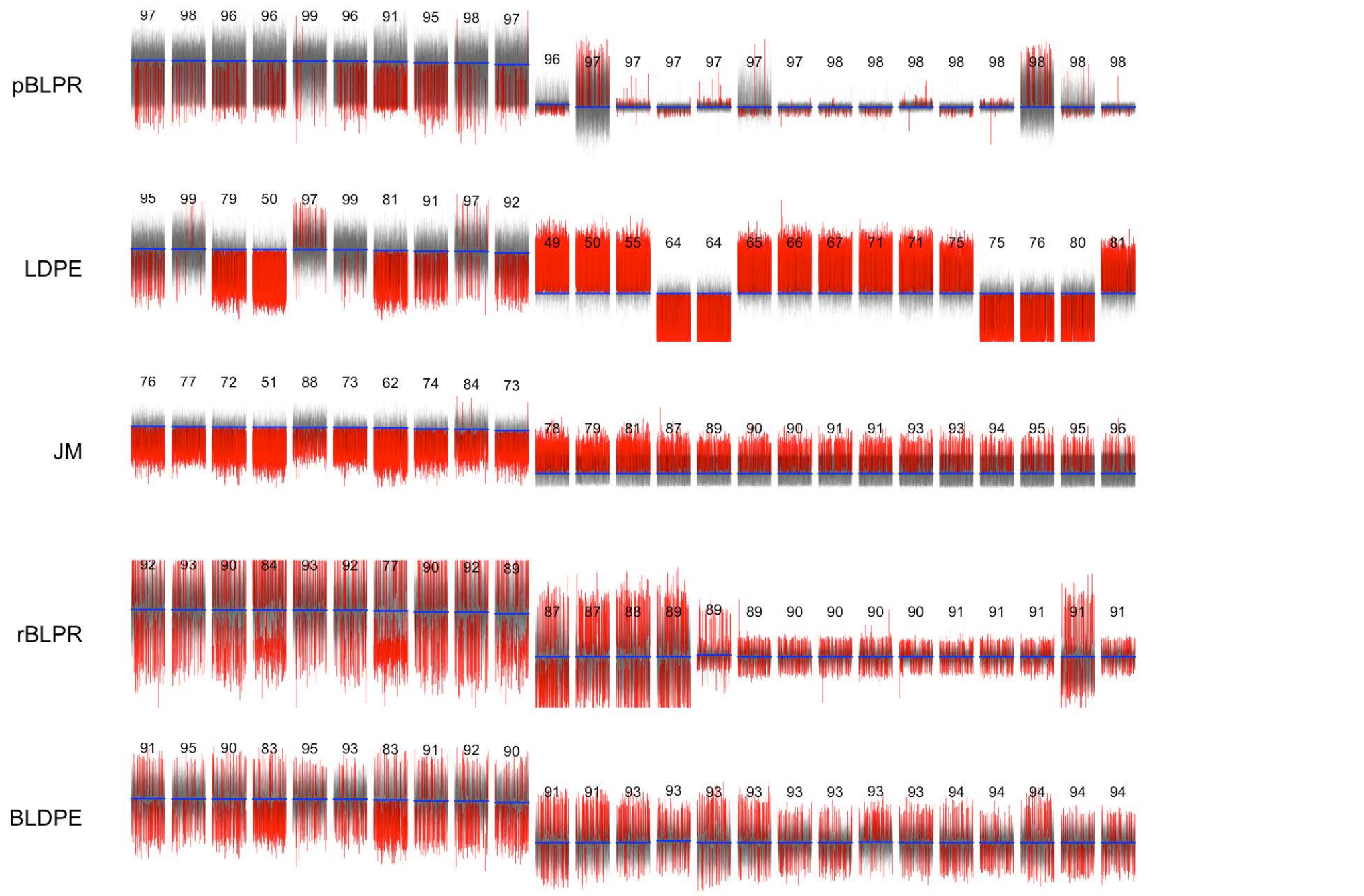}}
	\caption[Confidence intervals of pBLPR, LDPE, JM and BLDPE -- weak sparsity, Normal design, Toeplitz, $\rho=0.9$]{See caption of Figure~\ref{fig:paired-hdi-N1-1} with the only differences being weak sparsity and $\rho=0.9$.}\label{fig:paired-hdi-N1-4}
\end{figure}

\begin{figure}[ht]
	\centerline{\includegraphics[width=\textwidth]{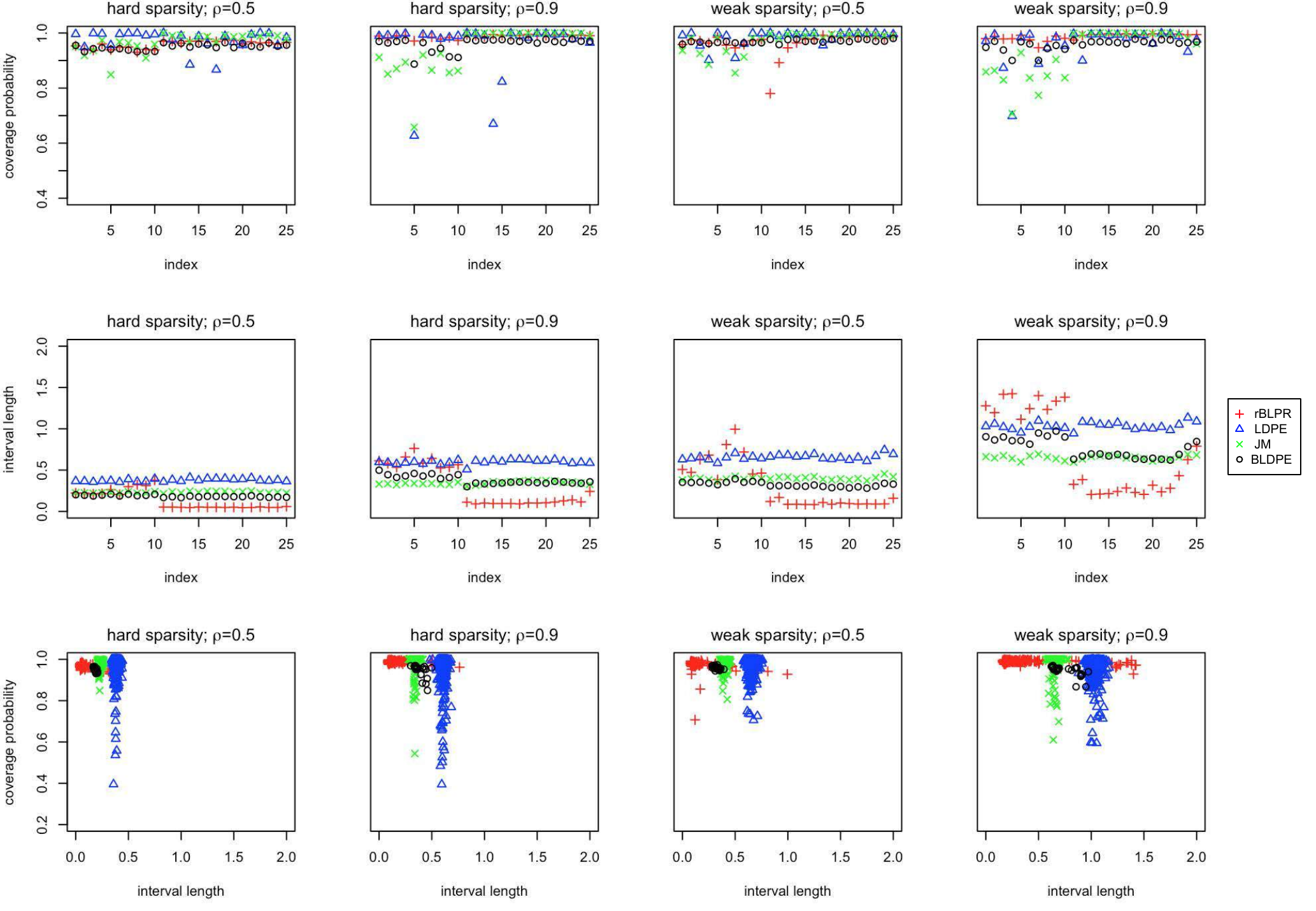}}
	\caption[Comparison of \rBLPR, \ZZ{}, \JM{} and BLDPE -- Normal design with a Toeplitz type covariance matrix]{Comparison of coverage probabilities (the first row) and mean interval lengths (the second row) produced by rBLPR, \ZZ{}, \JM{}, and BLDPE. The third \finalrevise{row} shows the coverage probabilities v.s. mean interval lengths. The design matrix is generated from a Normal distribution with a Toeplitz type covariance matrix.}\label{fig:paired-prob-length-N1-res}
\end{figure}

\begin{figure}[ht]
	\centerline{\includegraphics[width=\textwidth]{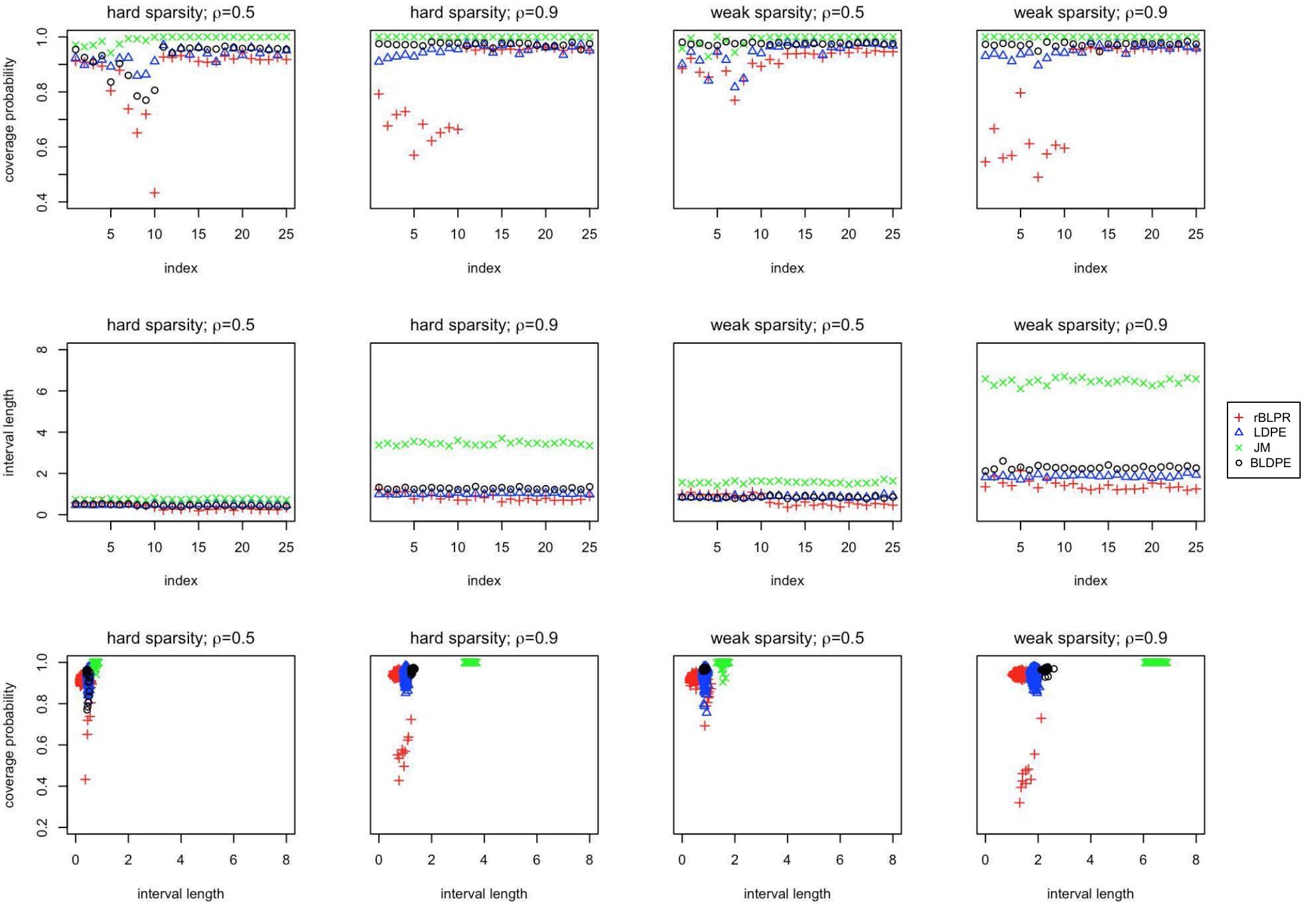}}
	\caption[Comparison of \rBLPR, \ZZ{}, \JM{} and BLDPE -- Normal design with an Equi.corr type covariance matrix]{See caption of Figure~\ref{fig:paired-prob-length-N1-res} with the only difference being that the covariance matrix is an Equi.corr type.}\label{fig:paired-prob-length-N3-res}
\end{figure}

\begin{figure}[ht]
	\centerline{\includegraphics[width=\textwidth]{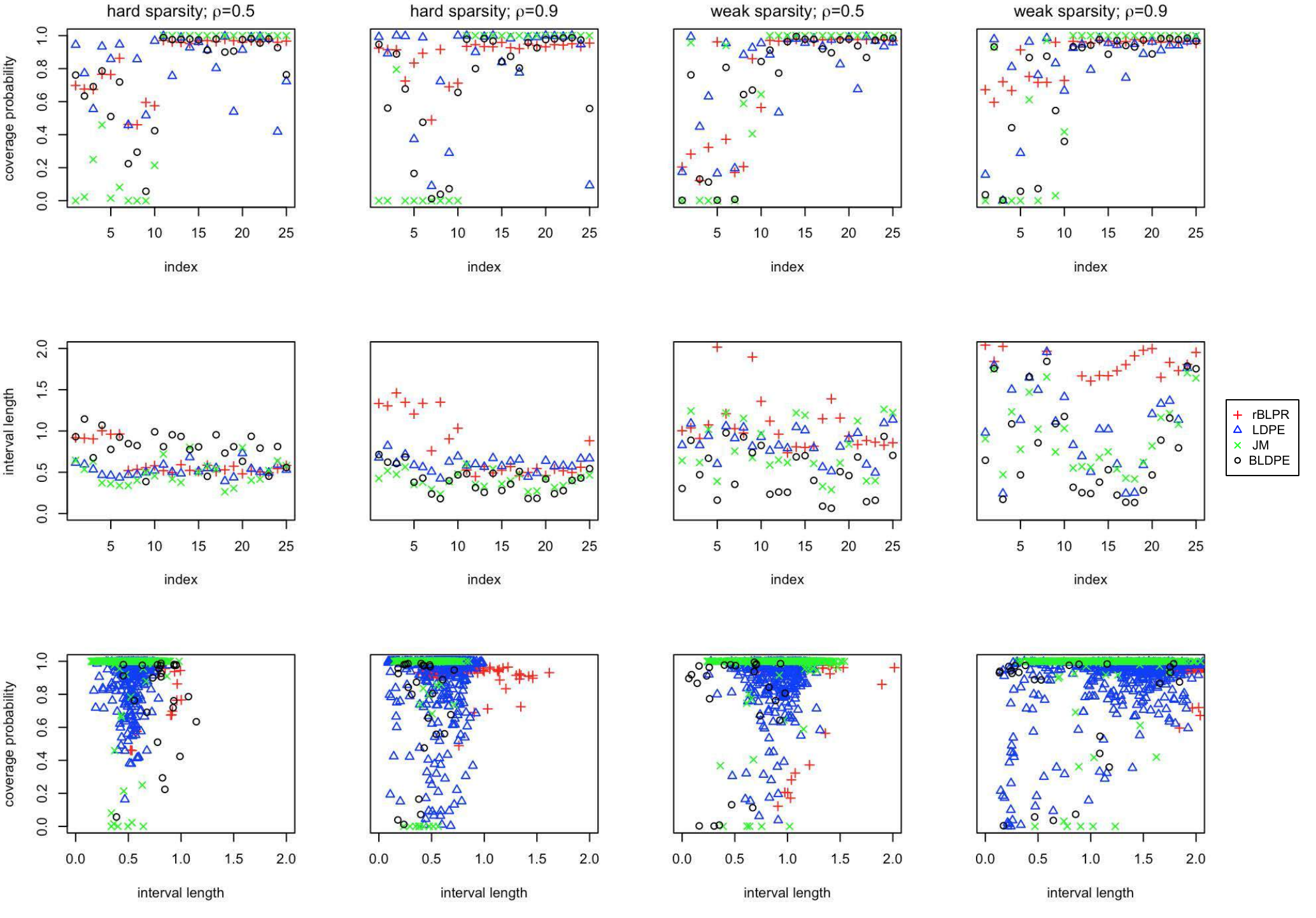}}
	\caption[Comparison of \rBLPR, \ZZ{}, \JM{} and BLDPE -- $t_2$ design with a Toeplitz type covariance matrix]{See caption of Figure~\ref{fig:paired-prob-length-N1-res} with the only difference being the type of design matrix. In this plot, the design matrix is generated from $t_2$ distribution with a Toeplitz type covariance matrix.}\label{fig:paired-prob-length-T-res}
\end{figure}

\begin{figure}[ht]
	\centerline{\includegraphics[width=\textwidth]{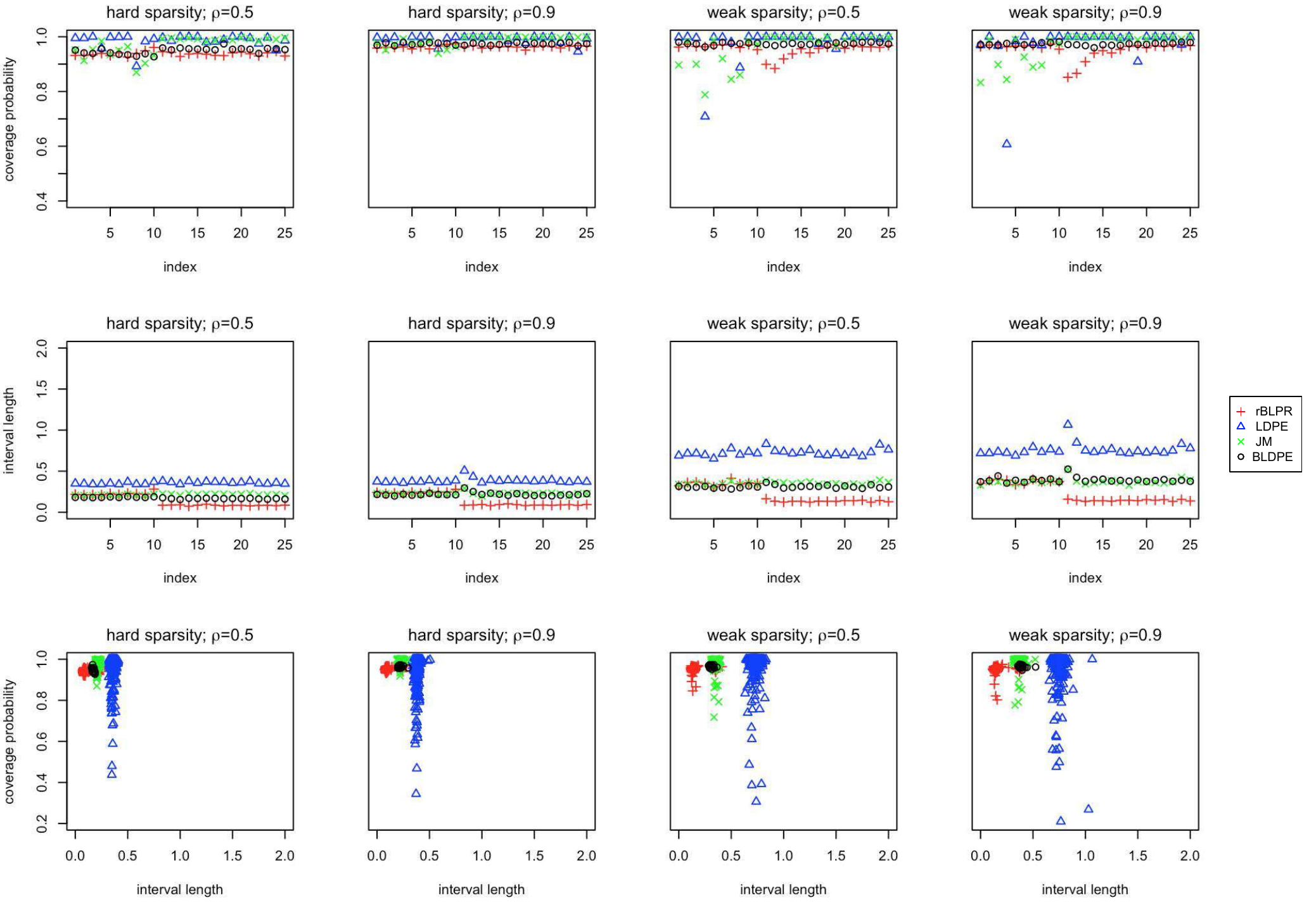}}
	\vspace*{0.05in}
	\caption[Comparison of \rBLPR, \ZZ{}, \JM{} and BLDPE -- Normal design with an Exp.decay type covariance matrix]{See caption of Figure~\ref{fig:paired-prob-length-N1-res} with the only difference being that the covariance matrix is Exp.decay type.}\label{fig:paired-prob-length-N2-res}
\end{figure}

\begin{figure}[ht]
	\centerline{\includegraphics[width=\textwidth]{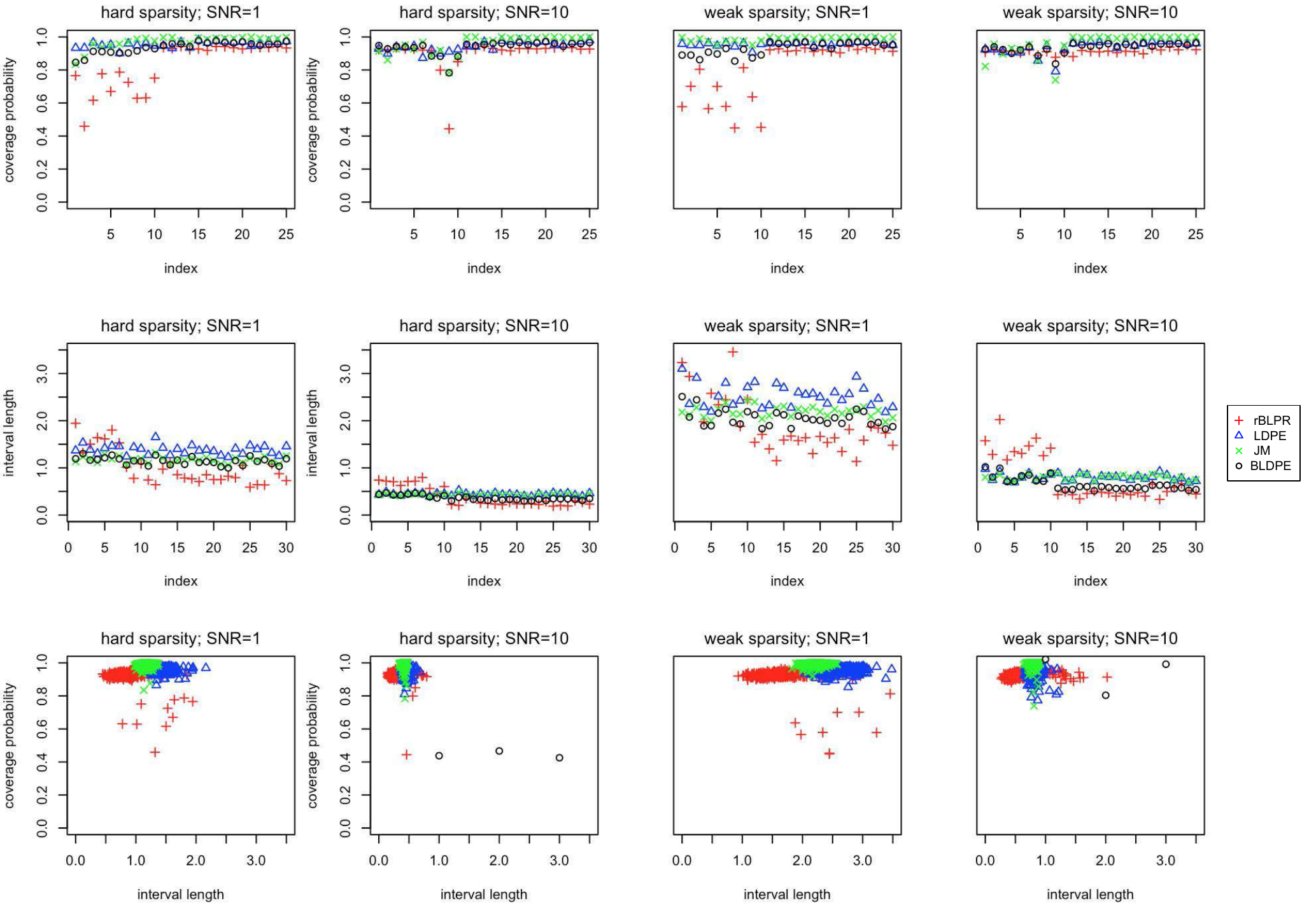}}
	\vspace*{0.05in}
	\caption[Comparison of \rBLPR, \ZZ{}, \JM{} and BLDPE -- fMRI design]{See caption of Figure~\ref{fig:paired-prob-length-N1-res} with the only difference being that the design matrix is generated from the fMRI data.}\label{fig:paired-prob-length-real-res}
\end{figure}

\clearpage


\begin{table}[ht]
	\renewcommand\arraystretch{0.75}
	\begin{center}
		\caption{\label{tab:cplarge} Mean coverage probabilities over large $\beta^0_j$'s (first $10$ largest in absolute value).}
		\begin{tabular}{rrrrrrrrrrr}
			& &  \multicolumn{8}{c}{Normal design, Toeplitz} \\ \hline
			$\beta^0$ & $\rho$ & \rBLPR{} & \pBLPR{} & rBlassoOLS & pBlassoOLS & rBlasso & pBlasso & LDPE & JM  \\ \hline
			
			hard   & .5 & .94 & .94 & .94 & .91 & .85 & .30 & .99 & .94 \\ 
			hard   & .9 & .90 & .97 & .90 & .87 & .83 & .31 & .92 & .77 \\ 
			weak   & .5 & .94 & .94 & .94 & .90 & .83 & .27 & .95 & .90 \\ 
			weak   & .9 & .89 & .96 & .87 & .84 & .82 & .32 & .88 & .73 \\  \hline
			
			& &  \multicolumn{8}{c}{Normal design, Exponential decay} \\ \hline
			hard   & .5 & .94 & .93 & .94 & .89 & .84 & .26 & .98 & .94 \\ 
			hard   & .9 & .94 & .93 & .94 & .88 & .83 & .25 & .99 & .95 \\ 
			weak   & .5 & .94 & .93 & .94 & .88 & .80 & .20 & .93 & .86 \\ 
			weak   & .9 & .94 & .94 & .94 & .88 & .80 & .19 & .91 & .87 \\  \hline
			
			& &  \multicolumn{8}{c}{Normal design, Equal correlation} \\ \hline
			hard   & .5 & .78 & .87 & .71 & .60 & .65 & .40 & .90 & .98 \\ 
			hard   & .9 & .46 & .66 & .20 & .40 & .19 & .33 & .90 & 1.00 \\ 
			weak   & .5 & .79 & .82 & .67 & .48 & .59 & .33 & .84 & .96 \\ 
			weak   & .9 & .34 & .57 & .15 & .34 & .15 & .28 & .88 & 1.00 \\ \hline
			
			& &  \multicolumn{8}{c}{$t_2$ design, Toeplitz} \\ \hline
			hard   & .5 & .65 & .53 & .39 & .45 & .23 & .03 & .78 & .10 \\ 
			hard   & .9 & .80 & .89 & .64 & .77 & .46 & .10 & .73 & .08 \\ 
			weak   & .5 & .41 & .53 & .33 & .47 & .16 & .09 & .63 & .35 \\ 
			weak   & .9 & .74 & .51 & .20 & .46 & .13 & .20 & .64 & .30 \\ \hline
			
			& &  \multicolumn{8}{c}{fMRI design} \\ \hline
			$\beta^0$ & SNR & \rBLPR{} & \pBLPR{} & rBlassoOLS & pBlassoOLS & rBlasso & pBlasso & LDPE & JM  \\ \hline
			hard   & 1  & .68 & .76 & .36 & .61 & .26 & .37 & .95 & .94 \\
			hard   & 5  & .78 & .90 & .73 & .75 & .63 & .45 & .92 & .91 \\ 
			hard   & 10 & .86 & .93 & .84 & .79 & .73 & .46 & .92 & .90 \\ 
			weak   & 1  & .63 & .75 & .30 & .59 & .22 & .36 & .95 & .97 \\ 
			weak   & 5  & .83 & .93 & .73 & .69 & .60 & .43 & .91 & .90 \\
			weak   & 10 & .91 & .96 & .88 & .75 & .79 & .44 & .90 & .90 \\ 
			\hline			
			
		\end{tabular}
	\end{center}
\end{table}

\begin{table}[ht]
	\renewcommand\arraystretch{0.75}
	\begin{center}
		\caption{\label{tab:lengthlarge} Mean confidence interval lengths over large $\beta^0_j$'s (first $10$ largest in absolute value).}
		\begin{tabular}{rrrrrrrrrrr}
			& &  \multicolumn{8}{c}{Normal design, Toeplitz} \\ \hline
			$\beta^0$ & $\rho$ & \rBLPR{} & \pBLPR{} & rBlassoOLS & pBlassoOLS & rBlasso & pBlasso & LDPE & JM  \\ \hline
			
			hard   & .5 & .24 & .27 & .18 & .27 & .15 & .20 & .37 & .23 \\ 
			hard   & .9 & .57 & .60 & .36 & .47 & .30 & .35 & .59 & .34 \\ 
			weak   & .5 & .44 & .61 & .33 & .65 & .28 & .39 & .64 & .39 \\ 
			weak   & .9 & 1.20 & 1.30 & .81 & 1.04 & .65 & .76 & 1.02 & .65 \\ \hline
			
			& &  \multicolumn{8}{c}{Normal design, Exponential decay} \\ \hline
			hard   & .5 & .23 & .26 & .17 & .27 & .15 & .20 & .35 & .21 \\ 
			hard   & .9 & .24 & .27 & .18 & .28 & .15 & .20 & .37 & .22 \\ 
			weak   & .5 & .35 & .60 & .26 & .67 & .24 & .34 & .71 & .34 \\ 
			weak   & .9 & .37 & .64 & .26 & .71 & .24 & .34 & .73 & .37 \\  \hline
			
			& &  \multicolumn{8}{c}{Normal design, Equal correlation} \\ \hline
			hard   & .5 & .53 & .61 & .38 & .51 & .34 & .44 & .48 & .75 \\ 
			hard   & .9 & .94 & .94 & .33 & .49 & .32 & .44 & 1.01 & 3.45 \\ 
			weak   & .5 & .99 & 1.18 & .72 & .94 & .62 & .81 & .86 & 1.55 \\ 
			weak   & .9 & 1.59 & 1.52 & .49 & .74 & .5 & .68 & 1.84 & 6.44 \\  \hline
			
			& &  \multicolumn{8}{c}{$t_2$ design, Toeplitz} \\ \hline
			hard   & .5 & .79 & .61 & .24 & .42 & .14 & .27 & .51 & .46 \\ 
			hard   & .9 & 1.20 & 1.11 & .45 & .64 & .23 & .39 & .62 & .41 \\ 
			weak   & .5 & 1.25 & 1.03 & .40 & .76 & .16 & .47 & .9 & .87 \\ 
			weak   & .9 & 2.63 & 1.89 & .54 & .96 & .22 & .51 & 1.33 & 1.11 \\ \hline
			
			& &  \multicolumn{8}{c}{fMRI design} \\ \hline
			$\beta^0$ & SNR & \rBLPR{} & \pBLPR{} & rBlassoOLS & pBlassoOLS & rBlasso & pBlasso & LDPE & JM  \\ \hline
			hard   & 1  & 1.42 & 1.32 & .57 & .69 & .38 & .48 & 1.40 & 1.18 \\ 
			hard   & 5  & .87 & .89 & .46 & .63 & .38 & .48 & .63 & .60 \\ 
			hard   & 10 & .66 & .71 & .37 & .53 & .32 & .42 & .44 & .43 \\ 
			weak   & 1  & 2.79 & 2.50 & .86 & 1.17 & .61 & .82 & 2.56 & 2.20 \\ 
			weak   & 5  & 1.89 & 1.89 & .89 & 1.15 & .72 & .91 & 1.15 & 1.12 \\
			weak   & 10 & 1.45 & 1.53 & .73 & 1.09 & .63 & .83 & .81 & .80 \\ 
			\hline			
			
		\end{tabular}
	\end{center}
\end{table}

\begin{table}[ht]
	\renewcommand\arraystretch{0.75}
	\begin{center}
		\caption{\label{tab:cpsmall} Mean coverage probabilities over small $\beta^0_j$'s (except for the first $10$ largest in absolute value).}
		\begin{tabular}{rrrrrrrrrrr}
			& &  \multicolumn{8}{c}{Normal design, Toeplitz} \\ \hline
			$\beta^0$ & $\rho$ & \rBLPR{} & \pBLPR{} & rBlassoOLS & pBlassoOLS & rBlasso & pBlasso & LDPE & JM  \\ \hline
			
			hard   & .5 & .94 & .97 & 1.00 & 1.00 & .96 & 1.00 & .98 & .99 \\ 
			hard   & .9 & .93 & .99 & 1.00 & 1.00 & .97 & 1.00 & .96 & 1.00 \\ 
			weak   & .5 & .94 & .98 & .01 & .06 & .36 & .33 & .98 & .99 \\ 
			weak   & .9 & .93 & .99 & .03 & .15 & .20 & .35 & .96 & 1.00 \\ \hline
			
			& &  \multicolumn{8}{c}{Normal design, Exponential decay} \\ \hline
			hard   & .5 & .94 & .97 & 1.00 & 1.00 & .96 & 1.00 & .98 & .99 \\ 
			hard   & .9 & .94 & .97 & 1.00 & 1.00 & .96 & 1.00 & .96 & 1.00 \\ 
			weak   & .5 & .94 & .98 & .00 & .06 & .42 & .32 & .98 & .99 \\ 
			weak   & .9 & .94 & .98 & .01 & .05 & .42 & .31 & .97 & 1.00 \\  \hline
			
			& &  \multicolumn{8}{c}{Normal design, Equal correlation} \\ \hline
			hard   & .5 & .92 & .98 & .98 & 1.00 & .98 & 1.00 & .95 & 1.00 \\ 
			hard   & .9 & .93 & .98 & .98 & 1.00 & .98 & 1.00 & .94 & 1.00 \\ 
			weak   & .5 & .91 & .99 & .16 & .37 & .07 & .46 & .95 & 1.00 \\ 
			weak   & .9 & .93 & .97 & .07 & .25 & .04 & .35 & .94 & 1.00 \\ \hline
			
			& &  \multicolumn{8}{c}{$t_2$ design, Toeplitz} \\ \hline
			hard   & .5 & .97 & .95 & .99 & 1.00 & .99 & 1.00 & .93 & 1.00 \\ 
			hard   & .9 & .94 & .98 & .99 & 1.00 & .98 & 1.00 & .91 & 1.00 \\ 
			weak   & .5 & .97 & .95 & .05 & .06 & .07 & .10 & .92 & 1.00 \\ 
			weak   & .9 & .96 & .97 & .04 & .09 & .05 & .13 & .9 & 1.00 \\  \hline
			
			& &  \multicolumn{8}{c}{fMRI design} \\ \hline
			$\beta^0$ & SNR & \rBLPR{} & \pBLPR{} & rBlassoOLS & pBlassoOLS & rBlasso & pBlasso & LDPE & JM  \\ \hline
			hard   & 1  & .93 & .98 & .99 & 1.00 & .97 & 1.00 & .96 & .99 \\
			hard   & 5  & .93 & .98 & .99 & 1.00 & .97 & 1.00 & .96 & .99 \\  
			hard   & 10 & .93 & .98 & .99 & 1.00 & .98 & 1.00 & .96 & .99 \\ 
			weak   & 1  & .93 & .98 & .05 & .22 & .05 & .35 & .96 & .99 \\ 
			weak   & 5  & .92 & .99 & .08 & .27 & .07 & .42 & .96 & 1.00 \\
			weak   & 10 & .92 & .99 & .07 & .26 & .07 & .44 & .96 & 1.00 \\ 
			\hline			
			
		\end{tabular}
	\end{center}
\end{table}

\begin{table}[ht]
	\renewcommand\arraystretch{0.75}
	\begin{center}
		\caption{\label{tab:lengthsmall} Mean confidence interval lengths over small $\beta^0_j$'s (except for the first $10$ largest in absolute value).}
		\begin{tabular}{rrrrrrrrrrr}
			& &  \multicolumn{8}{c}{Normal design, Toeplitz} \\ \hline
			$\beta^0$ & $\rho$ & \rBLPR{} & \pBLPR{} & rBlassoOLS & pBlassoOLS & rBlasso & pBlasso & LDPE & JM  \\ \hline
			
			hard   & .5 & .09 & .05 & .00 & .01 & .01 & .02 & .38 & .23 \\ 
			hard   & .9 & .20 & .12 & .02 & .03 & .03 & .03 & .61 & .35 \\ 
			weak   & .5 & .16 & .10 & .00 & .02 & .03 & .04 & .66 & .40 \\ 
			weak   & .9 & .39 & .24 & .04 & .06 & .05 & .06 & 1.04 & .66 \\  \hline
			
			& &  \multicolumn{8}{c}{Normal design, Exponential decay} \\ \hline
			hard   & .5 & .09 & .05 & .00 & .01 & .01 & .02 & .36 & .22 \\ 
			hard   & .9 & .09 & .05 & .00 & .01 & .01 & .02 & .38 & .23 \\ 
			weak   & .5 & .14 & .09 & .00 & .02 & .02 & .04 & .72 & .35 \\ 
			weak   & .9 & .14 & .09 & .00 & .02 & .02 & .04 & .74 & .37 \\  \hline
			
			& &  \multicolumn{8}{c}{Normal design, Equal correlation} \\ \hline
			hard   & .5 & .27 & .17 & .07 & .05 & .06 & .05 & .49 & .76 \\ 
			hard   & .9 & .75 & .46 & .13 & .10 & .12 & .09 & 1.03 & 3.45 \\ 
			weak   & .5 & .49 & .34 & .15 & .12 & .12 & .10 & .87 & 1.56 \\ 
			weak   & .9 & 1.36 & .83 & .22 & .18 & .23 & .17 & 1.86 & 6.44 \\   \hline
			
			& &  \multicolumn{8}{c}{$t_2$ design, Toeplitz} \\ \hline
			hard   & .5 & .53 & .24 & .02 & .03 & .01 & .02 & .54 & .51 \\ 
			hard   & .9 & .56 & .31 & .04 & .05 & .02 & .03 & .60 & .43 \\ 
			weak   & .5 & .88 & .39 & .03 & .05 & .02 & .03 & .90 & .90 \\ 
			weak   & .9 & 1.86 & .83 & .06 & .08 & .02 & .04 & 1.33 & 1.12 \\ \hline
			
			& &  \multicolumn{8}{c}{fMRI design} \\ \hline
			$\beta^0$ & SNR & \rBLPR{} & \pBLPR{} & rBlassoOLS & pBlassoOLS & rBlasso & pBlasso & LDPE & JM  \\ \hline
			hard   & 1  & .83 & .5 & .08 & .09 & .08 & .08 & 1.40 & 1.17 \\ 
			hard   & 5 & .37 & .23 & .04 & .05 & .05 & .05 & .63 & .58 \\
			hard   & 10 & .26 & .16 & .03 & .03 & .04 & .04 & .44 & .42 \\ 
			weak   & 1  & 1.63 & 1.01 & .19 & .18 & .15 & .15 & 2.54 & 2.21 \\ 
			weak   & 5  & .75 & .50 & .12 & .12 & .10 & .10 & 1.13 & 1.13 \\
			weak   & 10 & .52 & .36 & .07 & .09 & .08 & .08 & .80 & .81 \\ 
			\hline			
			
		\end{tabular}
	\end{center}
\end{table}

\clearpage

\section{Algorithm}

\begin{algorithm}[htp]
	\caption{\hspace{0.2cm} $K$-fold cross validation based on lasso+ols: cv(lasso+ols)}
	\label{alg:cv}	
	\begin{algorithmic}[1]\vspace{0.2cm}
		\REQUIRE
		Design matrix $X$, response $Y$, a sequence of tuning parameter values $\lambda_{1},\ldots,\lambda_{J}$, and number of folds $K$.\vspace{0.2cm}
		\ENSURE
		The optimal tuning parameter selected by cv(lasso+ols): $\lambda_{optimal}$. \vspace{0.2cm}
		\STATE Randomly divide the data $z=(X,Y)$ into $K$ roughly equal folds $\left\{ z_k,k=1,\ldots,K \right\};$\label{code:fram:extract}
		\STATE For each $k=1,\ldots,K$, denote $\hat S^{(k)}(\lambda_{0}) = \emptyset $ and $\hat \beta^{(k)}_\textnormal{lasso+ols}(\lambda_{0}) = 0$.
		\begin{itemize}
			\item Fit the model with parameters $\lambda_{j},j=1,\ldots,J$ to the other $K-1$ folds, $z_{-k} = z\setminus z_k$, of the data, giving the lasso solution path $\hat \beta^{(k)}(\lambda_{j}),j=1,\ldots,J$, and compute the sets of selected covariates on the path $$\hat S^{(k)}(\lambda_{j}) = \left\{l:  \hat \beta^{(k)}_l(\lambda_{j}) \neq 0 \right\},  \ \textnormal{for} \ j=1,\ldots,J;$$
			\item For each $j=1,\ldots,J$, compute the lasso+ols estimator:
			{\small
			\begin{equation}
			\hat \beta^{(k)}_\textnormal{lasso+ols}(\lambda_{j}) = \left\{
			\begin{aligned}
			& \argmin_{\beta: \ \beta_j=0, \  j \notin \hat S^{(k)}(\lambda_{j}) } \left\{ \frac{1}{2|z_{-k}|} \sum_{i \in z_{-k}}(y_i-x_i^T\beta)^2 \right\},  \ \ \ \textnormal{if} \ \ \ \hat S^{(k)}(\lambda_{j}) \neq  \hat S^{(k)}(\lambda_{j-1}), \\
			& \hat \beta^{(k)}_\textnormal{lasso+ols}(\lambda_{j-1}), \ \ \  \textnormal{otherwise};
			\end{aligned}
			\right. \nonumber
			\end{equation}
			}
			\item Compute the prediction error $PE^{(k)}$ on the $k$th fold of the data:
			\[ PE^{(k)}(\lambda_{j}) =  \frac{1}{|z_k|} \sum_{i \in z_k} \left( y_i - x_i^T \hat \beta^{(k)}_\textnormal{lasso+ols}(\lambda_{j}) \right)^2;\]
		\end{itemize}		
		\STATE Compute cross validated error $CVE(\lambda_{j})$, $j=1,\ldots,J$:
		\[ CVE(\lambda_{j}) = \frac{1}{K} \sum_{k=1}^{K} PE^{(k)}(\lambda_{j}); \]
		\STATE Compute the optimal $\lambda$ selected by cv(lasso+ols): $ \lambda_{optimal} = \argmin_{\lambda_{j}:\ j=1,\ldots,J} CVE(\lambda_{j}); $
		\RETURN $\lambda_{optimal}$.
	\end{algorithmic}
\end{algorithm}

\end{document}